\documentclass[prd,twocolumn,tightenlines,preprintnumbers,showpacs,superscriptaddress,notitlepage,nofootinbib,eqsecnum,floatfix,longbibliography]{revtex4-1}
\usepackage{graphicx}  % needed for figures
\usepackage{dcolumn}   % needed for some tables
\usepackage{bm}        % for math
\usepackage{amssymb}   % for math
\usepackage{standalone}
\usepackage{enumitem}
\usepackage[pdftex]{color}
\usepackage{xcolor}
\usepackage{slashed}
\usepackage{booktabs}
\usepackage{multirow}
\usepackage{amsmath}
\usepackage{bbm}
\usepackage{stackrel}
\usepackage{rotating}
\usepackage{CJKutf8}
\usepackage{pifont}
\usepackage{mathtools}
\usepackage{hyperref}
\hypersetup{
    colorlinks=true,       % false: boxed links; true: colored links
    linkcolor=blue,          % color of internal links
    citecolor=blue,        % color of links to bibliography
    filecolor=blue,      % color of file links
    urlcolor=blue           % color of external links
}
\usepackage{simplewick}
\usepackage{float} % Forces placement of figures
\usepackage{tikz} % adds /foreach (looping) command
\usepackage{xspace}

\def\tomub{{1.2050(87)^s(16)^\chi(46)^a(00)^V(21)^{\rm phys}(61)^M}}
\def\toub{{0.1422(10)^s(02)^\chi(05)^a(00)^V(02)^{\rm phys}(07)^M}}
\def\timub{{1.2052(77)^s(14)^\chi(46)^a(00)^V(21)^{\rm phys}(10)^M}}
\def\tiub{{0.1422(09)^s(02)^\chi(05)^a(00)^V(02)^{\rm phys}(01)^M}}

\def\womub{{1.4481(81)^s(14)^\chi(46)^a(00)^V(26)^{\rm phys}(11)^M}}
\def\woub{{}0.1709(10)^s(02)^\chi(05)^a(00)^V(03)^{\rm phys}(01)^M}
\def\wimub{{1.4485(82)^s(15)^\chi(44)^a(00)^V(25)^{\rm phys}(18)^M}}
\def\wiub{{0.1709(10)^s(02)^\chi(05)^a(00)^V(03)^{\rm phys}(02)^M}}

\def\tmub{{1.2051(82)^s(15)^\chi(46)^a(00)^V(21)^{\rm phys}(61)^M}}
\def\tm{{1.205(12)}}
\def\tfmub{{0.1422(09)^s(02)^\chi(05)^a(00)^V(02)^{\rm phys}(07)^M}}
\def\tfm{{0.1422(14)}}

\def\wmub{{1.4483(82)^s(15)^\chi(45)^a(00)^V(26)^{\rm phys}(18)^M}}
\def\wm{{1.4483(97)}}
\def\wfmub{{0.1709(10)^{s}(02)^{\chi}(05)^{a}(00)^V(03)^\text{phys}(02)^{M}}}
\def\wfm{{0.1709(11)}}

\def\tomuborig{{1.2107(78)^s(14)^\chi(40)^a(00)^V(22)^{\rm phys}(07)^M}}
\def\touborig{{0.1429(09)^s(02)^\chi(05)^a(00)^V(03)^{\rm phys}(01)^M}}
\def\timuborig{{1.2052(73)^s(13)^\chi(40)^a(00)^V(22)^{\rm phys}(08)^M}}
\def\tiuborig{{0.1422(09)^s(02)^\chi(05)^a(00)^V(03)^{\rm phys}(01)^M}}

\def\womuborig{{1.4481(81)^s(14)^\chi(46)^a(00)^V(26)^{\rm phys}(11)^M}}
\def\wouborig{{}0.1709(10)^s(02)^\chi(05)^a(00)^V(03)^{\rm phys}(01)^M}
\def\wimuborig{{1.4467(83)^s(15)^\chi(44)^a(00)^V(24)^{\rm phys}(10)^M}}
\def\wiuborig{{0.1707(10)^s(02)^\chi(05)^a(00)^V(03)^{\rm phys}(01)^M}}

\def\eqref#1{{(\ref{#1})}}
\newcommand{\eqnref}[1]{Eq.~\eqref{#1}}

\newcommand{\figref}[1]{Fig.~\ref{#1}}

\newcommand{\secref}[1]{Sec.~\ref{#1}}

\newcommand{\tabref}[1]{Table~\ref{#1}}

\definecolor{kngrey}{HTML}{A6AAA9}
\definecolor{knred}{HTML}{EC5D57}
\definecolor{knorange}{HTML}{F39019}
\definecolor{knyellow}{HTML}{F5D328}
\definecolor{kngreen}{HTML}{70BF41}
\definecolor{knblue}{HTML}{51A7F9}
\definecolor{knpurple}{HTML}{B36AE2}

\def\mc#1{{\mathcal #1}}

\def\nxlo#1{{N$^{#1}$LO}}

\DeclareMathOperator{\tr}{tr}

\def\one{\ensuremath{\mathbbm{1}}}

\def\a{{\alpha}}
\def\b{{\beta}}
\def\d{{\delta}}
\def\D{{\Delta}}
\def\e{{\epsilon}}
\def\g{{\gamma}}
\def\G{{\Gamma}}

\def\l{{\lambda}}
\def\L{{\Lambda}}

\def\O{{\Omega}}

\def\s{{\sigma}}

\def\tr{\text{tr}}

\def\chroma{\texttt{Chroma}\xspace}
\def\quda{\texttt{QUDA}\xspace}

\newcommand{\ithems}{
    Interdisciplinary Theoretical and Mathematical Sciences Program (iTHEMS),
    RIKEN, 2-1 Hirosawa,
    Wako, Saitama 351-0198, Japan
}
\newcommand{\cmu}{
    Department of Physics, Carnegie Mellon University,
    Pittsburgh, Pennsylvania 15213, USA
}
\newcommand{\arithmer}{
    Arithmer Inc., R\&D Headquarters,
    Minato, Tokyo 106-6040, Japan
}
\newcommand{\glasgow}{
 School of Physics and Astronomy,
    University of Glasgow,
    Glasgow G12 8QQ, UK
 }
\newcommand{\jlabt}{
	Theory Center,
	Thomas Jefferson National Accelerator Facility,
	Newport News, VA 23606, USA
	}

\newcommand{\julich}{
 Institut f\"{u}r Kernphysik and Institute for Advanced Simulation,
 Forschungszentrum J\"{u}lich, 54245 J\"{u}lich Germany
 }

\newcommand{\lblnsd}{
    Nuclear Science Division,
    Lawrence Berkeley National Laboratory,
	Berkeley, CA 94720, USA
	}

\newcommand{\llnl}{
	Physics Division,
	Lawrence Livermore National Laboratory,
	Livermore, CA 94550, USA
	}
\newcommand{\llnldesign}{
	Design Physics Division,
	Lawrence Livermore National Laboratory,
	Livermore, CA 94550, USA
	}
\newcommand{\nvidia}{
    NVIDIA Corporation,
    2701 San Tomas Expressway, Santa Clara, CA 95050, USA
    }

\newcommand{\ucb}{
	Department of Physics,
	University of California,
	Berkeley, CA 94720, USA
	}
\newcommand{\umd}{
	Department of Physics,
	University of Maryland,
	College Park, MD 20742, USA
}
\newcommand{\unc}{
	Department of Physics and Astronomy,
	University of North Carolina,
	Chapel Hill, NC 27516-3255, USA
	}

\newcommand{\wandm}{
	Department of Physics,
	The College of William \& Mary,
	Williamsburg, VA 23187, USA
	}
\newcommand{\ucr}{
        Escuela de F\'isca,
	Universidad de Costa Rica,
	11501 San Jos\'e, Costa Rica
	}

\newcount\hour \newcount\hourminute \newcount\minute
\hour=\time \divide \hour by 60
\hourminute=\hour \multiply \hourminute by 60
\minute=\time \advance \minute by -\hourminute
\newcommand{\mydate}{\ \today \ - \number\hour :\number\minute}

\begin{document}

\title{Scale setting the M{\"o}bius domain wall fermion on gradient-flowed HISQ action\\
using the omega baryon mass and the gradient-flow scales $t_0$ and $w_0$\\
}

\author{Nolan~Miller}
\affiliation{\unc}

\author{Logan Carpenter}
\affiliation{\cmu}

\author{Evan~Berkowitz}
\affiliation{\umd}
\affiliation{\julich}

\author{Chia~Cheng~Chang (\begin{CJK*}{UTF8}{bsmi}張家丞\end{CJK*})}
\affiliation{\ithems}
\affiliation{\lblnsd}
\affiliation{\ucb}

\author{Ben~H\"orz}
\affiliation{\lblnsd}

\author{Dean~Howarth}
\affiliation{\llnl}
\affiliation{\lblnsd}

\author{Henry~Monge-Camacho}
\affiliation{\ucr}
\affiliation{\unc}

\author{Enrico~Rinaldi}
\affiliation{\arithmer}
\affiliation{\ithems}

\author{David~A.~Brantley}
\affiliation{\llnl}

\author{Christopher~K\"orber}
\affiliation{\ucb}
\affiliation{\lblnsd}

\author{Chris Bouchard}
\affiliation{\glasgow}

\author{M.A.~Clark}
\affiliation{\nvidia}

\author{Arjun~Singh~Gambhir}
\affiliation{\llnldesign}
\affiliation{\lblnsd}

\author{Christopher~J.~Monahan}
\affiliation{\wandm}
\affiliation{\jlabt}

\author{Amy~Nicholson}
\affiliation{\unc}
\affiliation{\lblnsd}

\author{Pavlos~Vranas}
\affiliation{\llnl}
\affiliation{\lblnsd}

\author{Andr\'{e}~Walker-Loud}
\affiliation{\lblnsd}
\affiliation{\llnl}
\affiliation{\ucb}

\date{\mydate}

\begin{abstract}
We report on a subpercent scale determination using the omega baryon mass and gradient-flow methods.
The calculations are performed on 22 ensembles of $N_f=2+1+1$ highly improved, rooted staggered sea-quark configurations generated by the MILC and CalLat Collaborations.  The valence quark action used is M\"obius domain wall fermions solved on these configurations after a gradient-flow smearing is applied with a flowtime of $t_{\rm gf}=1$ in lattice units.  The ensembles span four lattice spacings in the range $0.06 \lesssim a \lesssim 0.15$~fm, six pion masses in the range $130 \lesssim m_\pi \lesssim 400$~MeV and multiple lattice volumes.  On each ensemble, the gradient-flow scales $t_0/a^2$ and $w_0/a$ and the omega baryon mass $a m_\O$ are computed.  The dimensionless product of these quantities is then extrapolated to the continuum and infinite volume limits and interpolated to the physical light, strange and charm quark mass point in the isospin limit, resulting in the determination of $\sqrt{t_0}=\tfm$~fm and $w_0 = \wfm$~fm with all sources of statistical and systematic uncertainty accounted for.  The dominant uncertainty in both results is the stochastic uncertainty, though for $\sqrt{t_0}$ there are comparable continuum extrapolation uncertainties.  For $w_0$, there is a clear path for a few-per-mille uncertainty just through improved stochastic precision, as recently obtained by the Budapest-Marseille-Wuppertal Collaboration.
\end{abstract}

\preprint{LLNL-JRNL-816949, RIKEN-iTHEMS-Report-20, JLAB-THY-20-3290}

\maketitle
%%%%%%%%%%%%%%%%%%%%%%%%%%%%%%%%%
%%%%%%%%%%%%%%%%%%%%%%%%%%%%%%%%%
%%%%%%%%%%%%%%%%%%%%%%%%%%%%%%%%%
%\tableofcontents
%----------------------------------------------------------
%    INTRODUCTION
\section{Introduction \label{sec:intro}}

Lattice QCD (LQCD) has become a prominent theoretical tool for calculations of hadronic quantities, and many calculations have reached a level of precision to be able to supplement and/or complement experimental determinations~\cite{Aoki:2019cca}. Precision calculations of Standard Model processes, for example, are crucial input for experimental tests of fundamental symmetries in searches for new physics.

Lattice calculations receive only dimensionless bare parameters as input, so the output is inherently dimensionless. In some cases, dimensionless quantities or ratios of quantities may be directly computed without the need to determine any dimensionful scale.
Calculations of $g_A$ and $F_K/F_\pi$ are examples for which a precise scale setting is not necessary to make a precise, final prediction.
However, there are many quantities for which a precise scale setting is desirable, such as the hadron spectrum, the nucleon axial radius, the hadronic contribution to the muon $g-2$~\cite{Aoyama:2020ynm} and many others.

In these cases, a quantity which is dimensionful (after multiplying or dividing by an appropriate power of the lattice spacing) is calculated and compared to experiment, following extrapolations to the physical point in lattice spacing, volume, and pion mass. Because the precision of any calculations of further dimensionful quantities is limited by the statistical and systematic uncertainties of this scale setting, quantities which have low stochastic noise and mild light quark mass dependence, such as the omega baryon mass $m_{\Omega}$, are preferred. The lattice spacing on each ensemble may then be determined by comparing the quantity calculated on a given ensemble to the continuum value.

However, the most precise quantities one may calculate are not necessarily accessible experimentally. For example,
the Sommer scale $r_0$~\cite{Sommer:1993ce} has been one of the most commonly used scales.  This scale requires a determination of the heavy-quark potential which is susceptible to fitting systematic uncertainties.
More recently, the gradient flow scales $t_0$~\cite{Luscher:2010iy} and $w_0$~\cite{Borsanyi:2012zs} have been used for a more precise determination of the lattice spacing~\cite{Deuzeman:2012jw,Bruno:2013gha,Dowdall:2013rya,Bornyakov:2015eaa,Bazavov:2014pvz,Blum:2014tka,Bazavov:2015yea,Bruno:2016plf,Borsanyi:2020mff}. In this case, a well-controlled extrapolation of these quantities to the physical point is also necessary.

In this paper we present a precision scale setting for our mixed lattice action~\cite{Berkowitz:2017opd} which uses $N_f=2+1+1$ highly improved, rooted staggered sea-quark (HISQ) configurations generated by the MILC~\cite{Bazavov:2012xda} and CalLat Collaborations and M\"obius domain wall fermions for the valence sector. We compute the dimensionless products $\sqrt{t_0} m_\O$ and $m_{\O} w_0$ on each ensemble and extrapolate them to the physical point resulting in the determinations
\begin{align}\label{eq:t0_result}
\sqrt{t_0} m_\O &= \tmub
\nonumber\\ &= \tm\, ,
\nonumber\\
\frac{\sqrt{t_0}}{\rm fm} &= \tfmub
\nonumber\\ &= \tfm\, ,
\end{align}
\begin{align}\label{eq:w0_result}
w_0 m_\O &= \wmub
\nonumber\\
    &= \wm
\nonumber\\
\frac{w_0}{\rm fm} &= \wfmub
\nonumber\\
    &= \wfm\, ,
\end{align}
with the statistical~($s$), chiral~($\chi$), continuum-limit~($a$), infinite volume ($V$), physical-point (phys), and model selection uncertainties~($M$).

We then perform an interpolation of the values of $t_0/a^2$ and $w_0/a$ to the physical quark-mass limit and extrapolation to infinite volume which allows us to provide a precise, quark mass independent scale setting for each lattice spacing, with our final results in \tabref{tab:a_fm}.
In the following sections we provide details of our lattice setup, our methods for extrapolation, and our results with uncertainty breakdown. We conclude with a discussion in the final section.

%----------------------------------------------------------
%    Lattice Calculation
\section{Details of the lattice calculation \label{sec:LQCD}}

%----------------------------------------------------------
%    Lattice Action
\subsection{MDWF on gradient-flowed HISQ \label{sec:action}}

%-------------------------------------------------------------------------------
% Table of lattice Parameters
\begingroup \squeezetable
\begin{table*}
\caption{\label{tab:lattice_params}
Input parameters for our lattice action.  The abbreviated ensemble name~\cite{Bhattacharya:2015wna} indicates the approximate lattice spacing in fm and pion mass in MeV.
The S, L, XL which come after an ensemble name denote a relatively small, large and extra-large volume with respect to $m_\pi L=4$.
}
\begin{ruledtabular}
\begin{tabular}{lccclll|cccllllcrr}
Ensemble& $\beta$& $N_{\rm cfg}$& Volume& $am_l$& $am_s$& $am_c$&
    $L_5/a$& $aM_5$& $b_5, c_5$& $am_l^{\rm val}$& $am^{\rm res}_l\hspace{-0.2em}\times\hspace{-0.2em}10^{4}$&
    $am_s^{\rm val}$& $am^{\rm res}_s\hspace{-0.2em}\times\hspace{-0.2em}10^{4}$& $\s$ &$N$& $N_{\rm src}$\\
%&&& [$L^3\times T$]&&&&&&&&[$\times10^{4}$]&&[$\times10^{4}$]\\
\hline
% a15
%a15m400\footnotemark&
a15m400\footnote{Additional ensembles generated by CalLat using the MILC code.  The m350 and m400 ensembles were made on the Vulcan supercomputer at LLNL while the a12m310XL, a12m180L, a15m135XL, a09m135, and a06m310L ensembles were made on the Sierra and Lassen supercomputers at LLNL and the Summit supercomputer at OLCF using \texttt{QUDA}~\cite{Clark:2009wm,Babich:2011np}. These configurations are available to any interested party upon request, and will be available for easy anonymous downloading---hopefully soon.}&
    5.80& 1000& $16^3\times48$& 0.0217& 0.065& 0.838&
    12& 1.3& 1.50, 0.50& 0.0278& 9.365(87)& 0.0902& 6.937(63)& 3.0& 30& 8\\
a15m350\footnotemark[1]&
    5.80& 1000& $16^3\times48$& 0.0166& 0.065& 0.838&
    12& 1.3& 1.50, 0.50& 0.0206& 9.416(90)& 0.0902& 6.688(62)& 3.0& 30& 16\\
a15m310&
    5.80& 1000& $16^3\times48$& 0.013& 0.065& 0.838&
    12& 1.3& 1.50, 0.50& 0.0158& 9.563(67)& 0.0902& 6.640(44)& 4.2& 45& 24\\
a15m310L\footnotemark[1]&
    5.80& 1000& $24^3\times48$& 0.013& 0.065& 0.838&
    12& 1.3& 1.50, 0.50& 0.0158& 9.581(50)& 0.0902& 6.581(37)& 4.2& 45& 4\\
a15m220&
    5.80& 1000& $24^3\times48$& 0.0064& 0.064& 0.828&
    16& 1.3& 1.75, 0.75& 0.00712& 5.736(38)& 0.0902& 3.890(25)& 4.5& 60& 16\\
a15m135XL\footnotemark[1]&
    5.80& 1000& $48^3\times64$& 0.002426& 0.06730& 0.8447&
    24& 1.3& 2.25, 1.25& 0.00237& 2.706(08)& 0.0945& 1.860(09)& 3.0& 30& 32\\
\hline
% a12
a12m400\footnotemark[1]&
    6.00& 1000& $24^3\times64$& 0.0170& 0.0509& 0.635&
    8&  1.2& 1.25, 0.25& 0.0219& 7.337(50)& 0.0693& 5.129(35)& 3.0& 30& 8\\
a12m350\footnotemark[1]&
    6.00& 1000& $24^3\times64$& 0.0130& 0.0509& 0.635&
    8&  1.2& 1.25, 0.25& 0.0166& 7.579(52)& 0.0693& 5.062(34)& 3.0& 30& 8\\
a12m310&
    6.00& 1053& $24^3\times64$& 0.0102& 0.0509& 0.635&
    8&   1.2& 1.25, 0.25& 0.0126& 7.702(52)& 0.0693& 4.950(35)& 3.0& 30& 8\\
a12m310XL\footnotemark[1]&
    6.00& 1000& $48^3\times64$& 0.0102& 0.0509& 0.635&
    8&   1.2& 1.25, 0.25& 0.0126& 7.728(22)& 0.0693& 4.927(21)& 3.0& 30& 8\\
a12m220S&
    6.00& 1000& $24^4\times64$& 0.00507& 0.0507& 0.628&
    12&  1.2& 1.50, 0.50& 0.00600& 3.990(42)& 0.0693& 2.390(24)& 6.0& 90& 4\\
a12m220&
    6.00& 1000& $32^3\times64$& 0.00507& 0.0507& 0.628&
    12&  1.2& 1.50, 0.50& 0.00600& 4.050(20)& 0.0693& 2.364(15)& 6.0& 90& 4\\
a12m220ms&
    6.00& 1000& $32^3\times64$& 0.00507& 0.0304& 0.628&
    12&  1.2& 1.50, 0.50& 0.00600& 3.819(26)& 0.0415& 2.705(20)& 6.0& 90& 8\\
a12m220L&
    6.00& 1000& $40^3\times64$& 0.00507& 0.0507& 0.628&
    12&  1.2& 1.50, 0.50& 0.00600& 4.040(26)& 0.0693& 2.361(19)& 6.0& 90& 4\\
a12m180L\footnotemark[1]&
    6.00& 1000& $48^3\times64$& 0.00339& 0.0507& 0.628&
    14&  1.2& 1.75, 0.75& 0.00380& 3.038(13)& 0.0693& 1.888(11)& 3.0& 30& 16 \\
a12m130&
    6.00& 1000& $48^3\times64$& 0.00184& 0.0507& 0.628&
    20& 1.2& 2.00, 1.00& 0.00195& 1.642(09)& 0.0693& 0.945(08)& 3.0& 30& 32\\
\hline
% a09
a09m400\footnotemark[1]&
    6.30& 1201& $32^3\times64$& 0.0124& 0.037& 0.44&
    6&  1.1& 1.25, 0.25& 0.0160& 2.532(23)& 0.0491& 1.957(17)& 3.5& 45& 8\\
a09m350\footnotemark[1]&
    6.30& 1201& $32^3\times64$& 0.00945& 0.037& 0.44&
    6&  1.1& 1.25, 0.25& 0.0121& 2.560(24)& 0.0491& 1.899(16)& 3.5& 45& 8\\
a09m310&
    6.30& 780& $32^3\times96$& 0.0074& 0.037& 0.44&
    6&  1.1& 1.25, 0.25& 0.00951& 2.694(26)& 0.0491& 1.912(15)& 6.7& 167& 8\\
a09m220&
    6.30& 1001& $48^3\times96$& 0.00363& 0.0363& 0.43&
    8&  1.1& 1.25, 0.25& 0.00449& 1.659(13)& 0.0491& 0.834(07)& 8.0& 150& 6\\
a09m135\footnotemark[1]&
    6.30& 1010& $64^3\times96$& 0.001326& 0.03636& 0.4313&
    12& 1.1& 1.50, 0.50& 0.00152& 0.938(06)& 0.04735& 0.418(04)& 3.5& 45& 16\\
\hline
% a06
a06m310L\footnotemark[1]&
    6.72& 1000& $72^3\times96$& 0.0048& 0.024& 0.286&
    6&  1.0& 1.25, 0.25& 0.00617& 0.225(03)& 0.0309& 0.165(02)& 3.5& 45& 8
\end{tabular}
\end{ruledtabular}
\end{table*}
\endgroup
%-------------------------------------------------------------------------------

The lattice action we use is the mixed-action~\cite{Renner:2004ck,Bar:2002nr} with M\"obius~\cite{Brower:2012vk} domain wall fermions~\cite{Kaplan:1992bt,Shamir:1993zy,Furman:1994ky} solved on $N_f=2+1+1$ highly improved staggered quarks~\cite{Follana:2006rc} after they are gradient-flow smeared~\cite{Narayanan:2006rf,Luscher:2011bx,Luscher:2013cpa} (corresponding to an infinitesimal stout-smearing procedure~\cite{Morningstar:2003gk}) to a flow-time of $t_{\rm gf}/a^2=1$~\cite{Berkowitz:2017opd}.
The choice to hold the flow-time fixed in lattice units is important to ensure that as the continuum limit is taken, effects arising from finite flow-time also extrapolate to zero.

This action has been used to compute the nucleon axial coupling, $g_A$, with a 1\% total uncertainty~\cite{Bouchard:2016heu,Berkowitz:2017gql,Chang:2018uxx,Berkowitz:2018gqe}, the $\pi^-\rightarrow\pi^+$ matrix elements relevant to neutrinoless double beta-decay~\cite{Nicholson:2018mwc} and most recently, $F_K/F_\pi$~\cite{Miller:2020xhy}.  Our calculation of $F_K/F_\pi$ was obtained with a total uncertainty of 0.4\% which provides an important benchmark for our action, as the result is consistent with other determinations in the literature~\cite{Follana:2007uv,Blossier:2009bx,Durr:2010hr,Dowdall:2013rya,Blum:2014tka,Carrasco:2014poa,Bazavov:2014wgs,Durr:2016ulb,Bornyakov:2016dzn,Bazavov:2017lyh} (and the FLAG average~\cite{Aoki:2019cca}), and also contributes to the test of the universality of lattice QCD results in the continuum limit.

Our plan to compute the axial and other elastic form factors of the nucleon with this mixed-action, as well as other quantities, leads to a desire to have a scale setting with sufficiently small uncertainty that it does not increase the final uncertainty of such quantities.
It has been previously observed that both $w_0$~\cite{Borsanyi:2012zs,Bazavov:2015yea} and the omega baryon mass~\cite{Allton:2008pn,WalkerLoud:2008bp,Lin:2008pr,Durr:2008zz,Alexandrou:2009qu,Alexandrou:2014sha,Brantley:2016our,Borsanyi:2020mff} have mild quark mass dependence and that they can be determined with high statistical precision with relatively low computational cost.  The input parameters of our action on all ensembles are provided in \tabref{tab:lattice_params}.

%----------------------------------------------------------
%    Correlation functions
\subsection{Correlation function construction and analysis \label{sec:two_points}}

%-------------------------------------------------------------------------------
% Omega stability
\begin{figure*}
\includegraphics[width=0.49\textwidth]{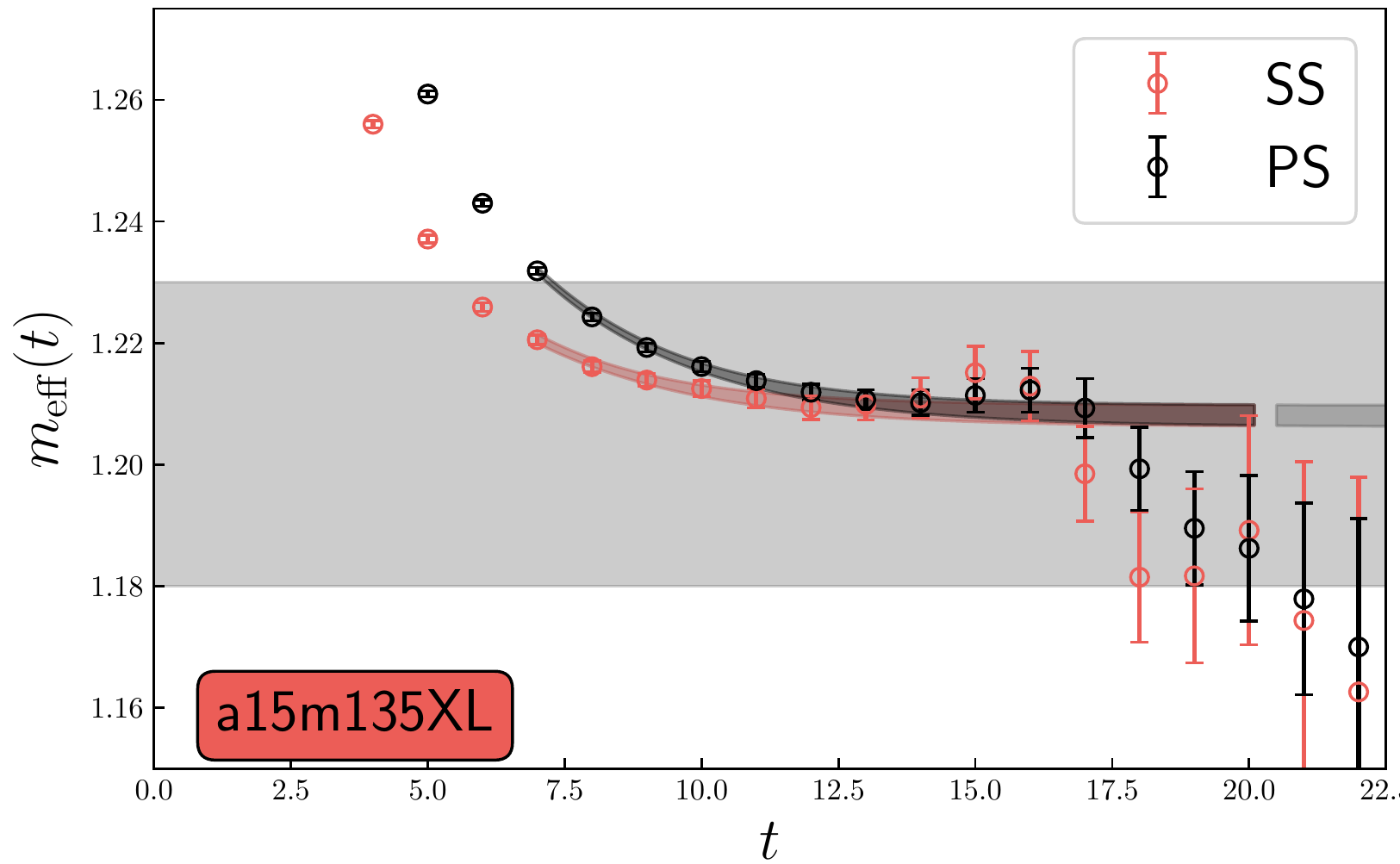}
\includegraphics[width=0.49\textwidth]{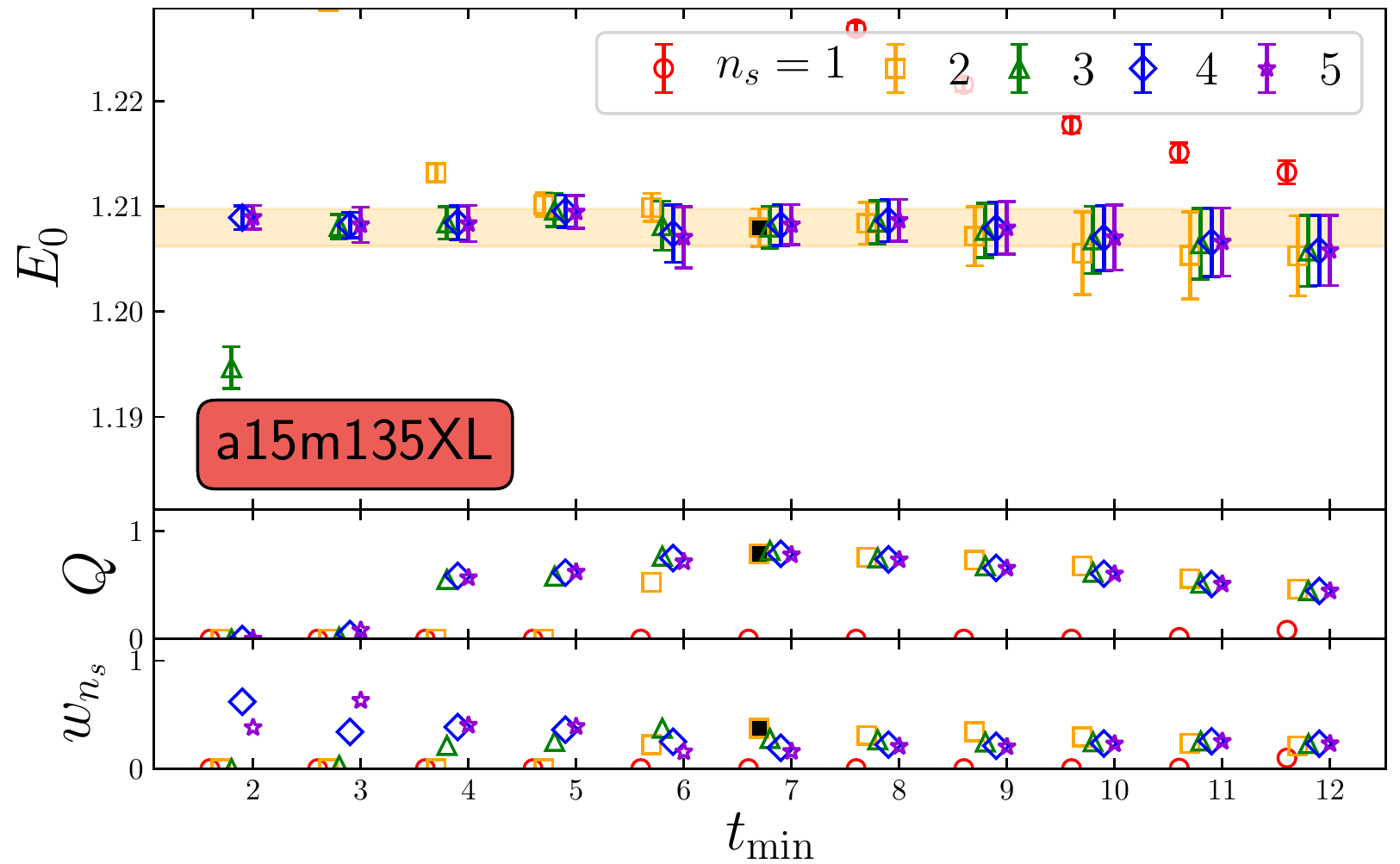}
\includegraphics[width=0.49\textwidth]{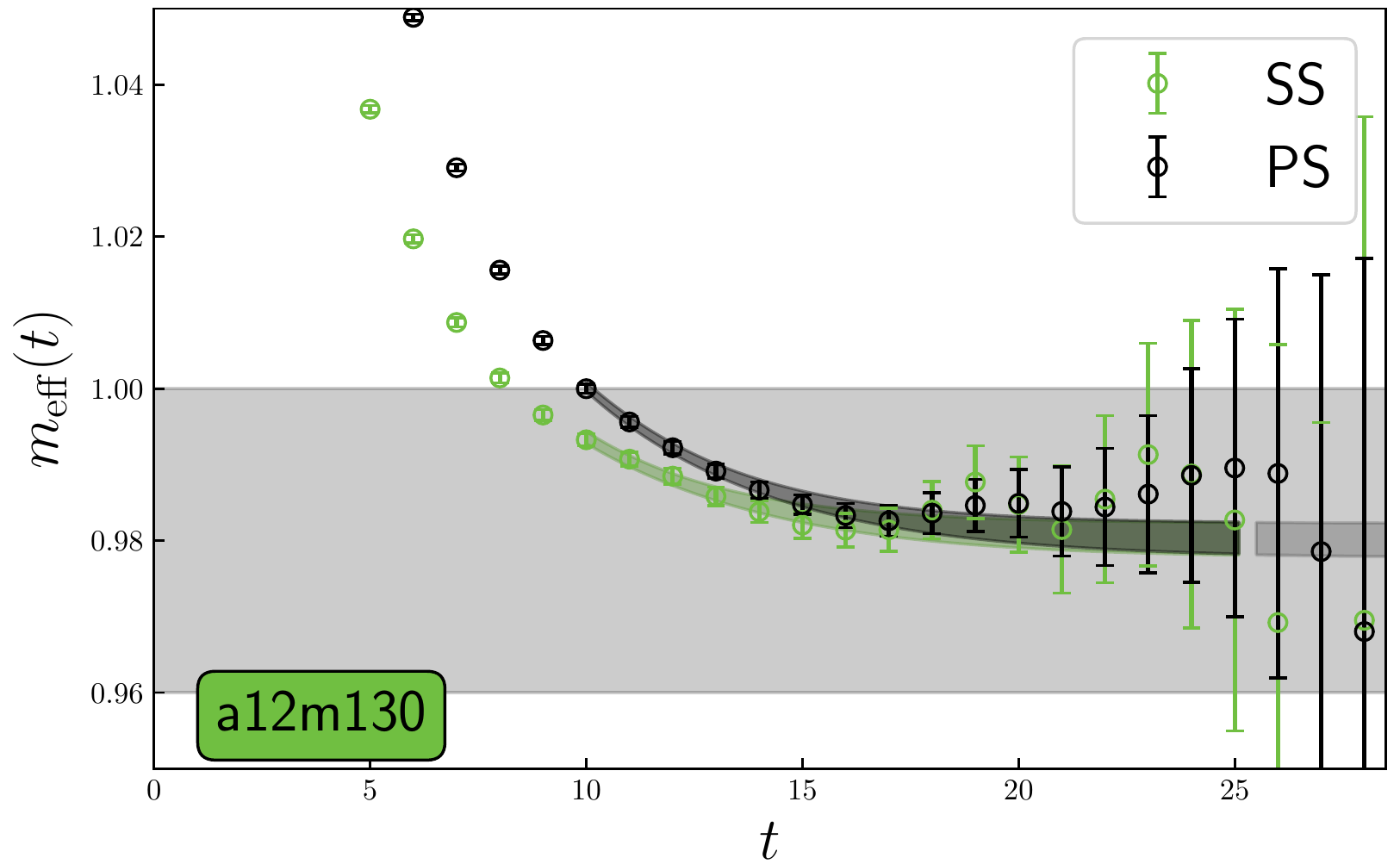}
\includegraphics[width=0.49\textwidth]{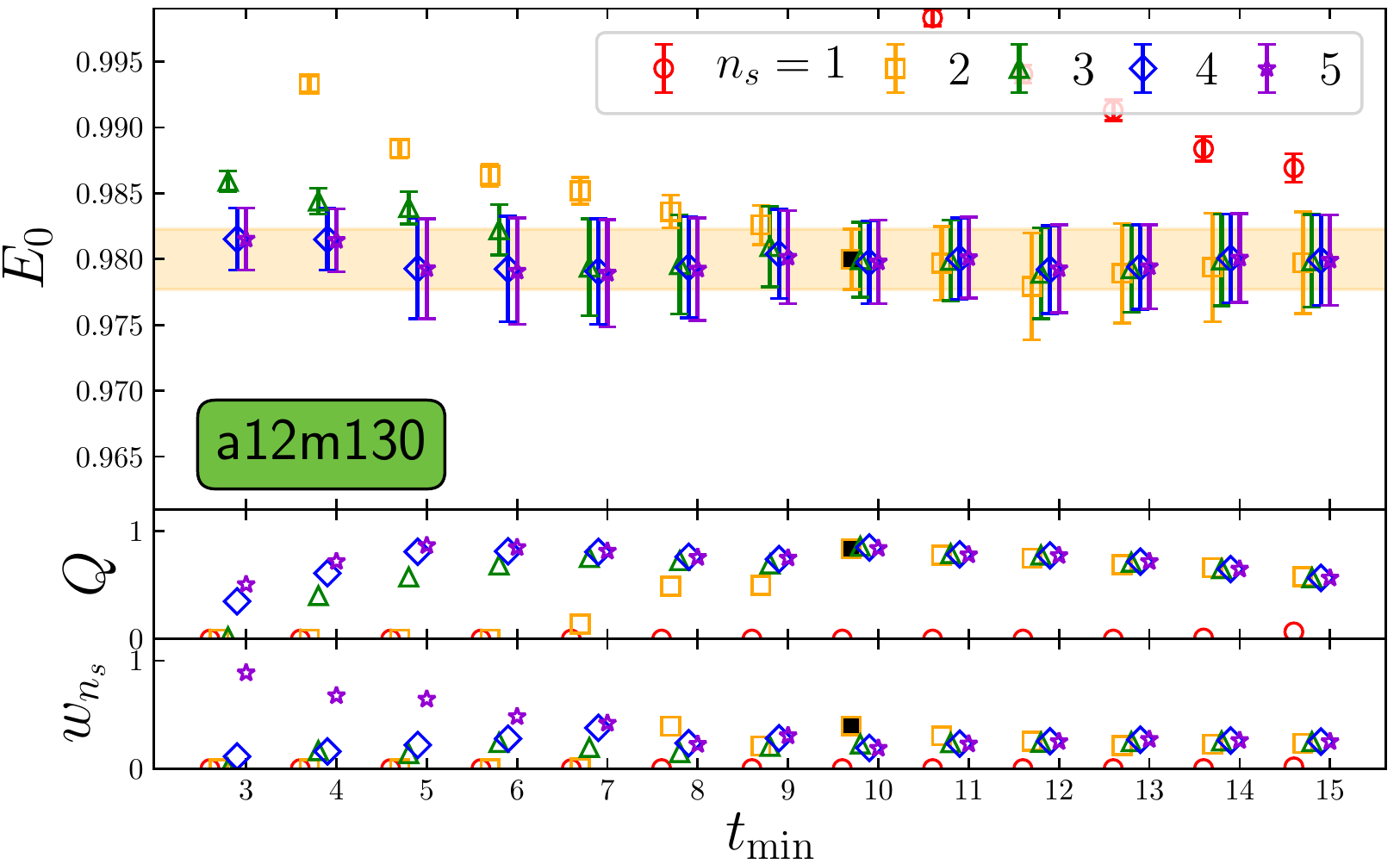}
\includegraphics[width=0.49\textwidth]{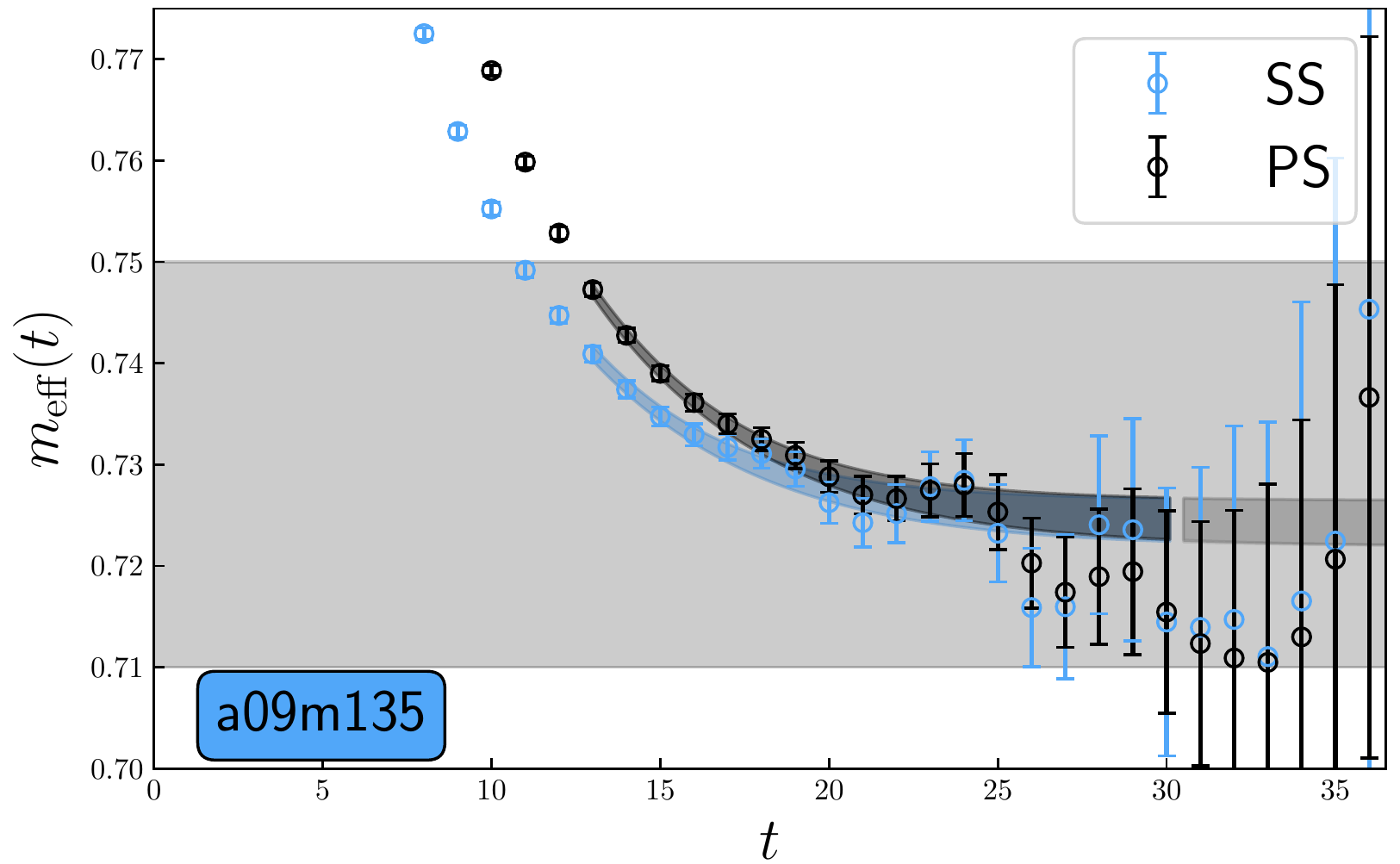}
\includegraphics[width=0.49\textwidth]{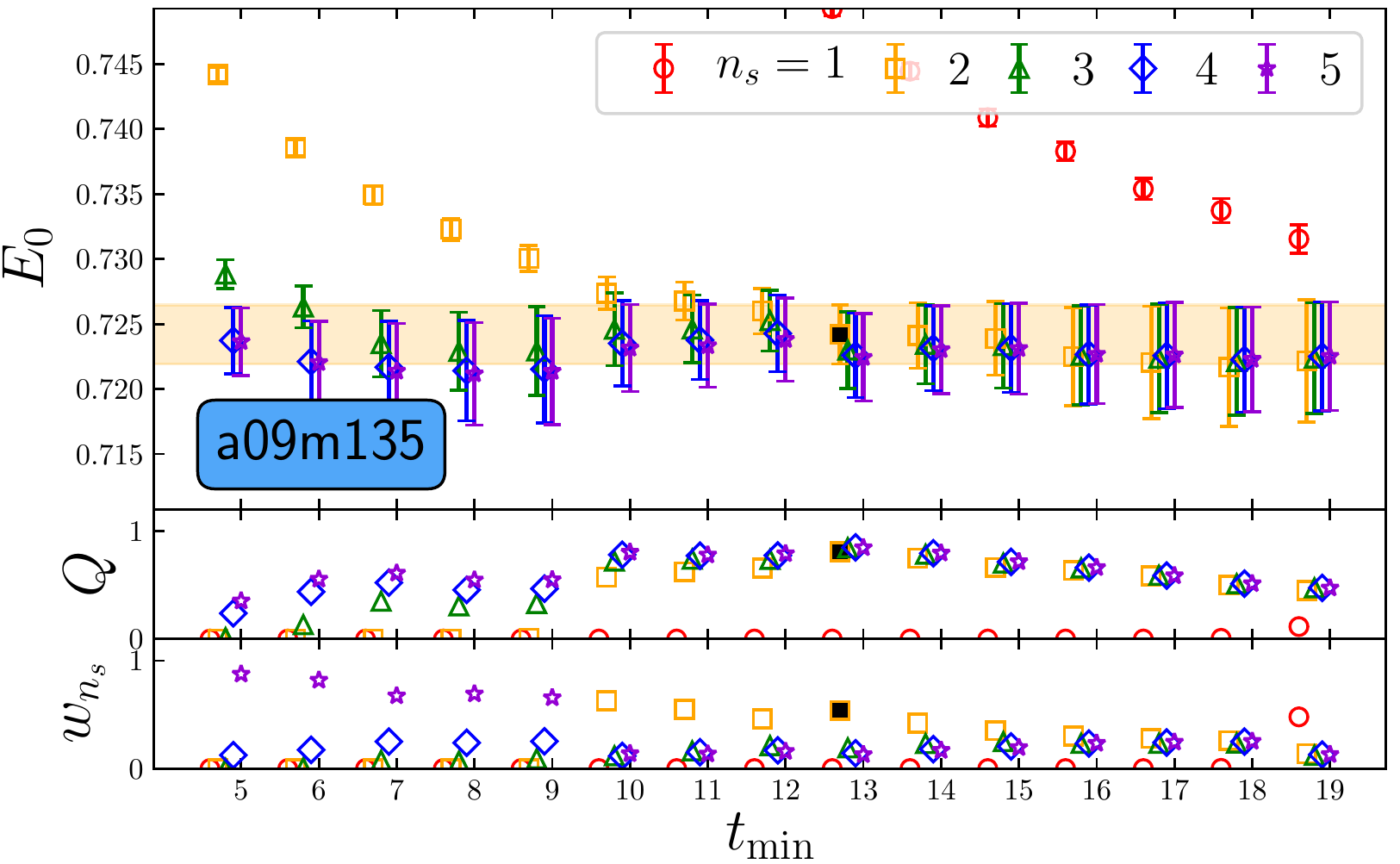}
\caption{\label{fig:stability_m135}
Stability plots of the ground state omega baryon mass on the three physical pion mass ensembles.
The left plots show the effective mass data (in lattice units) and reconstructed effective mass from the chosen fit for both the SS and PS correlation functions.  The dark gray and colored band are displayed for the region of time used in the analysis and an extrapolation beyond $t_{\rm max}$ is shown after a short break in the fit band.
The horizontal gray band is the prior used for the ground state mass.
The right plots show the corresponding value of $E_0$ as a function of $t_{\rm min}$ and the number of states $n_s$ used in the analysis, as well as the corresponding $Q$ value and relative weight as a function of $n_s$ for a given $t_{\rm min}$, where the weight is set by the Bayes Factor.  See Appendix \ref{app:reweighting} for more detail on the selection of the final fit.
The chosen fit is denoted with a filled black symbol and the horizontal band is the value of $E_0$ from the chosen fit.
The $y$-range of the upper panel of the stability plots is equal to the prior of the ground state energy (the horizontal gray band in the left plot).
}
\end{figure*}
%-------------------------------------------------------------------------------

For the scale setting computation, we have to determine four or five quantities on each ensemble, the pion, kaon and omega masses, the gradient-flow scale $w_0$ and the pion decay constant $F_\pi$.
For $m_\pi$, $m_K$ and $F_\pi$, we take the values from our $F_K/F_\pi$ computation for the 18 ensembles in common.  For the four new ensembles in this work (a15m310L, a12m310XL, a12m220ms, a12m180L), we follow the same analysis strategy described in Ref.~\cite{Miller:2020xhy}.

The a12m220ms ensemble is identical to a12m220 except that the strange quark mass is roughly 60\% of the physical value rather than being near the physical value.  The a15m310L ensemble has identical input parameters as the a15m310 ensemble but $L=24$ (3.6 fm) instead of $L=16$ (2.4 fm), while the a12m310XL ensemble is identical to the a12m310 ensemble but with $L=48$ (5.8 fm) instead of $L=24$ (2.9 fm).
The a12m180L and a12m310XL ensembles have a lattice volume that is the same size as a12m130 but pion masses of roughly $m_\pi\simeq 180$ and 310~MeV.  These new ensembles provide important lever arms for the various extrapolations.  The a12m220ms provides a unique lever arm for varying the strange quark mass significantly from its physical value, the a15m310L and a12m310XL provide other pion masses where we can perform a volume study and the a12m180L ensemble provides an additional light pion mass ensemble to help with the physical pion mass extrapolation.  The first of these is important for this scale setting while the latter three will be more important for future work.

The omega baryon correlation functions are constructed similarly to the pion and kaon.
A source for the propagator is constructed with the gauge invariant Gaussian smearing routine in \chroma~\cite{Edwards:2004sx} (\texttt{GAUGE\_INV\_GAUSSIAN}).  Then, correlation functions are constructed using both a point sink as well as the same gauge invariant Gaussian smearing routine with the same parameters as the source.  The values of the ``smearing width'' ($\s$) and the number of iterations ($N$) used to approximate the exponential smearing profile are provided in \tabref{tab:lattice_params}.  The correlation functions constructed with the point sink are referred to as PS and those with the smeared sink as SS.

Local spin wave functions are constructed following Refs.~\cite{Basak:2005aq,Basak:2005ir}.  Both positive- and negative-parity omega-baryon correlation functions are constructed with the upper and lower spin components of the quark propagators in the Dirac basis.  The negative-parity correlation functions are time-reversed with an appropriate sign flip of the correlation function, effectively doubling the statistics with no extra inversions.  The four different spin projections of the omega are averaged as well to produce the final spin and parity averaged two-point correlation functions.

The reader will notice that the values of $\s$ and $N$ do not follow an obvious pattern.  This is because in our first computations of $g_A$~\cite{Bouchard:2016heu,Berkowitz:2017gql}, we applied an ``aggressive'' smearing with a larger value of $\s$ and correspondingly larger number of iterations, which led to a large suppression of excited states, but also showed evidence of ``over smearing'' such that the non-positive-definite PS correlation functions displayed symptoms of having a relatively large negative overlap factor for excited states (there were wiggles in the PS effective masses).  In a subsequent paper studying the two-nucleon system on the a12m350 ensemble~\cite{Berkowitz:2017smo}, where we utilized matrix Prony~\cite{Beane:2009kya} to form linear combinations of PS and SS nucleons to construct a ``calm'' nucleon which is ground-state dominated earlier in time, we observed that using a milder smearing with smaller width and fewer iterations provided a much more stable extraction of the ground state and did not show signs of large negative overlap factors.  Hence, many but not all of the ensembles have been rerun with our improved choices of $\s$ and $N$.  We have observed the choice $\s=3.0$ and $N=30$ works well for the a15 and a12 ensembles and that $\s=3.5$ with $N=45$ works well for the a09 and a06 ensembles.

In order to determine the omega baryon mass on each ensemble, we perform a stability analysis of the extracted ground state mass as a function of $t_{\rm min}$ used in the fit as well as the number of states used in the analysis.  The correlation functions are analyzed in a Bayesian framework with constraints~\cite{Lepage:2001ym}.
We choose normally distributed priors for the ground-state energy and all overlap factors, and log-normal distributions for excited-state energy priors. The ground-state energy and overlap factors are motivated by the plateau values of the effective masses with the priors taken to be roughly 10 times larger than the stochastic uncertainty of the respective effective mass data in the plateau region.  The excited-state energy splittings are set to the value of two pion masses with a width allowing for fluctuations down to one pion mass within one standard deviation.

In \figref{fig:stability_m135} we show sample extractions of the ground state mass on our three physical pion mass ensembles.
In the left plot, we show the effective mass data from the two correlation functions.
%The horizontal gray band is the prior of the ground state mass.  The dark gray and colored band are the reconstructed effective mass based upon the chosen fit and shown in the window of time used in the analysis.  The light gray band at late time is the extrapolated effective mass from this fit.
%The right panels show the value of the ground state mass, the corresponding fit quality $Q$ and the relative weight $w_{n_s}$ as a function of the number of states ($n_s$) used in the analysis and the minimum time separation used, $t_{\rm min}$.
The weights are normalized on a given time slice by the largest Bayes factor at that $t_{\rm min}$ value.  We have not implemented a more thorough algorithm to weight fits against each other that utilize different amounts of data, as described for example in Ref.~\cite{Jay:2020jkz}.
Rather, we have chosen a fit for a given ensemble (the filled black symbol in the right panels highlighted by the horizontal colored band) that has a good fit quality, the maximum or near maximum relative weight, and consistency with the late-time data.
We tried to optimize this choice over all ensembles simultaneously, with $t_{\rm min}$ held nearly fixed for a given lattice spacing, rather than hand-picking the optimal fit on each ensemble separately, in order to
minimize the possible bias introduced by analysis choices.
Good fits are obtained on all ensembles with $n_s=2$, simplifying the model function and reducing the chance of overfitting the correlation functions, which is most relevant on ensembles with the more aggressive choices of smearing parameters.
In Appendix \ref{app:stability}, we show the corresponding stability plots for all remaining ensembles.
In \tabref{tab:lattice_fits} we show the resulting values of $am_\O$ on all ensembles used in this work.

%----------------------------------------------------------
%    Correlation functions
\subsection{Calculation of $t_0$ and $w_0$ \label{sec:w0}}
\begin{figure}
\includegraphics[width=\columnwidth]{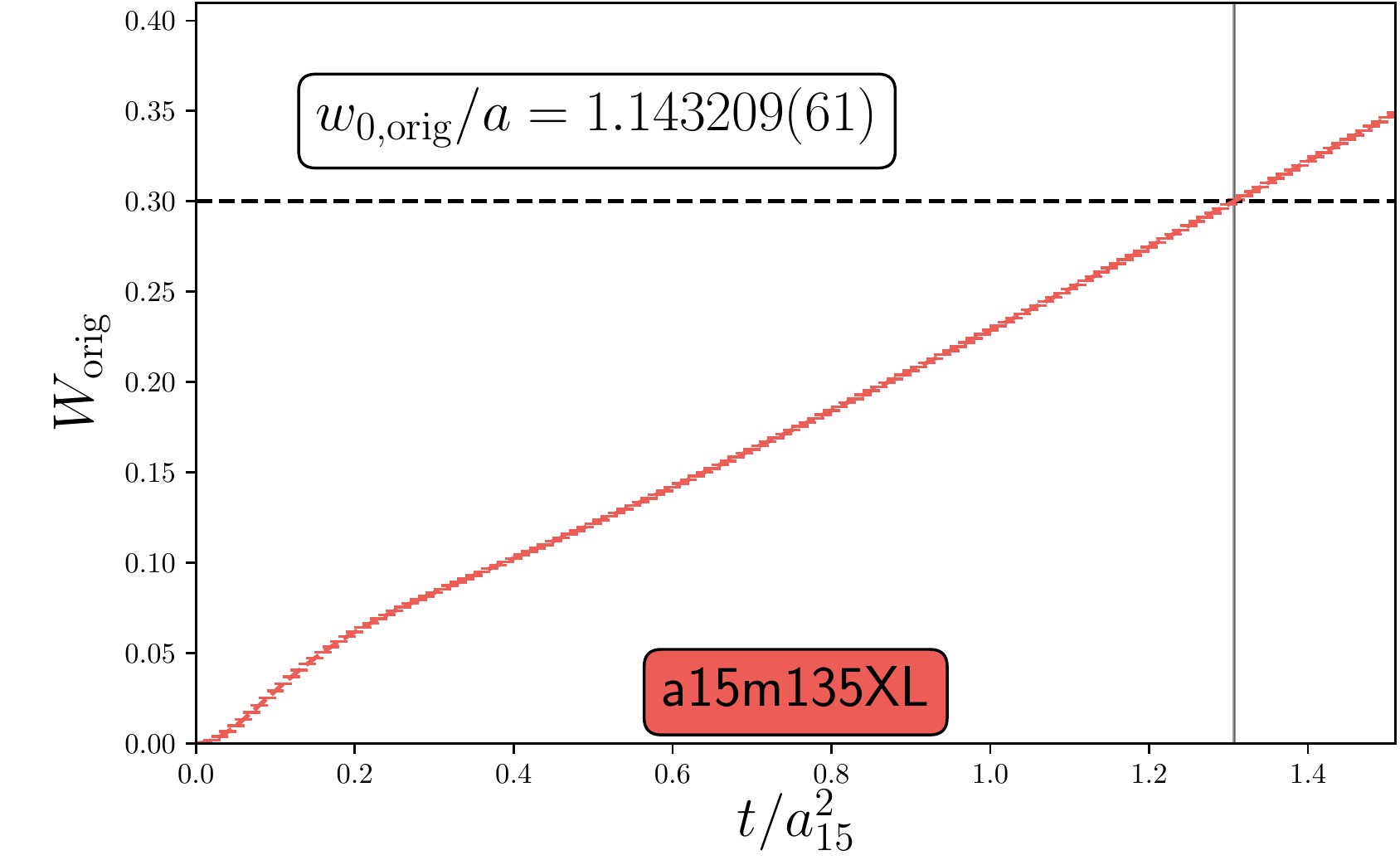}
\includegraphics[width=\columnwidth]{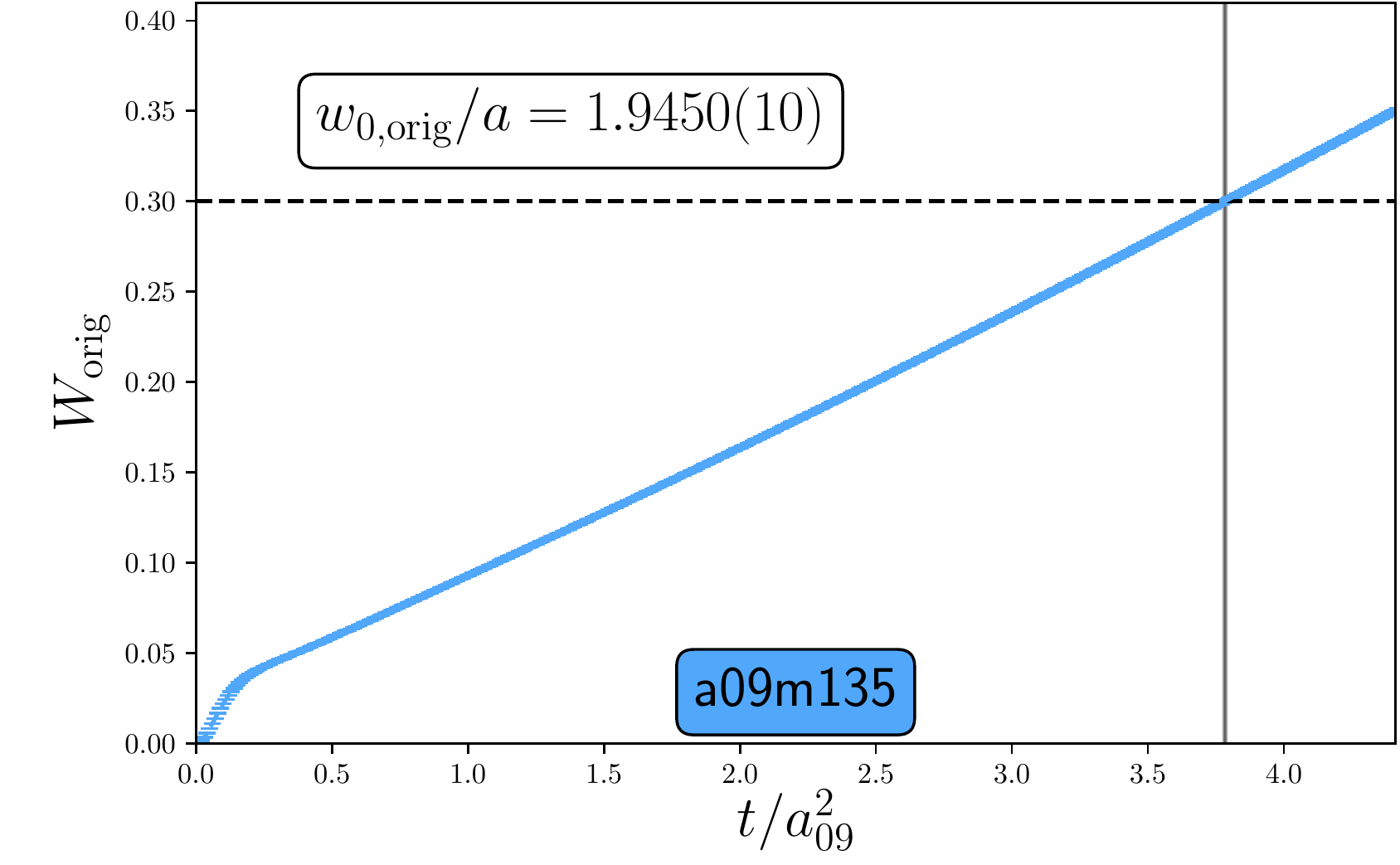}
\caption{\label{fig:w0_m135}
Determination of $w_{0,\rm orig}$ on the two new physical pion mass ensembles.}
\end{figure}

In order to efficiently compute the value of $t_0/a^2$ and $w_0/a$ on each ensemble, we have implemented the Symanzik flow in the \quda library~\cite{quda:code,Clark:2009wm,Babich:2011np}.  We used the tree-level improved action and the symmetric, cloverleaf definition of the field-strength tensor, following the MILC implementation~\cite{milc:code}.  We used a step size of $\e=0.01$ in the Runge-Kutta algorithm proposed by L{\"u}scher~\cite{Luscher:2010iy}, which leads to negligibly small integration errors.
The scales $t_0$ and $w_0$ are defined by the equations
\begin{align}
t^2 \langle E(t) \rangle \Big|_{t=t_{0,\rm orig}} = 0.3\, ,
\nonumber\\
W_{\rm orig}\equiv t \frac{d}{dt} (t^2 \langle E(t)\rangle)\Big|_{t=w_{0,\rm orig}^2} = 0.3\, ,
\end{align}
where $\langle E(t)\rangle$ is the gluonic action density at flow time $t$.
In \figref{fig:w0_m135}, we show the determination of the $w_{0,\rm orig}/a$ on the two physical pion mass ensembles that we have generated.
The uncertainties are determined by observing a saturation of the uncertainty as the bin-size is increased when binning the results from configurations close in Monte Carlo time.  These uncertainties were cross-checked with an autocorrelation study using the $\G$-method~\cite{Wolff:2003sm} implemented in the \texttt{unew} Python package~\cite{DePalma:2017lww}.

%-------------------------------------------------------------------------------
% Omega stability
%-------------------------------------------------------------------------------

%-------------------------------------------------------------------------------
% Table of lattice Parameters
\begin{table*}
\caption{\label{tab:lattice_fits}
The omega baryon mass ($m_\O$) and gradient flow scales ($t_{0,\rm orig}$, $t_{0,\rm imp}$, $w_{0,\rm orig}$ and $w_{0,\rm imp}$) determined on each ensemble are listed as well as their dimensionless products.
Additionally, in the bottom panel, we list the parameters used to control the physical point extrapolation:
$l_\O^2 = m_\pi^2 / m_\O^2$, $s_\O^2 = (2m_K^2 - m_\pi^2)/m_\O^2$, $l_F^2 = m_\pi^2 / (4\pi F_\pi)^2$, $s_F^2 = (2m_K^2 - m_\pi^2)/(4\pi F_\pi)^2$, $m_\pi L$, $\e_a = a / (2w_{0,\rm orig})$ and $\a_S$.  The values of $\a_S$ are taken from Table III of Ref.~\cite{Bazavov:2015yea} which were determined with a heavy-quark potential method~\cite{Davies:2002mv}.
An HDF5 file is provided with this publication which includes the resulting bootstrap samples of all these quantities which can be used to construct the correlated uncertainties.
}
\begin{ruledtabular}
\begin{tabular}{lccccccccc}
Ensemble& $am_\O$& $t_{0,\rm orig}/a^2$& $t_{0,\rm imp}/a^2$& $\sqrt{t_{0,\rm orig}}m_\O$& $\sqrt{t_{0,\rm imp}}m_\O$& $w_{0,\rm orig}/a$& $w_{0,\rm imp}/a$& $w_{0,\rm orig} m_\O$& $w_{0,\rm imp} m_\O$ \\
\hline
  a15m400& 1.2437(43)& 1.1905(15)& 0.9563(11)& 1.3570(47)& 1.2162(42)& 1.1053(12)& 1.0868(14)& 1.3747(49)& 1.3516(50)\\
  a15m350& 1.2331(31)& 1.2032(14)& 0.9660(11)& 1.3526(35)& 1.2120(32)& 1.1154(11)& 1.0988(13)& 1.3754(37)& 1.3550(38)\\
 a15m310L& 1.2287(31)& 1.2111(07)& 0.9720(05)& 1.3521(34)& 1.2113(30)& 1.1219(06)& 1.1066(07)& 1.3784(35)& 1.3596(35)\\
  a15m310& 1.2312(36)& 1.2121(09)& 0.9727(07)& 1.3555(40)& 1.2143(36)& 1.1230(07)& 1.1079(09)& 1.3826(41)& 1.3640(41)\\
  a15m220& 1.2068(26)& 1.2298(08)& 0.9861(07)& 1.3383(30)& 1.1984(27)& 1.1378(07)& 1.1255(08)& 1.3731(31)& 1.3582(31)\\
a15m135XL& 1.2081(19)& 1.2350(04)& 0.9897(03)& 1.3425(21)& 1.2018(19)& 1.1432(03)& 1.1319(04)& 1.3811(22)& 1.3674(22)\\
\hline
  a12m400& 1.0279(25)& 1.6942(11)& 1.4172(10)& 1.3380(33)& 1.2237(30)& 1.3616(07)& 1.3636(08)& 1.3997(35)& 1.4017(35)\\
  a12m350& 1.0139(26)& 1.7091(14)& 1.4298(12)& 1.3255(35)& 1.2124(32)& 1.3728(10)& 1.3755(11)& 1.3918(38)& 1.3946(38)\\
a12m310XL& 1.0072(41)& 1.7221(07)& 1.4407(06)& 1.3217(54)& 1.2089(50)& 1.3831(05)& 1.3865(06)& 1.3930(58)& 1.3964(58)\\
  a12m310& 1.0112(32)& 1.7213(19)& 1.4398(17)& 1.3267(42)& 1.2134(39)& 1.3830(13)& 1.3863(12)& 1.3985(46)& 1.4019(46)\\
a12m220ms& 0.8896(92)& 1.7891(15)& 1.4977(13)& 1.190(12)\phantom{0}& 1.089(11)\phantom{0}& 1.4339(12)& 1.4406(12)& 1.276(13)\phantom{0}& 1.282(13)\phantom{0}\\
 a12m220S& 0.9970(26)& 1.7466(20)& 1.4614(17)& 1.3177(35)& 1.2053(32)& 1.4021(15)& 1.4069(16)& 1.3980(39)& 1.4027(39)\\
 a12m220L& 0.9944(30)& 1.7489(09)& 1.4633(08)& 1.3150(40)& 1.2028(37)& 1.4041(06)& 1.4090(07)& 1.3962(43)& 1.4010(43)\\
  a12m220& 0.9924(60)& 1.7498(14)& 1.4641(12)& 1.3127(80)& 1.2007(73)& 1.4047(10)& 1.4096(11)& 1.3940(85)& 1.3988(86)\\
 a12m180L& 0.9924(26)& 1.7553(05)& 1.4686(05)& 1.3148(35)& 1.2026(32)& 1.4093(05)& 1.4145(05)& 1.3985(38)& 1.4037(38)\\
  a12m130& 0.9801(26)& 1.7628(07)& 1.4749(06)& 1.3013(34)& 1.1903(31)& 1.4155(05)& 1.4211(06)& 1.3873(37)& 1.3928(37)\\
\hline
  a09m400& 0.7716(23)& 2.9158(42)& 2.6040(33)& 1.3176(41)& 1.2451(38)& 1.8602(26)& 1.8686(27)& 1.4353(48)& 1.4418(48)\\
  a09m350& 0.7561(35)& 2.9455(37)& 2.6301(34)& 1.2977(61)& 1.2262(58)& 1.8810(25)& 1.8900(26)& 1.4222(69)& 1.4291(70)\\
  a09m310& 0.7543(36)& 2.9698(32)& 2.6521(29)& 1.2998(63)& 1.2283(59)& 1.8970(22)& 1.9066(22)& 1.4308(71)& 1.4381(71)\\
  a09m220& 0.7377(30)& 3.0172(16)& 2.6952(15)& 1.2814(53)& 1.2111(50)& 1.9282(12)& 1.9388(12)& 1.4224(59)& 1.4302(60)\\
  a09m135& 0.7244(25)& 3.0390(12)& 2.7147(11)& 1.2629(44)& 1.1936(42)& 1.9450(10)& 1.9563(11)& 1.4091(50)& 1.4172(50)\\
\hline
 a06m310L& 0.5069(21)& 6.4079(45)& 6.0606(44)& 1.2830(54)& 1.2478(52)& 2.8958(20)& 2.9053(20)& 1.4678(62)& 1.4726(62)\\
\end{tabular}
\begin{tabular}{lccccccc}
    Ensemble& $l_F^2$& $s_F^2$& $l_\O^2$& $s_\O^2$& $m_\pi L$& $\e_a^2$& $\a_S$\\
    \hline
      a15m400& 0.09216(33)& 0.2747(10)& 0.05928(42)& 0.1767(12)& 4.85& 0.20462(44)& 0.58801\\
      a15m350& 0.07505(28)& 0.2915(09)& 0.04609(25)& 0.1790(09)& 4.24& 0.20096(39)& 0.58801\\
     a15m310L& 0.06018(23)& 0.2984(11)& 0.03630(18)& 0.1800(09)& 5.62& 0.19864(20)& 0.58801\\
      a15m310& 0.06223(17)& 0.3035(09)& 0.03675(23)& 0.1792(11)& 3.78& 0.19825(26)& 0.58801\\
      a15m220& 0.03269(11)& 0.3253(09)& 0.01877(09)& 0.1868(08)& 3.97& 0.19311(23)& 0.58801\\
    a15m135XL& 0.01319(05)& 0.3609(11)& 0.00726(03)& 0.1986(06)& 4.94& 0.19129(10)& 0.58801\\
    \hline
      a12m400& 0.08889(30)& 0.2648(10)& 0.05610(28)& 0.1671(08)& 5.84& 0.13484(15)& 0.53796\\
      a12m350& 0.07307(37)& 0.2810(13)& 0.04454(24)& 0.1713(09)& 5.14& 0.13266(19)& 0.53796\\
    a12m310XL& 0.05904(23)& 0.2909(12)& 0.03506(29)& 0.1727(14)& 9.05& 0.13069(10)& 0.53796\\
      a12m310& 0.05984(25)& 0.2933(12)& 0.03482(23)& 0.1707(11)& 4.53& 0.13071(24)& 0.53796\\
    a12m220ms& 0.03400(16)& 0.2000(09)& 0.02229(46)& 0.1311(27)& 4.25& 0.12159(20)& 0.53796\\
     a12m220S& 0.03384(19)& 0.3210(19)& 0.01849(13)& 0.1754(09)& 3.25& 0.12716(27)& 0.53796\\
     a12m220L& 0.03289(15)& 0.3195(14)& 0.01816(12)& 0.1765(11)& 5.36& 0.12681(11)& 0.53796\\
      a12m220& 0.03314(15)& 0.3202(15)& 0.01831(23)& 0.1769(22)& 4.30& 0.12670(18)& 0.53796\\
     a12m180L& 0.02277(09)& 0.3319(13)& 0.01220(07)& 0.1779(10)& 5.26& 0.12588(08)& 0.53796\\
      a12m130& 0.01287(08)& 0.3429(14)& 0.00687(05)& 0.1832(10)& 3.90& 0.12477(09)& 0.53796\\
    \hline
      a09m400& 0.08883(32)& 0.2638(09)& 0.05512(34)& 0.1637(10)& 5.80& 0.07225(20)& 0.43356\\
      a09m350& 0.07256(32)& 0.2827(11)& 0.04358(42)& 0.1698(16)& 5.05& 0.07066(19)& 0.43356\\
      a09m310& 0.06051(22)& 0.2946(10)& 0.03481(34)& 0.1695(16)& 4.50& 0.06947(16)& 0.43356\\
      a09m220& 0.03307(14)& 0.3278(13)& 0.01761(15)& 0.1746(14)& 4.70& 0.06724(08)& 0.43356\\
      a09m135& 0.01346(08)& 0.3500(17)& 0.00674(05)& 0.1752(12)& 3.81& 0.06608(07)& 0.43356\\
    \hline
     a06m310L& 0.06141(35)& 0.2993(17)& 0.03481(29)& 0.1696(14)& 6.81& 0.02981(04)& 0.29985\\
\end{tabular}
\end{ruledtabular}
\end{table*}
%-------------------------------------------------------------------------------

Reference~\cite{Fodor:2014cpa} determined the tree-level in lattice perturbation theory improvement coefficients for the determination of these gradient flow scales through $\mathrm{O}(a^8/t^4)$ for various choices of the gauge action, the gradient flow action and the definition of the field-strength tensor.  As defined in Ref.~\cite{Fodor:2014cpa}, we have implemented the Symanzik-Symanzik-Clover (SSC) scheme with the relevant improvement coefficients (see Table 1 of Ref.~\cite{Fodor:2014cpa})
\begin{align}
&C_2 = -\frac{19}{72}\, ,&
&C_4 = \frac{145}{1536}\, ,&
\nonumber\\
&C_6 = -\frac{12871}{276480}\, ,&
&C_8 = \frac{52967}{1769472}\, .&
\end{align}
One can then determine improved scales, $t_{0,\rm imp}$ and $w_{0,\rm imp}$ in which one has perturbatively removed the leading discretization effects in these flow observables,
\begin{align}\label{eq:t0_w0_improved}
\frac{t^2 \langle E(t) \rangle}{1 + \sum_n C_{2n} \frac{a^{2n}}{t^n}} \bigg|_{t=t_{0,\rm imp}} = 0.3\, ,
\nonumber\\
t \frac{d}{dt} \left( \frac{t^2 \langle E(t)\rangle}{1 + \sum_n C_{2n} \frac{a^{2n}}{t^n}}\right)\bigg|_{t=w_{0,\rm imp}^2} = 0.3\, ,
\end{align}

In the present work, we explore using both the original and improved versions of $t_0$ and $w_0$ when performing our scale setting analysis.  For the improved versions, we have implemented the fourth order improvement [up to and including the $C_8 (a^2/t)^4$ correction].
This is the same implementation performed by MILC in Ref.~\cite{Bazavov:2015yea}.

In \tabref{tab:lattice_fits}, we list the values of $t_0/a^2$ and $w_0/a$ for the original and improved definitions as well as the dimensionless products of $\sqrt{t_0}m_\O$ and $w_0 m_\O$.

%----------------------------------------------------------
%    Extrapolation Analysis
\section{Extrapolation Functions}
This work utilizes 22 different ensembles, each with $\mathrm{O}(1000)$ configurations (\tabref{tab:lattice_params}), to control the systematic uncertainties in the LQCD calculation of the scale.  This allows us to address:
\begin{enumerate}[leftmargin=*, font=\itshape]
\item The physical light and strange quark mass limit;
\item The physical charm quark mass limit;
\item The continuum limit;
\item The infinite volume limit.
\end{enumerate}

\subsubsection{Physical light and strange quark mass limit \label{sec:light_strange_mass}}
The ensembles have a range of light quark masses which correspond roughly to $130 \lesssim m_\pi \lesssim 400$~MeV.  We have three lattice spacings at $m_\pi\simeq m_\pi^{\rm phys}$ such that the light quark mass extrapolation is really an interpolation.  On all but one of the 22 ensembles, the strange quark mass is close to its physical value, allowing us to perform a simple interpolation to the physical strange quark mass point.  One ensemble has a strange quark mass of roughly 2/3 its physical value (a12m220ms), allowing us to explore systematics in this strange quark mass interpolation.

\begin{figure*}
\includegraphics[width=0.49\textwidth]{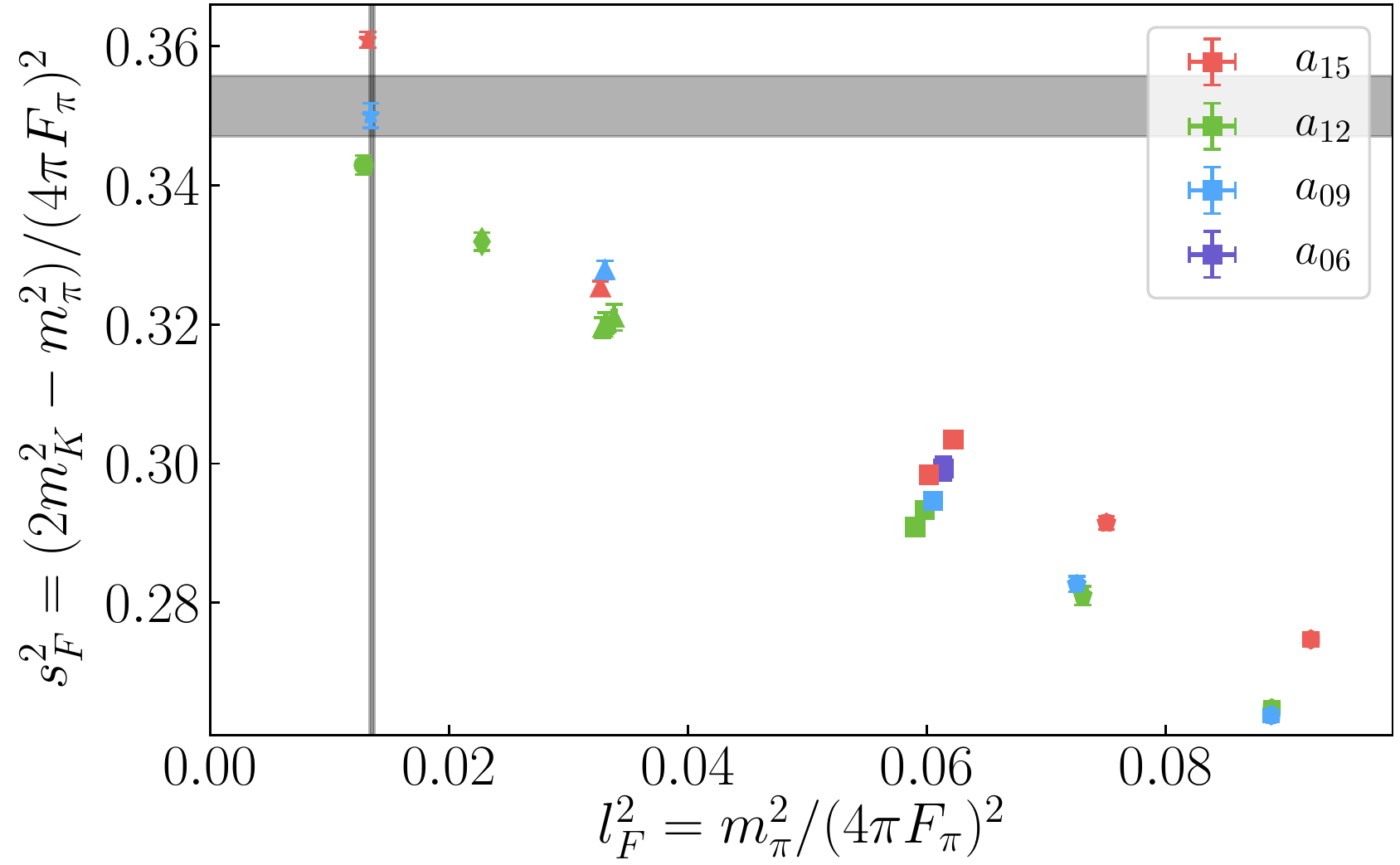}
\includegraphics[width=0.49\textwidth]{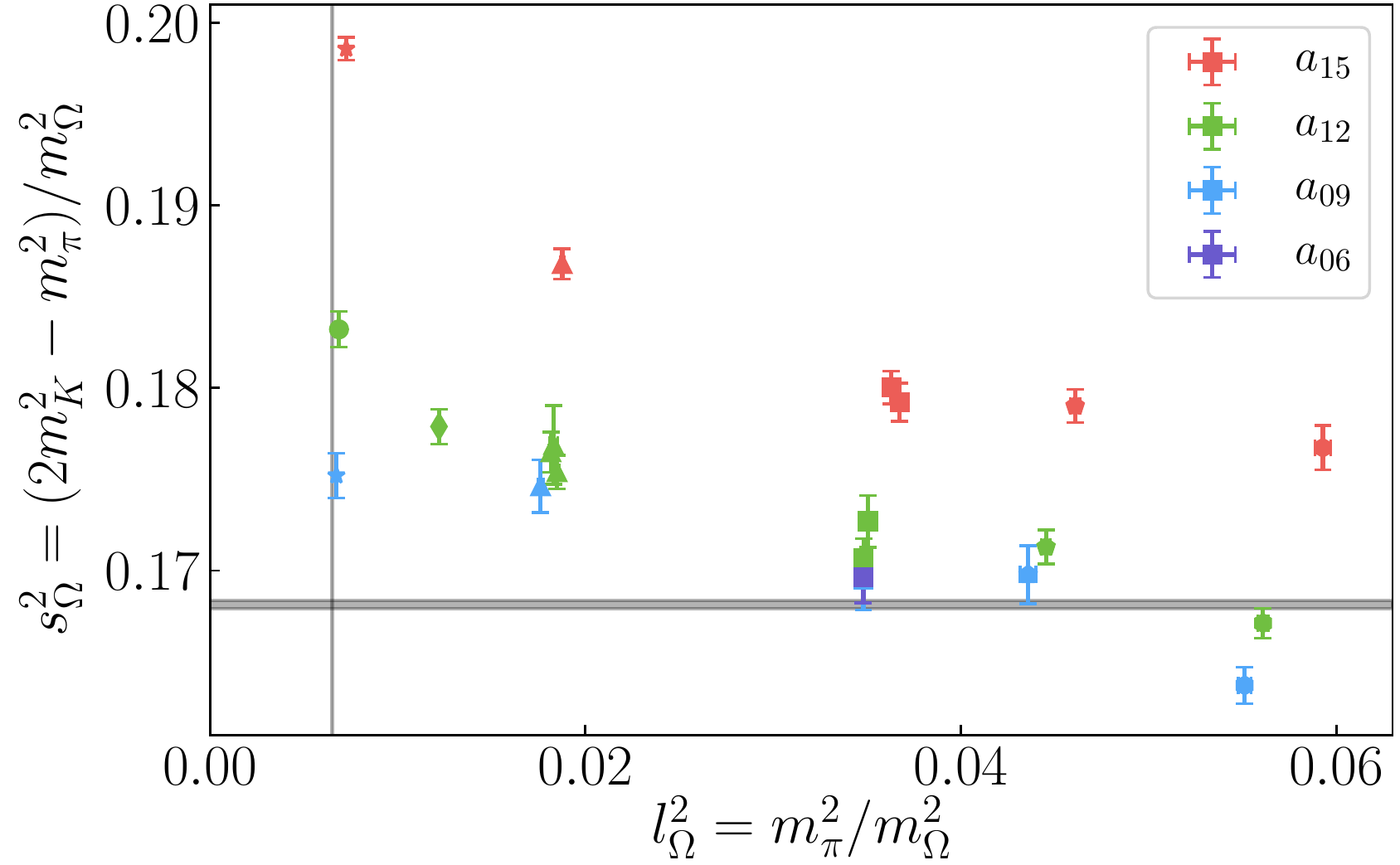}
\caption{\label{fig:ls_space}
The parameter space of $s_F^2$ versus $l_F^2$ (left) and $s_\O^2$ versus $l_\O^2$ used in this calculation.
The vertical and horizontal gray lines represent the physical point defined by \eqnref{eq:phys_point}.
}
\end{figure*}

To parametrize the light and strange quark mass dependence, we utilize two sets of small parameters
\begin{align}\label{eq:ls_F}
&\L=F:&
&l_F^2 = \frac{m_\pi^2}{\L_\chi^2},&
&s_F^2 = \frac{2m_K^2 -m_\pi^2}{\L_\chi^2},&
\\\label{eq:ls_O}
&\L=\O:&
&l_\O^2 = \frac{m_\pi^2}{m_\O^2},&
&s_\O^2 = \frac{2m_K^2 -m_\pi^2}{m_\O^2},&
\end{align}
where we have defined
\begin{equation}\label{eq:Lam_chi}
    \L_\chi \equiv 4\pi F_\pi\, .
\end{equation}
Using the Gell-Mann--Oakes--Renner relation~\cite{GellMann:1968rz}, the numerators in these parameters correspond roughly to the light and strange quark mass, $m_\pi^2 = 2B\hat{m}$ and $2m_K^2 - m_\pi^2 = 2Bm_s$, where $\hat{m} = \frac{1}{2}(m_u+m_d)$.  The first set of parameters, \eqnref{eq:ls_F}, is inspired by $\chi$PT and commonly used as a set of small expansion parameters in extrapolating LQCD results.
The second set of small parameters, \eqnref{eq:ls_O}, is inspired by Ref.~\cite{Lin:2008pr}.
In \figref{fig:ls_space}, we plot the values of these parameters in comparison with the physical point.
Since we are working in the isospin limit in this work, we define the physical point as
\begin{align}\label{eq:phys_point}
m_\pi^{\rm phys} &= \bar{M}_\pi = 134.8(3) \textrm{ MeV},
\nonumber\\
m_K^{\rm phys}   &= \bar{M}_K   = 494.2(3) \textrm{ MeV},
\nonumber\\
F_\pi^{\rm phys} &= F_{\pi^+}^{\rm phys} = 92.07(57) \textrm{ MeV},
\nonumber\\
m_\O^{\rm phys} &= 1672.43(32) \textrm{ MeV}\, ,
\end{align}
with the first three values from the FLAG report~\cite{Aoki:2019cca} and the omega baryon mass from the PDG~\cite{Tanabashi:2018oca}.
The values of $l_{F,\O}$ and $s_{F,\O}$ are given in \tabref{tab:lattice_fits} for all ensembles.

%-------------------------------------------------------------------------------
% Figure of s_O vs l_O and s_F vs l_F
%-------------------------------------------------------------------------------

\subsubsection{Physical charm quark mass limit \label{sec:charm_mass}}

The FNAL and MILC Collaborations have provided a determination of the input value of the charm quark mass that reproduces the ``physical'' charm quark mass for each of the four lattice spacings used in this work.  The mass of the $D_s$ meson was used to tune the input charm quark mass until the physical $D_s$ mass was reproduced (with the already tuned values of the input strange quark mass), defining the ``physical'' charm quark masses~\cite{Bazavov:2014wgs},
\begin{align}\label{eq:mc_phys}
&a_{\rm 15} m_c^{\rm phys} = 0.8447(15) ,&
&a_{\rm 12} m_c^{\rm phys} = 0.6328(8) ,&
\nonumber\\
&a_{\rm 09} m_c^{\rm phys} = 0.4313(6) ,&
&a_{\rm 06} m_c^{\rm phys} = 0.2579(4) .&
\end{align}
Comparing to \tabref{tab:lattice_params}, the simulated charm quark mass is mistuned by less than 2\% of the physical charm quark mass for all ensembles used in this work except the a06m310L ensemble, whose simulated charm quark is almost 10\% heavier than its physical value.

In order to test the sensitivity of our results to this mistuning of the charm quark mass, we perform a reweighting~\cite{Ferrenberg:1988yz} study of the a06m310L correlation functions and extracted pion, kaon and omega baryon masses.
While the relative shift of the charm quark mass is small, this shift is approximately equal to the value of the physical strange quark mass,
\begin{align}
\d a_{06} m_c &= a_{06}m_c^{\rm phys} - a_{06}m_c^{\rm HMC}
\nonumber\\&
    = -0.0281(4)\, .
\end{align}
As the reweighting factor is provided by a ratio of the charm quark fermion determinant, it is an extensive quantity, and the relatively large volume we have used to generate the a06m310L ensemble causes some challenges in accurately determining the reweighting factor.

The summary of our study is that our scale setting is not sensitive to this mistuning of the charm quark mass, in line with prior expectation.  For example, we find
\begin{align}\label{eq:mO_reweighting}
a_{\rm 06} m_\O = \left\{
    \begin{array}{cl}
        0.5069(21)\, ,& \textrm{unweighted}\\
        0.5065(29)\, ,& \textrm{reweighted}\\
    \end{array}
    \right.\, ,
\nonumber\\
a_{\rm 06} (m_\O^{\rm reweighted} - m_\O^{\rm unweighted})
    = -0.0004(34)\, ,
\end{align}
where the splitting is determined under bootstrap.
We provide extensive details in Appendix \ref{app:reweighting}.

\subsubsection{Continuum limit \label{sec:continuum}}

\begin{figure}
    \includegraphics[width=\columnwidth]{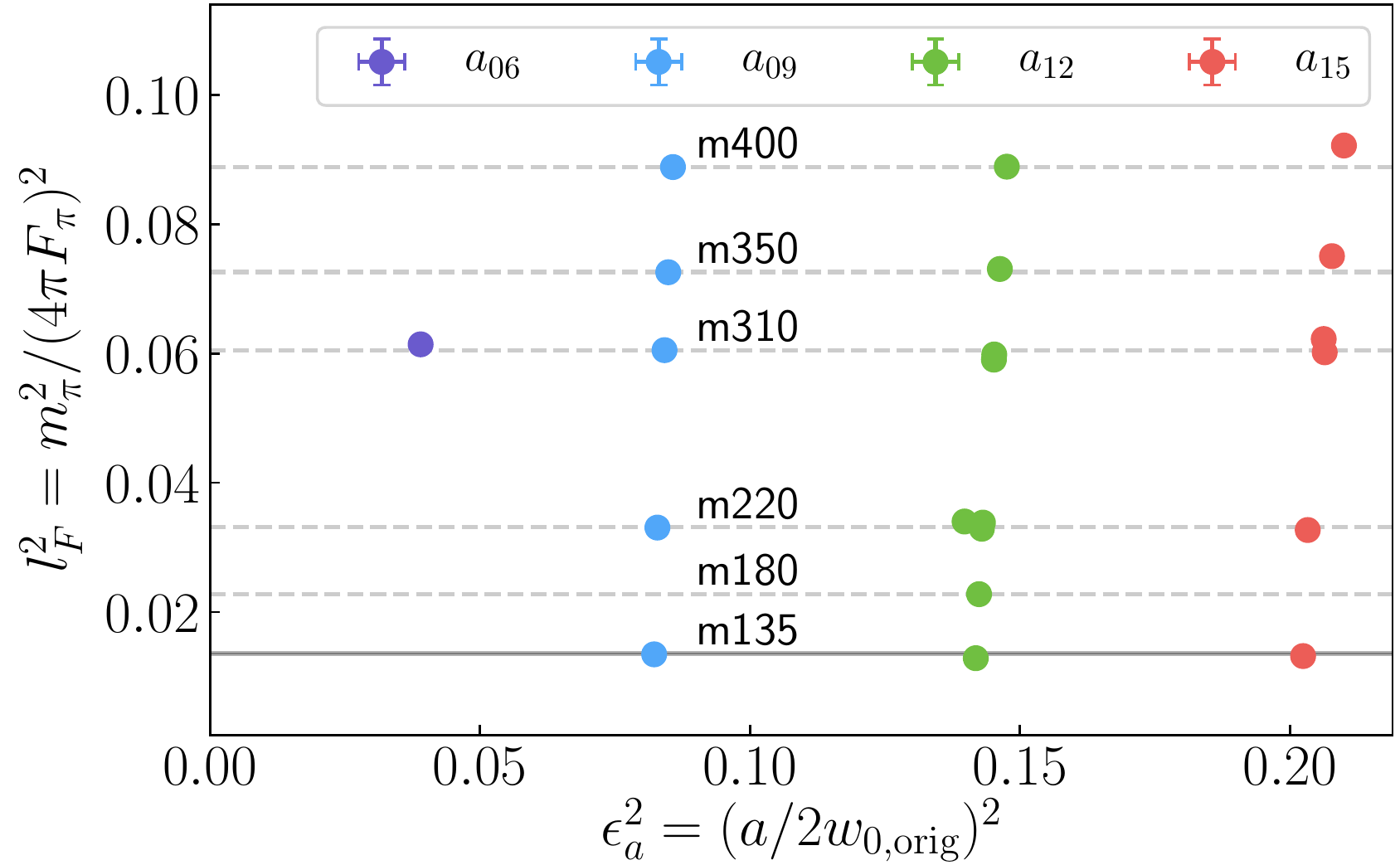}
\caption{\label{fig:l_F_vs_a}
Parameter space of pion mass and lattice spacing utilized in this work expressed in terms of $l_F^2$ and $\e_a^2$.}
\end{figure}

In order to control the continuum extrapolation, we utilize four lattice spacings ranging from $0.057\lesssim a \lesssim 0.15$~fm.  For most of the pion masses, we have three values of $a$ with four values at $m_\pi\sim310$~MeV and one value at $m_\pi\sim180$~MeV.  The parameter space is depicted in \figref{fig:l_F_vs_a}.  The small dimensionless parameter we utilize to extrapolate to the continuum limit is
\begin{equation}\label{eq:eps2a_w0_orig}
    \e_a^2 = \frac{a^2}{(2w_{0,\rm orig})^2}\, .
\end{equation}
As noted in Ref.~\cite{Miller:2020xhy}, this choice is convenient as the values of $\e_a^2$ span a similar range as $l_F^2$.
This allows us to test the ansatz of our assumed power counting that treats corrections of $\mathrm{O}(l_F^2) = \mathrm{O}(\e_a^2)$ which we found to be the case for $F_K/F_\pi$~\cite{Miller:2020xhy}.

An equally valid way to define the small parameter characterizing the discretization corrections is to utilize the gradient flow scale that is also used to define the observable $y$ being extrapolated.
The following normalizations are comparable to our standard choice, \eqnref{eq:eps2a_w0_orig}
\begin{equation}\label{eq:eps2a_gf}
\e_a^2 = \left\{
    \begin{array}{rl}
        \frac{a^2}{(2w_{0,\rm orig})^2}\, ,& y = w_{0,\rm orig} m_\O \\ \vspace{1pt}
        \frac{a^2}{(2w_{0,\rm imp})^2}\, , & y = w_{0,\rm imp} m_\O  \\ \vspace{1pt}
        \frac{a^2}{4 t_{0,\rm orig}}\, ,   & y = \sqrt{t_{0,\rm orig}} m_\O\\ \vspace{1pt}
        \frac{a^2}{4 t_{0,\rm imp}}\, ,    & y = \sqrt{t_{0,\rm imp}} m_\O\\
    \end{array}
\right. \, .
\end{equation}
While these choices do not exhaust the possibilities, they are used to explore possible systematic uncertainties arising from this choice.
Unless otherwise noted, the fixed choice in \eqnref{eq:eps2a_w0_orig} is used in subsequent results and plots.

%-------------------------------------------------------------------------------
% l_F vs eps_a
%-------------------------------------------------------------------------------

\subsubsection{Infinite volume limit \label{sec:infinite_volume}}

The leading sensitivity of $m_\O$, $t_0$ and $w_0$ to the size of the volume is exponentially suppressed for sufficiently large $m_\pi L$~\cite{Luscher:1985dn}.
We have ensembles with multiple volumes at a15m310, a12m310 and a12m220 to test the predicted finite volume corrections against the observed ones.
We derive the predicted volume dependence of $w_0 m_\O$ to the first two nontrivial orders in \secref{sec:FV}.

%----------------------------------------------------------
%    quark mass
\subsection{Light and strange quark mass dependence \label{sec:mq_dependence}}

The light and strange quark mass dependence of the omega baryon has been derived in $\mathrm{SU}(3)$ heavy baryon $\chi$PT (HB$\chi$PT)~\cite{Jenkins:1990jv,Jenkins:1991es} to next-to-next-to-leading order (\nxlo{2}) which is $\mathrm{O}(m_{\pi,K,\eta}^4)$~\cite{Lebed:1993yu,Lebed:1994gt,Tiburzi:2004rh}.
It has been shown that $\mathrm{SU}(3)$ HB$\chi$PT does not produce a converging expansion at the physical quark masses~\cite{WalkerLoud:2008bp,Torok:2009dg,Ishikawa:2009vc,Jenkins:2009wv,WalkerLoud:2011ab}, and so using these formulas to obtain a precise, let alone subpercent, determination, at the physical pion mass is not possible when incorporating systematic uncertainties associated with the truncation of $\mathrm{SU}(3)$ HB$\chi$PT.

However, many LQCD calculations, including this one, keep the strange quark mass fixed near its physical value.
Therefore, a simple interpolation in the strange quark mass is possible.
Further, as the omega is an isosinglet, it will have a simpler, and likely more rapidly converging chiral expansion of the light-quark mass dependence than baryons with one or more light valence quarks.
This has motivated the construction of an $\mathrm{SU}(2)$ HB$\chi$PT for hyperons which considers only the pion as a light degree of freedom~\cite{Beane:2006gf,Tiburzi:2008bk,Jiang:2009sf,Jiang:2009fa,Jiang:2009jn}.
In particular, the chiral expansion for the omega baryon mass was determined to $\mathrm{O}(m_\pi^6)$~\cite{Tiburzi:2008bk}
\begin{align}\label{eq:mO_xpt}
m_\O &= m_0 + \a_2 \frac{m_\pi^2}{\L_\chi}
    +\frac{m_\pi^4}{\L_\chi^3}\left[
        \a_4 \l_\pi
        +\b_4
        \right]
\nonumber\\&\phantom{=}
    +\frac{m_\pi^6}{\L_\chi^5}\left[
        \a_6 \l_\pi^2
        +\b_6 \l_\pi
        +\g_6
    \right]\, ,
\end{align}
where
\begin{equation}\label{eq:ln_mpi}
    \l_\pi = \l_\pi(\mu) = \ln\left( \frac{m_\pi^2}{\mu^2} \right)\, ,
\end{equation}
$\a_n$, $\b_n$ and $\g_6$ are linear combinations of $\mu$-dependent dimensionless low energy constants (LECs) of the theory, and $m_0$ is the mass of the omega baryon in the $\mathrm{SU}(2)$ chiral limit at the physical strange quark mass.
The renormalization group~\cite{Weinberg:1978kz} restricts the coefficient of the $\ln^2$ term to be linearly dependent on $\a_2$ and $\a_4$, with the relation provided in Ref.~\cite{Tiburzi:2008bk}.
In standard HB$\chi$PT power counting, in which the expansion includes odd powers of the pion mass, this order would be called next-to-next-to-next-to-next-to-leading order (\nxlo{4}), where leading order (LO) is the $\mathrm{O}(m_\pi^2 / \L_\chi)$ contribution, next-to-leading order (NLO) would be an $\mathrm{O}(m_\pi^3/\L_\chi^2)$ contribution, which vanishes for $m_\O$, etc.

The light quark mass dependence for $t_0$ and $w_0$ has also been determined in $\chi$PT through $\mathrm{O}(m_\pi^4)$~\cite{Bar:2013ora} which is \nxlo{2} in the meson chiral power counting.  For example
\begin{multline}\label{eq:w0_xpt}
w_0 = w_{0,\rm{ch}} \bigg\{
    1 + k_1 \frac{m_\pi^2}{\L_\chi^2}
    +k_3 \frac{m_\pi^4}{\L_\chi^4}
%\\
    +k_2 \frac{m_\pi^4}{\L_\chi^4} \l_\pi
\bigg\}\, ,
\end{multline}
where the LO term, $w_{0,\rm{ch}}$, is the value in the chiral limit and the $k_i$ are linear combinations of dimensionless LECs.  The expression for $t_0$ is identical in form and will have different numerical values of the LECs.

From these expressions, we can see both $m_\O$ and $t_0$ and $w_0$ depend only upon even powers of the pion mass through the order we are working: $m_\O$ receives a chiral correction that scales as $\mathrm{O}(m_\pi^7)$ from a double-sunset two-loop diagram~\cite{Tiburzi:2008bk} and the next correction to $t_0$ and $w_0$ will appear at $\mathrm{O}(m_\pi^6)$.
We can multiply these expressions together, Eqs.~\eqref{eq:mO_xpt} and \eqref{eq:w0_xpt}, in order to form an expression describing the light-quark mass dependence of $w_0 m_\O$.
As the characterization of the order of the expansion with respect to the order of $m_\pi^2$ is not the same for $w_0$ and $m_\O$, we define the contributions to $w_0 m_\O$ as
\begin{align}
w_0 m_\O &= c_0
    + \d_{ls,\L}^{\rm NLO}
    + \d_{ls,\L}^{\textrm{\nxlo{2}}}
    + \d_{ls,\L}^{\textrm{\nxlo{3}}}\, ,
\end{align}
with a similar expression for $\sqrt{t_0}m_\O$.
We add polynomial terms in $s_{\L,\O}^2$ such that
\begin{align}\label{eq:w0mO_ls}
\d_{ls,\L}^{\rm NLO} &=
    l_\L^2 c_{l} + s_\L^2 c_{s}\, ,
\nonumber\\
\d_{ls,\L}^{\textrm{\nxlo{2}}} &=
    l_\L^4 (c_{ll} + c_{ll}^{ln} \l_\pi) +l_\L^2 s_\L^2 c_{ls} + s_\L^4 c_{ss}\, ,
\nonumber\\
\d_{ls,\L}^{\textrm{\nxlo{3}}} &=
    l_\L^6 ( c_{lll} + c_{lll}^{ln} \l_\pi + c_{lll}^{ln^2} \l_\pi^2)
    +l_\L^4 s_\L^2 \l_\pi c_{lls}^{ln}
\nonumber\\&\phantom{=}
    +l_\L^4 s_\L^2 c_{lls} +l_\L^2 s_\L^4 c_{lss} + s_\L^6 c_{sss}\, .
\end{align}
We will consider both $\L=F$ and $\L=\O$ for the two choices of small parameters.
For convenience, we set $\mu=\L_\chi$ and $\mu=m_\O$ respectively for these choices.  For a detailed discussion how one can track the consequence of such a quark mass dependent choice for the dim-reg scale, see Ref.~\cite{Miller:2020xhy}.

%----------------------------------------------------------
%    Finite Volume Corrections
\subsection{Finite volume corrections \label{sec:FV}}

The finite-volume (FV) corrections for $m_\O$ are determined at one loop through the modification to the tadpole integral~\cite{Gasser:1986vb,Colangelo:2004xr}
\begin{equation}
\frac{(4\pi)^2}{m^2} \mc{I}^{\rm FV} = \ln\left(\frac{m^2}{\mu^2}\right)
    +4 k_1(m_\pi L)
\end{equation}
where $k_1(x)$ is given by
\begin{align}
k_1(x) = \sum_{|\mathbf{n}| \neq 0} c_n \frac{K_1(x|\mathbf{n}|)}{x|\mathbf{n}|}\, .
\end{align}
$K_1(x)$ is a modified Bessel function of the second kind and $c_n$ are multiplicity factors for the number of ways the integers $(n_x, n_y, n_z)$ can form a vector of length $|\mathbf{n}|$; see \tabref{tab:cn_weigths} for the first few.

%------   FV multiplicity tables  ---------------------------------------------------
\begin{table}
\caption{\label{tab:cn_weigths}
Multiplicity factors for the finite volume corrections of the first ten vector lengths, $|\mathbf{n}|$.
}
\begin{ruledtabular}
\begin{tabular}{c|cccccccccc}
$|\mathbf{n}|$& 1 & $\sqrt{2}$& $\sqrt{3}$& $\sqrt{4}$& $\sqrt{5}$& $\sqrt{6}$& $\sqrt{7}$& $\sqrt{8}$& $\sqrt{9}$& $\sqrt{10}$\\
\hline
$c_n$& 6&12& 8& 6& 24& 24& 0& 12& 30& 24
\end{tabular}
\end{ruledtabular}
\end{table}
%------------------------------------------------------------

At \nxlo{3}, the finite volume corrections for $m_\O$ are also trivially determined, as the only two-loop integral that contributes is a double-tadpole with un-nested momentum integrals; see Fig.~2 of Ref.~\cite{Tiburzi:2008bk}.
The \nxlo{3} correction to $w_0$ is not known.  However, the isoscalar nature of $w_0$ means that at the two-loop order, just like the correction to $m_\O$, it will only receive contributions from trivial two-loop integrals with factorizable momentum integrals.  Therefore, the \nxlo{3} FV correction can also be determined from the square of the tadpole integral
\begin{multline}
\frac{(4\pi)^4}{m^4} \left[
    \left(\mc{I}^{\rm FV}\right)^2 - \left( \mc{I}^{\infty} \right)^2
    \right]
\\    =
    8 \l_\pi k_1(m_\pi L)
    +16 k_1^2(m_\pi L)\, ,
\end{multline}
resulting in
\begin{align}\label{eq:w0mO_fv}
    \d_{L,F}^{\textrm{\nxlo{2}}}(l_F,m_\pi L) &= c_{ll}^{ln} l_F^4\, 4k_1(m_\pi L)
    \nonumber\\
    \d_{L,F}^{\textrm{\nxlo{3}}}(l_F,m_\pi L) &=
        c_{lll}^{ln^2} l_F^6 \ln(l_F^2)\, 8k_1(m_\pi L)
    \nonumber\\&\phantom{=}
        +c_{lll}^{ln} l_F^6\,  16k_1(m_\pi L)\, .
    \end{align}
The FV correction through \nxlo{3} arising from loop corrections to $w_0$ and $m_O$ is given by
\begin{equation}\label{eq:w0mO_fv_n3lo}
\d_{L,F} = \d_{L,F}^{\textrm{\nxlo{2}}} + \d_{L,F}^{\textrm{\nxlo{3}}}\, ,
\end{equation}
with a similar expression for $\d_{L,\O}$.

We are neglecting a few FV corrections at $\d_{L,F}^{\textrm{\nxlo{3}}}$.
The NLO $l_{F,\O}^2$ correction to the omega mass is from a quark mass operator, which has been converted to an $m_\pi^2$ correction, $2B\hat{m}_l = m_\pi^2 +\mathrm{O}(m_\pi^4/\L_\chi^2)$.  This choice for organizing the perturbative expansion induces corrections in what we have called \nxlo{2} and \nxlo{3}.  At \nxlo{2}, the corrections arise from single tadpole diagrams and so the FV corrections are accounted for through \eqnref{eq:w0mO_fv}.  At the next order, the corrections to the pion self-energy involve more complicated two-loop diagrams~\cite{Bijnens:2014dea} and so the FV corrections arising from these are not captured in our parametrization.
Similar corrections arise from expressing $4\pi F_0 = 4\pi F_\pi +\mathrm{O}(m_\pi^2/\L_\chi^2)$ when using $l_F^2$ to track the light-quark mass corrections [$F_0$ is $F_\pi$ in the $\mathrm{SU}(2)$ chiral limit].

\begin{figure}
    \includegraphics[width=\columnwidth]{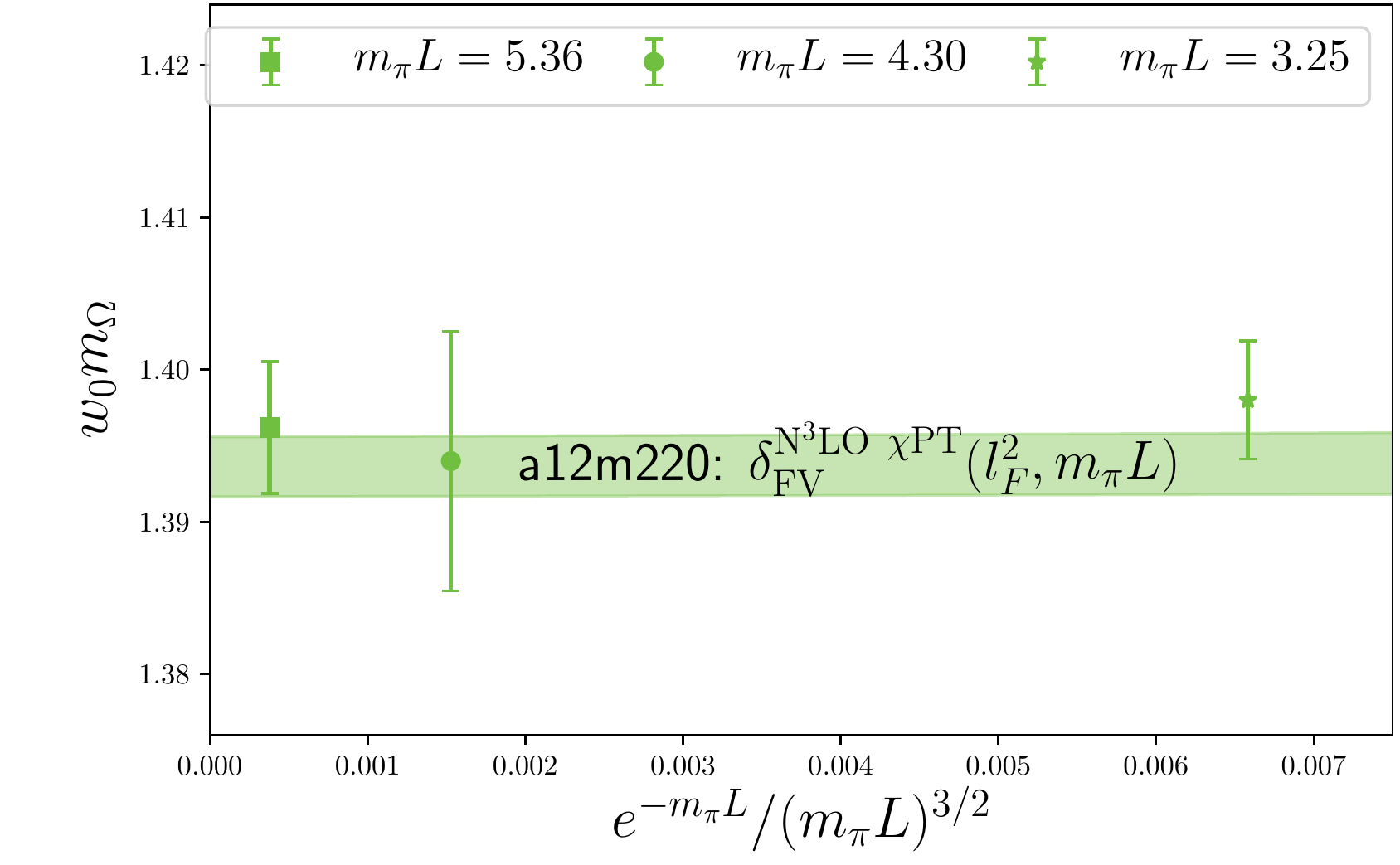}
\caption{\label{fig:vs_mL}
Predicted finite volume corrections (band) compared with the results on the a12m220 ensembles.
The \nxlo{3} result is from \eqnref{eq:w0mO_fv_n3lo} using $l_F^2$ as the small parameter.
}
\end{figure}

While we have neglected these contributions, the FV corrections to $m_\O$ are suppressed by an extra power in the chiral power counting compared with many observables, beginning with an $m_\pi^4/\L_\chi^4$ prefactor, \eqnref{eq:w0mO_fv}.
In \figref{fig:vs_mL}, we show the predicted FV correction along with the results at three volumes on the a12m220 ensembles.  As can be observed, the predicted FV corrections are very small and consistent with the numerical results.

%-------------------------------------------------------------------------------
% l_F vs eps_a
%-------------------------------------------------------------------------------

%----------------------------------------------------------
%    Discretization Corrections
\subsection{Discretization corrections \label{sec:discretization}}
A standard method of incorporating discretization effects into the extrapolation formula used for hadronic observables is to follow the strategy of Sharpe and Singleton~\cite{Sharpe:1998xm}:
\begin{enumerate}[leftmargin=*]
\item For a given lattice action, one first constructs the Symanzik effective theory (SET) by expanding the discretized action about the continuum limit.  This results in a local effective action in terms of quark and gluon fields~\cite{Symanzik:1983dc,Symanzik:1983gh};

\item With this continuum effective theory, one builds a chiral effective theory by using spurion analysis to construct not only operators with explicit quark mass dependence, but also operators with explicit lattice spacing dependence.
\end{enumerate}
Such an approach captures the leading discretization effects in a local hadronic effective theory.
At the level of the SET, radiative corrections generate logarithmic dependence upon the lattice spacing.  The leading corrections can be resummed such that, for an $\mathrm{O}(a)$ improved action, the leading discretization effects then scale as $a^2 \a_S^{n+\hat{\gamma}_1}$~\cite{Balog:2009yj,Balog:2009np} where $n=0$ for an otherwise unimproved action, $n=1$ for a tree-level improved action and $n=2$ for a one-loop improved action.  The coefficient $\hat{\gamma}_1$ is an anomalous dimension which has been determined for Yang-Mills and Wilson actions~\cite{Husung:2019ytz}.%
\footnote{See also the presentation by N.~Husung at the MIT Virtual Lattice Field Theory Colloquium Series, \url{http://ctp.lns.mit.edu/latticecolloq/}.}

For mixed-action setups~\cite{Renner:2004ck,Bar:2002nr} such as the one used in this work, a low-energy mixed-action effective field theory (MAEFT)~\cite{Bar:2002nr,Bar:2003mh,Bar:2005tu,Golterman:2005xa,Tiburzi:2005is,Chen:2005ab,Chen:2006wf,Orginos:2007tw,Chen:2007ug,Chen:2009su} can be constructed to capture the manifestation of infrared radiative corrections from the discretization%
% FOOTNOTE
\footnote{While this might seem counterintuitive, it is analogous to the infrared sensitivity of hadronic quantities to the Higgs vacuum expectation value (vev): hadronic quantities have infrared (logarithmic) sensitivity to the pion mass from radiative pion loops, and the squared pion mass is proportional to the light quark mass which is proportional to the Higgs vev~\cite{higgsvev}.}.
%-------------------------------------------------------------------------------
Corrections come predominantly from a modification of the pseudoscalar meson spectrum as well as from ``hairpin'' interactions~\cite{Bernard:1992mk} that are proportional to the lattice spacing in rooted-staggered~\cite{Bernard:1993sv} and mixed-action theories~\cite{Golterman:2005xa}; in partially quenched theories, these hairpins are proportional to the difference in the valence and sea quark masses~\cite{Sharpe:1997by,Sharpe:2000bc,Sharpe:2001fh}.

In our analysis of $F_K/F_\pi$, we observed that the use of continuum chiral perturbation theory with corrections polynomial in $\e_a^2$ was highly favored over the use of the MAEFT expression, as measured by the Bayes-Factor, though the results from both were consistent within a fraction of 1 standard deviation~\cite{Miller:2020xhy}.
Similar findings have been observed by other groups for various quantities; see for example Refs.~\cite{Borsanyi:2012zv,Colquhoun:2015mfa,Dowdall:2019bea}.
Therefore, in this work, we restrict our analysis to a continuum-like expression enhanced by polynomial discretizaton terms.

The dynamical HISQ ensembles have a perturbatively improved action such that the leading discretization effects (before resumming the radiative corrections~\cite{Balog:2009yj,Balog:2009np,Husung:2019ytz}) scale as $\mathrm{O}(\a_S a^2)$~\cite{Follana:2006rc}.
The MDWF action, in the limit of infinite extent in the fifth dimension, has no chiral symmetry breaking other than that from the quark mass. Consequently, the leading discretization corrections begin at $\mathrm{O}(a^2)$~\cite{Sharpe:2007yd,Aoki:2010dy}.  For finite $L_5$, the $\mathrm{O}(a)$ corrections are proportional to $am_{\rm res}$ which is sufficiently small that these terms are numerically negligible.
Therefore, we parametrize our discretizaton corrections with the following terms where we count $\e_a^2 \sim l_\L^2 \sim s_\L^2$:
\begin{align}\label{eq:w0mO_a}
\d_{a,\L} &= \d_{a,\L}^{\rm NLO} + \d_{a,\L}^{\textrm{\nxlo{2}}} + \d_{a,\L}^{\textrm{\nxlo{3}}}\, ,
\nonumber\\
\d_{a,\L}^{\rm NLO} &= d_a \e_a^2 + d^\prime_a \a_S \e_a^2\, ,
\nonumber\\
\d_{a,\L}^{\textrm{\nxlo{2}}} &= d_{aa} \e_a^4
        + \e_a^2 \left( d_{al} l_\L^2 + d_{as} s_\L^2 \right)\, ,
\nonumber\\
\d_{a,\L}^{\textrm{\nxlo{3}}} &= d_{aaa} \e_a^6
        +\e_a^4 ( d_{aal} l_\L^2 + d_{aas} s_\L^2)
\nonumber\\&\phantom{=}\quad
    +\e_a^2 ( d_{all} l_\L^4 +d_{als} l_\L^2 s_\L^2 + d_{ass} s_\L^4)\, .
\end{align}

%----------------------------------------------------------
%    Analysis
\section{Extrapolation Details and Uncertainty Analysis \label{sec:analysis}}

We perform our extrapolation analysis under a Bayesian model-averaging framework as described in detail in Refs.~\cite{Chang:2018uxx,Miller:2020xhy,BMA}, which is more extensively discussed for lattice QFT analysis in Ref.~\cite{Jay:2020jkz}.
We consider a variety of extrapolation functions by working to different orders in the power counting, using the $l_F, s_F$ or $l_\O, s_\O$ small parameters, by including or excluding the chiral logarithms associated with pion loops, and by including or excluding discretization corrections scaling as $\a_S a^2$.
The resulting Bayes factors are then used to weight the fits with respect to each other and perform a model averaging.
In this section, we discuss the selection of the priors for the various LECs and then present an uncertainty analysis of the results.

%----------------------------------------------------------
%    Prior widths
\subsection{Prior widths of LECs\label{sec:prior_widths}}

In our $F_K/F_\pi$ analysis~\cite{Miller:2020xhy}, we observed that using $\e_\pi^2=l_F^2$, $\e_K^2$
and $\e_a^2$ as the small parameters in the expansion,%
\footnote{The small parameter $\e_K^2 = \frac{m_K^2}{\L_\chi^2} = \frac{1}{2}(s_F^2 + l_F^2)$.}
the LECs were naturally of $\mathrm{O}(1)$.  We therefore have a prior expectation that this may hold for $\sqrt{t_0}m_\O$ and $w_0 m_\O$ as well.

Let us use $w_{0,\rm orig} m_\O$ to guide the discussion.
The $a\sim0.12$~fm ensembles were simulated with a fixed strange quark mass.
Therefore, the entire change in $w_{0,\rm orig} m_\O$ between the a12m130 and a12m180L ensembles can be attributed to the change in $l_F$.  This allows us to ``eyeball'' the $c_l$ prior to be $c_l\simeq1$ if we assume the dominant contribution comes from the NLO $c_l l_F^2$ term.%
%   FOOTNOTE
\footnote{This is analogous to using the effective mass and effective overlap factors to choose conservative priors for the ground state parameters in the correlation function analysis.}
Motivated by $\mathrm{SU}(3)$ flavor symmetry considerations, we can roughly expect $c_{s}\sim c_{l}$.
In order to be conservative, we set the prior for these LECs as
\begin{equation}
    \tilde{c}_{l} = \tilde{c}_{s} = N(\mu=1,\sigma=1)\, ,
\end{equation}
where $N(\mu,\sigma)$ denotes a normal distribution with mean $\mu$ and width $\sigma$.
A similar observation is made for $w_{0,\rm imp} m_\O$ and the original and improved values of $\sqrt{t_0}m_\O$.
We observe (with a full analysis) that the log-Bayes-Factor (logGBF) prefers even tighter priors, with logGBF continuing to increase as the width is taken down to 0.1 on these NLO LECs.%
\footnote{For fixed data, $\exp\{\rm logGBF\}$ provides a relative weight of the likelihood of one model versus another.}

The observation that $m_\O$ increases with increasing values of $l_F$ and $s_F$ (normalized by any and all gradient flow scales) allows us to conservatively estimate the LO prior,
\begin{equation}
    \tilde{c}_0 = N(1,1)\, .
\end{equation}
We then conservatively estimate the priors for all of the higher order $l_F$ and $s_F$ LECs to be
\begin{equation}
    \tilde{c}_i = N(0,1)\, .
\end{equation}
We observe, with a full analysis, that this choice is near the optimal value as measured by the logGBF weighting.

For the discretization corrections, see \figref{fig:extrapolation} in \secref{sec:full_analysis}, as we change the gradient flow scale from $w_{0,\rm orig}$ to the improved version to using the original and improved versions of $\sqrt{t_0}$, the approach to the continuum limit can change sign.
We also observe, the convexity of the approach to the continuum limit (the $\e_a^4$ contributions) can change sign.  Therefore, we perform a prior-optimization study for the discretization LECs in which we change the prior width of the NLO and \nxlo{2} LECs in concert, such that the priors are given by
\begin{equation}
    \tilde{d}_i = N(0, \s_a^{\rm opt})\, ,
\end{equation}
with the optimized values
\begin{equation}\label{eq:sig_a_opt}
\s_a^{\rm opt} = \left\{
    \begin{array}{cl}
        1.2,& w_{0,\rm orig} m_\O\\
        1.4,& w_{0,\rm imp} m_\O\\
        1.8,& \sqrt{t_{0,\rm orig}} m_\O\\
        1.4,& \sqrt{t_{0,\rm imp}} m_\O\\
    \end{array}
    \right. \, ,
\end{equation}
maximizing the respective logGBF factors.
We set the \nxlo{3} priors for discretization LECs to
\begin{equation}
    \tilde{d}_i = N(0, 1)\, .
\end{equation}
Although we find using tighter priors increases the logGBF, the final results are unchanged.

When we add the $\a_S \e_a^2$ term in the analysis, this introduces a fourth class of discretization corrections.  As we only have four lattice spacings in this work, we perform an independent prior-width optimization for this LEC.  For all four choices of gradient-flow scales, we find the choice,
\begin{equation}
    \tilde{d}_a^\prime = N(0,0.5)\, ,
\end{equation}
to be near-optimal, with three of the analyses preferring an even tighter prior width.
An empirical Bayes study~\cite{Lepage:2001ym} in which the widths of all the chiral and all the discretization priors are varied together at a given order leads to similar choices of all the priors.

In \tabref{tab:prior_widths} we list the values of all the priors used in the final analysis.  The full analysis demonstrates that these choices also result in no tension between the priors and the final posterior values of the LECs, further indication that our choices are reasonable.

When we use $l_\O^2$ and $s_\O^2$ as the small parameters instead of $l_F^2$ and $s_F^2$, we note that since $(m_\O / \L_\chi)^2\sim2$, we can use the same prescription, except to double the mean and width of all the NLO priors (which scale linearly in $l_\O^2$ and $s_\O^2$), set the widths to be 4 times larger for the \nxlo{2} priors and 8 times larger for the \nxlo{3} LECs.  The mixed contributions which scale with some power of $\e_a^2$ and $l_\O^2$ and $s_\O^2$ are scaled accordingly; see also \tabref{tab:prior_widths}.

%-------------------------------------------------------------------------------
\begin{table}
\caption{\label{tab:prior_widths}
The values of the priors used in the analysis.
}
\begin{ruledtabular}
\begin{tabular}{ccc}
LEC& $\L=F$& $\L=\O$\\
\hline
$c_0$                                 & $N(1,1)$& $N(1,1)$\\
$c_{l}, c_{s}$                    & $N(1,1)$& $N(2,2)$\\
$c_{ll}, c_{ll}^{ln}, c_{ls}, c_{ss}$& $N(0,1)$& $N(0,4)$\\
$c_{lll}, c_{lll}^{ln}, c_{lll}^{ln^2}, c_{lls}, c_{lls}^{ln}, c_{lss}, c_{sss}$& $N(0,1)$& $N(0,8)$\\
$d_a, d_{aa}$                        & $N(0,\s_a^{\rm opt})$& $N(0,\s_a^{\rm opt})$\\
$d^\prime_a$         & $N(0,0.5)$& $N(0,0.5)$\\
$d_{al}, d_{as}$     & $N(0,\s_a^{\rm opt})$& $N(0,2\s_a^{\rm opt})$\\
$d_{aaa}$                     & $N(0,1)$& $N(0,1)$\\
$d_{aal}, d_{aas}$    & $N(0,1)$& $N(0,2)$\\
$d_{all}, d_{als}, d_{ass}$           & $N(0,1)$& $N(0,4)$
\end{tabular}
\end{ruledtabular}
\end{table}
%-------------------------------------------------------------------------------

%----------------------------------------------------------
%    Full analysis
\subsection{Extrapolation analysis\label{sec:full_analysis}}
\begin{figure*}
\includegraphics[width=\textwidth]{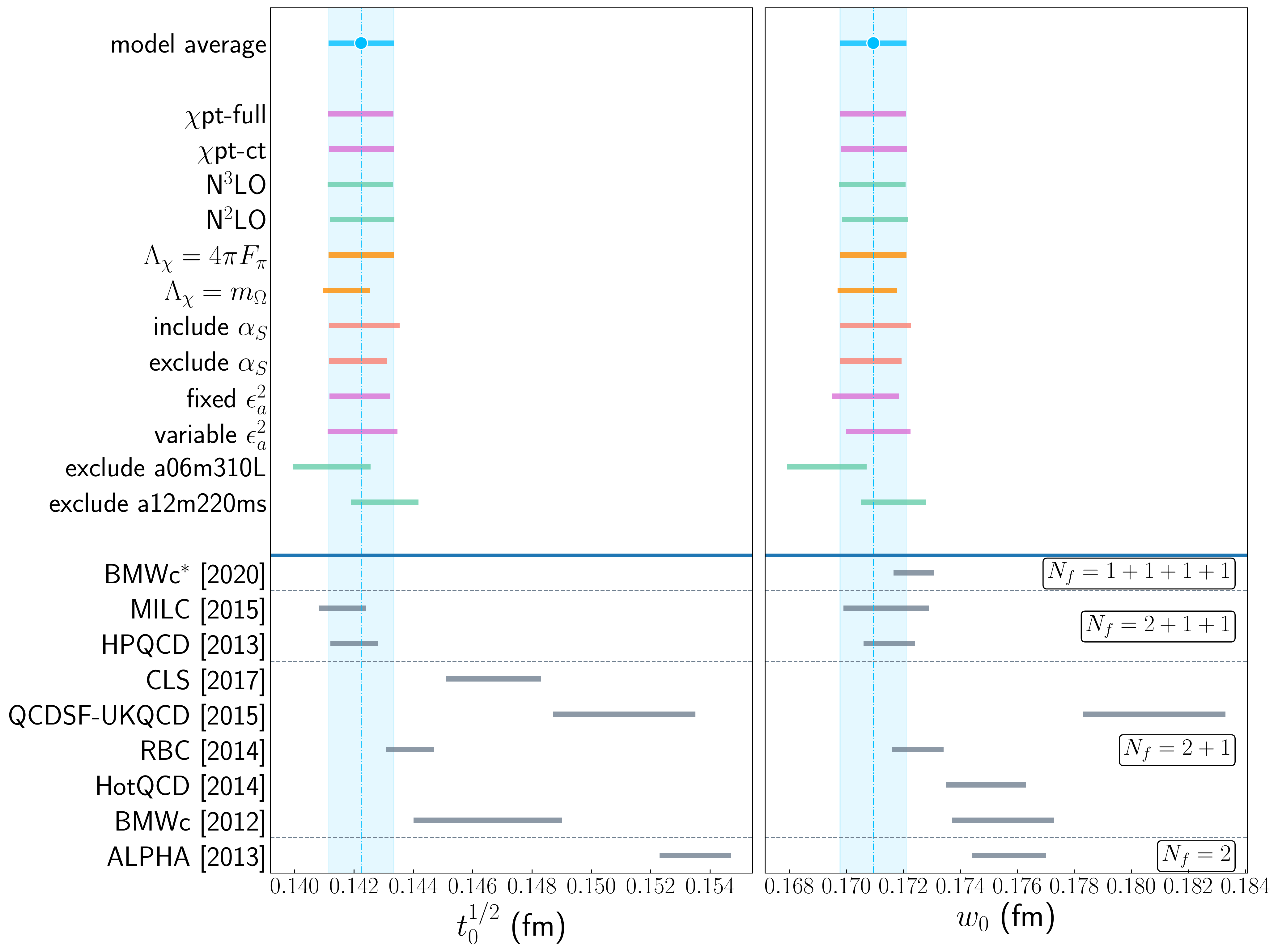}
\caption{\label{fig:model_breakdown}
Model breakdown and comparison using the scale-improved, gradient-flow derived quantities (i.e., $\sqrt{t_{0, \text{imp}}} m_\Omega$, $w_{0, \text{imp}} m_\Omega$).  The vertical band is our model average result to guide the eye.
\pmb{$\chi$pt-full}:~$\chi$PT model average, including $\ln(m_\pi^2/\mu^2)$ corrections.
\pmb{$\chi$pt-ct}:~$\chi$PT model average with counterterms only, excluding $\ln(m_\pi^2/\mu^2)$ corrections.
\pmb{N$^3$LO/N$^2$LO}:~model average restricted to specified order.
\pmb{$\Lambda_\chi$}:~model average with specified chiral cutoff.
\pmb{incl./excl. $\alpha_S$}:~model average with/without $\alpha_S$ corrections.
\pmb{fixed/variable $\epsilon_a^2$}:~model average with Eqs.~\eqref{eq:eps2a_w0_orig} or \eqref{eq:eps2a_gf}, respectively.
\pmb{excl. a06m310L}:~model average excluding $a=0.06$ fm ensemble (\texttt{a06m310L}).
\pmb{excl. a12m220ms}:~model average excluding small strange quark mass ensemble (\texttt{a12m220ms}).
\pmb{Below solid line}:~results from other collaborations for various numbers of dynamical fermions:
BMWc [2020]~\cite{Borsanyi:2020mff},
MILC [2015]~\cite{Bazavov:2015yea},
HPQCD [2013]~\cite{Dowdall:2013rya},
CLS [2017]~\cite{Bruno:2016plf},
QCDSF-UKQCD [2015]~\cite{Bornyakov:2015eaa},
RBC [2014]~\cite{Blum:2014tka},
HotQCD [2014]~\cite{Bazavov:2014pvz},
BMWc [2012]~\cite{Deuzeman:2012jw}
and ALPHA [2013]~\cite{Bruno:2013gha}.
}
\end{figure*}

%-------------------------------------------------------------------------------
For each of the four quantities, $w_{0,\rm orig} m_\O$, $w_{0,\rm imp}m_\O$, $\sqrt{t_{0,\rm orig}}m_\O$ and $\sqrt{t_{0,\rm imp}}m_\O$, we consider several reasonable choices of extrapolation functions to perform the continuum, infinite volume and physical quark-mass limits.
The final result for each extrapolation is then determined through a model average in which the relative weight of each model is given by the exponential of the corresponding logGBF value.
The various choices we consider in the extrapolations consists of
\begin{align*}
\begin{array}{rl}
\text{Include the $\ln(m_\pi)$ terms or counterterm only} : & \times 2 \\
\text{Expand to \nxlo{2}, or \nxlo{3}}                    : & \times 2 \\
\text{Include/exclude finite volume corrections}          : & \times 2 \\
\text{Include/exclude the $\alpha_Sa^2$ term}             : & \times 2 \\
\text{Use the $\L=F$ or $\L=\O$ expansion}                : & \times 2 \\
\hline
\text{total choices} : & 32
\end{array}
\end{align*}
We find that there is very little dependence upon the particular model chosen.
In \figref{fig:model_breakdown}, we show the stability of the final result of $\sqrt{t_{0,\rm imp}}m_\O$ and $w_{0,\rm imp}m_\O$ as various options from the above list are turned on and off.
In addition, we show the impact of including or excluding the a12m220ms ensemble, whose strange quark mass is $m_s\sim0.6\times m_s^{\rm phys}$, as well as the impact of including the a06m310L ensemble.
We observe a small variation of the result when either of these ensembles is dropped, but the results are still consistent with our final result (top of the figure).
Using the fixed definition of $\e_a^2$, \eqnref{eq:eps2a_w0_orig}, we find
\begin{widetext}
\begin{align}\label{eq:final_t0w0_orig}
%\textrm{with } \e_a^2 &= \frac{a^2}{(2 w_{0,\rm orig})^2}\, ,
%\nonumber\\
\sqrt{t_{0,\rm orig}} m_\O &= \tomuborig
    \rightarrow \frac{\sqrt{t_0}}{\rm fm} = \touborig\, ,
\nonumber\\
\sqrt{t_{0,\rm imp}} m_\O &= \timuborig
    \rightarrow \frac{\sqrt{t_0}}{\rm fm} = \tiuborig\, ,
\nonumber\\
w_{0,\rm orig} m_\O &= \womuborig
    \rightarrow \frac{w_0}{\rm fm} = \wouborig\, ,
\nonumber\\
w_{0,\rm imp} m_\O &= \wimuborig
    \rightarrow \frac{w_0}{\rm fm} = \wiuborig\, ,
\end{align}
\end{widetext}
with the statistical~($s$), chiral interpolation~($\chi$), continuum-limit~($a$), infinite-volume ($V$), physical-point (phys), and model selection uncertainties~($M$).  The conversion to physical units is performed with \eqnref{eq:phys_point}.

% Figure of model breakdown

%-------------------------------------------------------------------------------
\begin{figure*}
\includegraphics[width=0.49\textwidth]{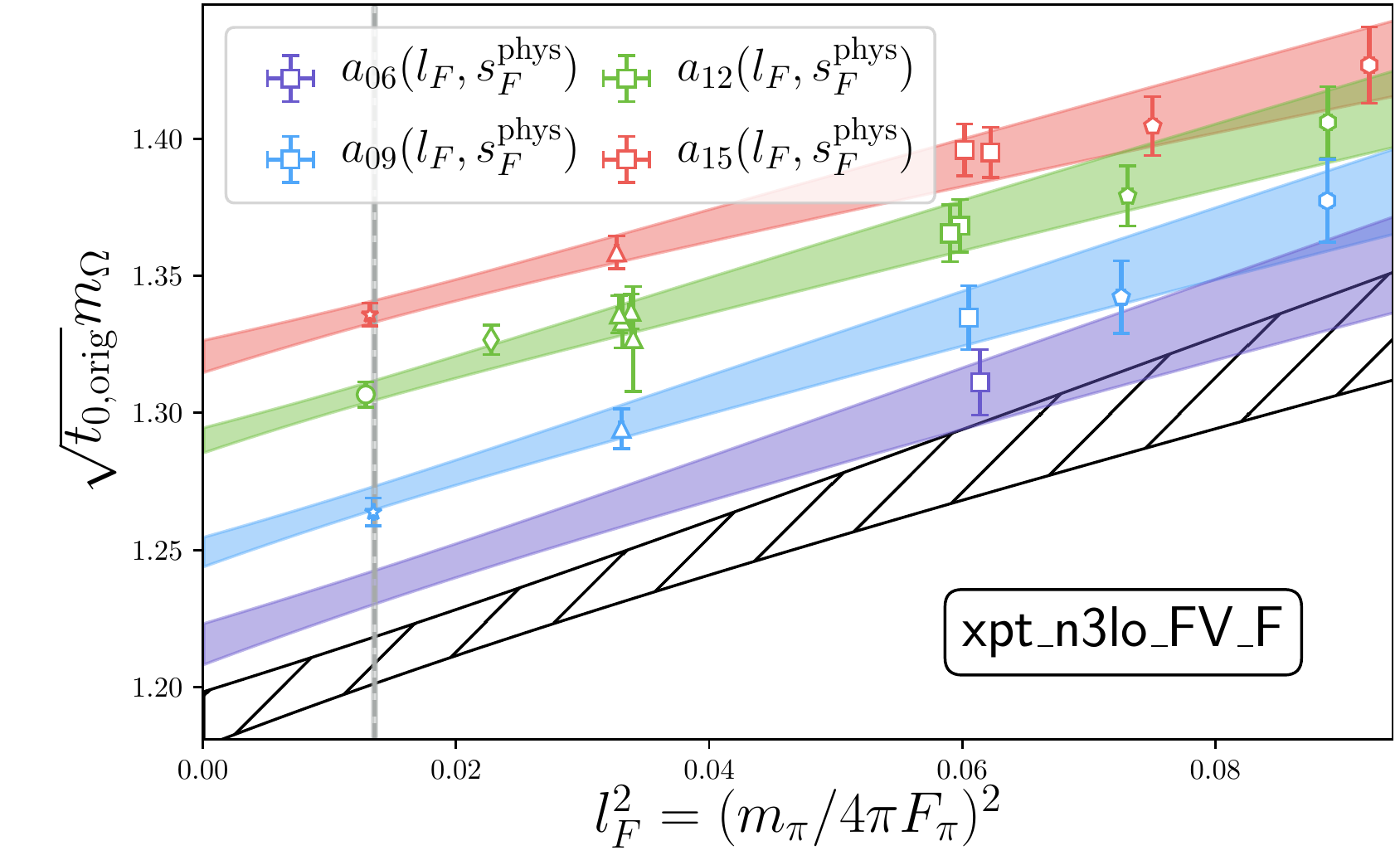}
\includegraphics[width=0.49\textwidth]{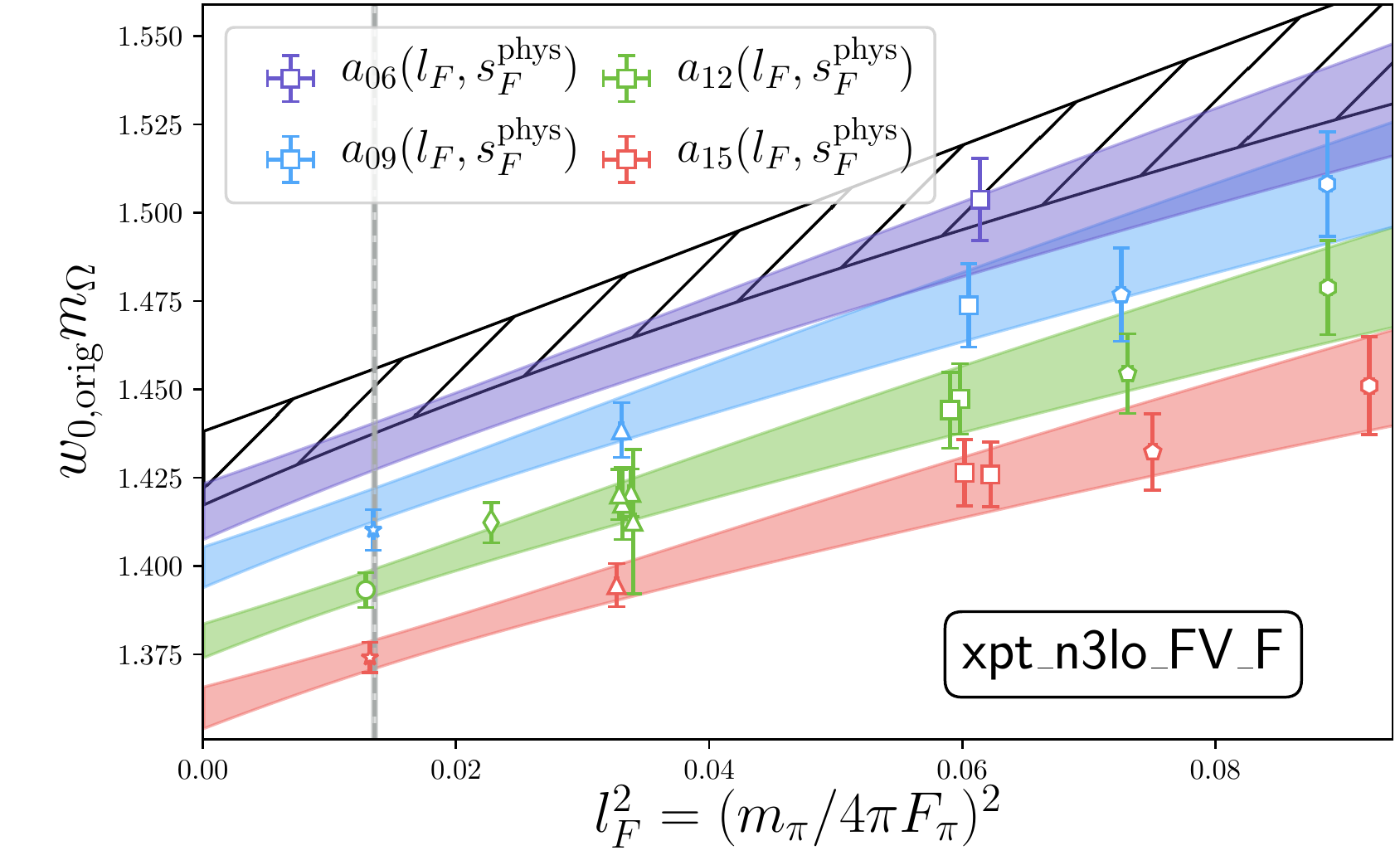}
\includegraphics[width=\columnwidth]{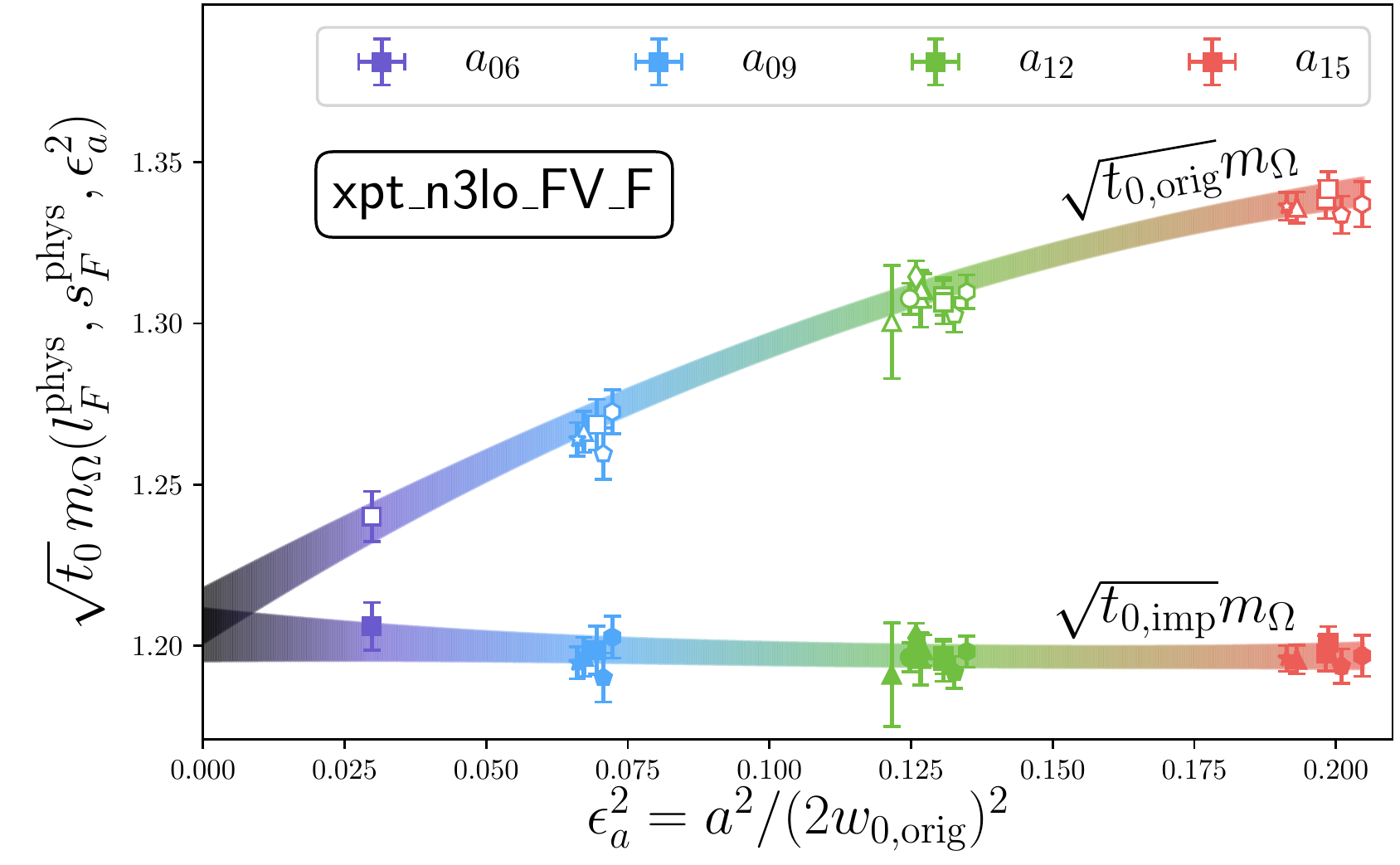}
\includegraphics[width=\columnwidth]{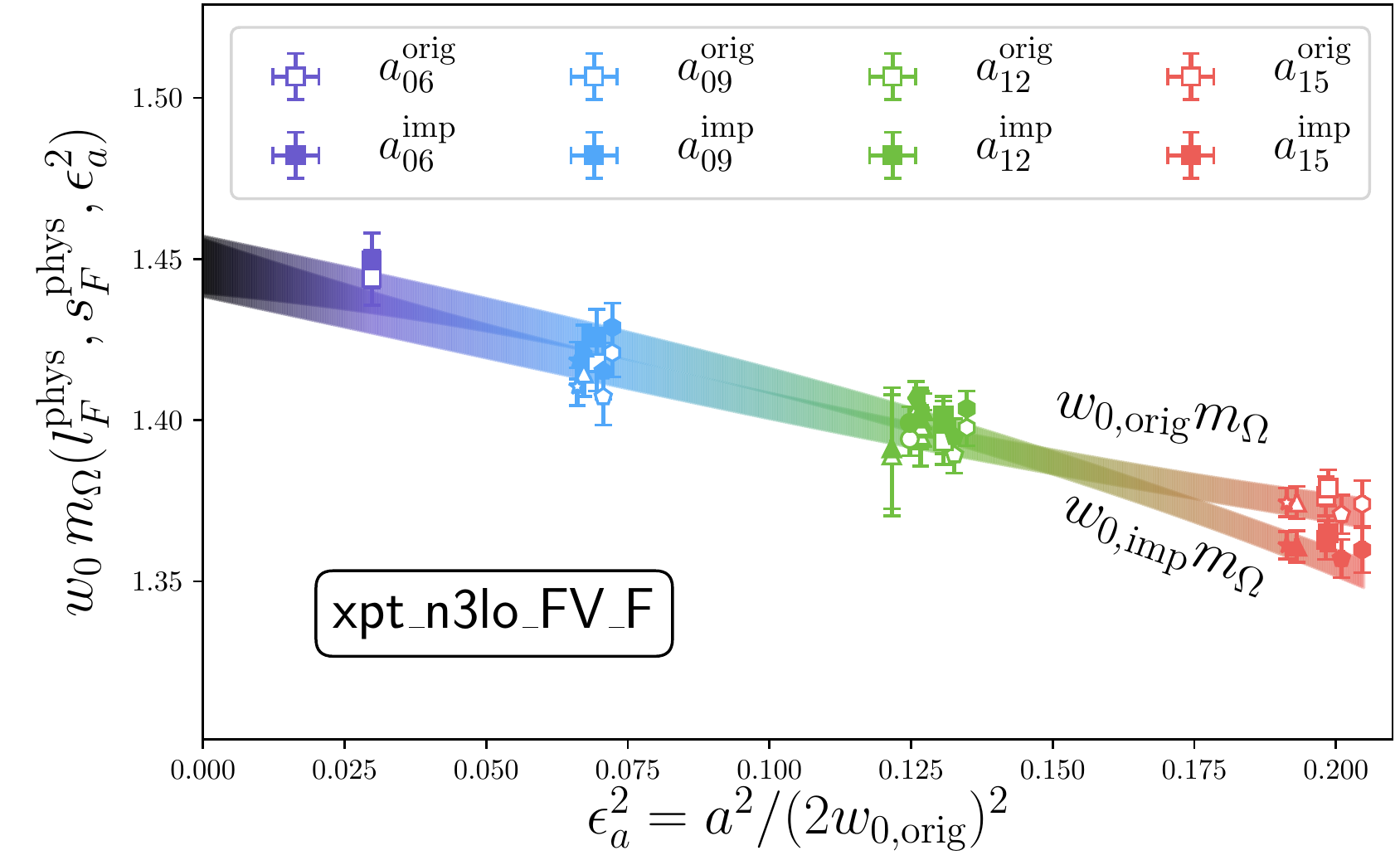}
\caption{\label{fig:extrapolation}
Example extrapolations of $\sqrt{t_{0,\rm orig}}m_\O$ and $w_{0,\rm orig}m_\O$ versus $l_F^2$ (top) and $\e_a^2$ (bottom) using the \nxlo{3} $\chi$PT analysis.
For the continuum extrapolation plots (bottom) we also show the results using the ``improved'' determinations of the scales, \eqnref{eq:t0_w0_improved}.
The numerical results have been shifted from the original data points as described in the text.}
\end{figure*}

As discussed in \secref{sec:continuum}, we explore potential systematics in the use of one definition of the small parameter used to characterize the discretization corrections, \eqnref{eq:eps2a_w0_orig}, versus another equally valid choice, \eqnref{eq:eps2a_gf}.
For each choice of $w_0 m_\O$ and $\sqrt{t_0} m_\O$ that is extrapolated to the physical point, we repeat the above model averaging procedure, but we also include the variation of using these two definitions of $\e_a^2$ for which we find
\begin{widetext}
\begin{align}\label{eq:final_t0w0_avg_eps2a}
\nonumber\\
\sqrt{t_{0,\rm orig}} m_\O &= \tomub
    \rightarrow \frac{\sqrt{t_0}}{\rm fm} = \toub\, ,
\nonumber\\
\sqrt{t_{0,\rm imp}} m_\O &= \timub
    \rightarrow \frac{\sqrt{t_0}}{\rm fm} = \tiub\, ,
\nonumber\\
w_{0,\rm orig} m_\O &= \womub
    \rightarrow \frac{w_0}{\rm fm} = \woub\, ,
\nonumber\\
w_{0,\rm imp} m_\O &= \wimub
    \rightarrow \frac{w_0}{\rm fm} = \wiub\, .
\end{align}
\end{widetext}
For all choices of the gradient-flow scale besides $\sqrt{t_{0,\rm orig}}$, the average over the choice of how to define $\e_\a^2$ has minimal impact on the final result, as can be seen comparing \eqnref{eq:final_t0w0_orig} and \eqref{eq:final_t0w0_avg_eps2a}.  In the case of $\sqrt{t_{0,\rm orig}}$, the two choices for the cutoff-effect expansion parameter lead to a slight difference in the continuum-extrapolated value, which is reflected in the model-averaging uncertainty.
\footnote{Rather than performing a model average over the two definitions of $\epsilon^2_a$ as defined in Eqs.~\eqref{eq:eps2a_w0_orig} and \eqref{eq:eps2a_gf}, one might instead consider a model average over the choices for fixed $\epsilon^2_a$, i.e.
\begin{equation*}
\epsilon^2_a =
\left\{
\left(\frac{a}{2w_{0,\text{orig}}}\right)^2 \, ,
\left(\frac{a}{2w_{0,\text{impr}}}\right)^2 \, ,
\frac{a^2}{4 t_{0,\text{orig}}} \, ,
\frac{a^2}{4 t_{0,\text{impr}}}
\right\} \, .
\end{equation*}
Performing the model average in this manner instead yields results for $\sqrt{t_0}, w_0$ consistent within a fraction of a sigma of Eqs.~\eqref{eq:t0_result} and \eqref{eq:w0_result}.
}
In all cases, the dominant uncertainty is statistical, suggesting a straightforward path to reducing the uncertainty to a few per-mille.

To arrive at our final determination of $\sqrt{t_0}$, \eqnref{eq:t0_result} and $w_0$, \eqnref{eq:w0_result}, we perform an average of the results in \eqnref{eq:final_t0w0_avg_eps2a}.
As the data between the two choices differ slightly, we can not perform this final averaging step under the Bayes model-averaging procedure; instead we treat each result with equal weight.  We would add half the difference between the central values as an additional discretization uncertainty, but as is evident from \eqnref{eq:final_t0w0_avg_eps2a}, the central values are essentially the same.

In \figref{fig:model_breakdown}, we also compare our result with other values in the literature.
All the results, except the most recent one from BMWc~\cite{Borsanyi:2020mff}, have been determined in the isospin symmetric limit.
Our results are in good agreement with the more recent and precise results, though one notes, there is some tension in the values of $\sqrt{t_0}$ and $w_0$ reported.

In \figref{fig:extrapolation}, we show the resulting extrapolation of $\sqrt{t_{0,\rm orig}}m_\O$ and $w_{0,\rm orig} m_\O$ projected into the $l_F^2$ plane using the \nxlo{3} analysis including the $\ln(m_\pi)$ type corrections.
The finite lattice spacing bands are plotted with a value of $\e_a^2$ taken from the near-physical pion mass ensembles from \tabref{tab:lattice_fits}, a15m135XL, a12m130 and a09m135.  For the $a\sim0.06$~fm band, we use the value,
\begin{equation}\label{eq:w0_a06_milc}
\frac{w_{0,{\rm orig}}}{a_{06}}=3.0119(19)\, ,
\end{equation}
from Table IV of Ref.~\cite{Bazavov:2015yea} with $m^\prime_l/m^\prime_s=1/27$ and use this to construct $\e_a^2$.
The data points are plotted after being shifted to the extrapolated values of all the parameters using the posterior values of the LECs from the \nxlo{3} fit.

The lower panel of \figref{fig:extrapolation} is similarly constructed by shifting all the data points to the infinite volume limit, $l_F^{\rm phys}$, $s_F^{\rm phys}$ and the value of $\e_a^2$ from the particular ensemble with the corresponding band in this same limit and only varying $\e_a^2$.
We plot the continuum extrapolation of both the original and improved values to demonstrate the impact of the improvement at finite lattice spacing, noting the agreement in the continuum limit.

For $w_0$, there is very little difference between the original and improved values with very similar continuum extrapolations.  In contrast, there is a striking difference between the original and improved values using $\sqrt{t_0}$, though they agree in the continuum limit.
We also observe that the use of $\sqrt{t_{0,\rm orig}}$ is susceptible to larger model-extrapolation uncertainties arising from different choices of parametrizing the continuum extrapolation; see \eqnref{eq:final_t0w0_avg_eps2a}.
Additional results at $a\lesssim0.06$~fm will be required to control the continuum extrapolation using $\sqrt{t_{0}}$ in order to obtain a few-per-mille level of precision.

%-------------------------------------------------------------------------------
%       w0/a interpolation
\subsection{Interpolation of $t_0$ and $w_0$ \label{sec:gf_interpolation}}
With our determination of $t_0$ and $w_0$, Eqs.~\eqref{eq:t0_result} and \eqref{eq:w0_result}, we can determine the lattice spacing for each bare coupling.
We could use the near-physical pion mass ensemble values of the gradient-flow scales, or alternatively, we could interpolate the results to the physical quark mass point using the predicted quark mass dependence~\cite{Bar:2013ora}.
The interpolation can be performed for each lattice spacing separately, or in a combined analysis of all lattice spacings simultaneously.  The latter is preferable in order for us to determine the lattice spacing $a_{06}$ as we only have results at a single pion mass at this lattice spacing.
To perform the global analysis, we use an \nxlo{2} extrapolation function (which has the same form for $t_0$),
\begin{align}\label{eq:w0_xpt_a}
\frac{w_0}{a} &= \frac{w_{0,{\rm ch}}}{a} \bigg\{
    1 + k_l l_F^2 + k_s s_F^2 +k_a \e_{a,{\rm ch}}^2
\nonumber\\&\qquad
    +k_{ll} l_F^4 +k_{lln} l_F^4 \ln(l_F^2) +k_{ls}l_F^2 s_F^2 +k_{ss}s_F^4
    \nonumber\\&\qquad
    +k_{aa} \e_{a,{\rm ch}}^4
    +k_{al} l_F^2 \e_{a,{\rm ch}}^2 +k_{as} s_F^2 \e_{a,{\rm ch}}^2
    \bigg\}\, ,
\\\label{eq:eps_a_0}
\e_{a,{\rm ch}} &\equiv \frac{1}{(2 w_{0,{\rm ch}}/a)}\, .
\end{align}

\bigskip

This global analysis treats the value of LO parameter $\frac{w_{0,{\rm ch}}}{a}$ for each lattice spacing as a separate unknown parameter, and then assumes that the remaining dimensionless LECs are shared between all lattice spacings.  We use this LO parameter to also construct $\e_{a,{\rm ch}}$ which controls the discretization corrections rather than using $\e_a$, as $\e_a$ is half the inverse of the left-hand side of \eqnref{eq:w0_xpt_a}.

It is tempting to think of this as a combined chiral and continuum extrapolation analysis of $w_0$, but it is not as one normally thinks of them.
Because we do not know the lattice spacings already, there remains an ambiguity in the interpretation of $w_{0,\rm ch}/a$ and the LECs $k_a$, $k_{aa}$, $k_{la}$ and $k_{sa}$: we are not able to interpret $w_{0,\rm ch}/a$ as the chiral limit value of $w_0$ divided by the lattice spacing.
%While the LECs accompanying the terms which are only proportional to the pure $l_F$ and/or $s_F$ contributions describe the quark mass dependence of $w_0$, here we are focussed on using this functional form to interpolate the value of $w_0/a$ from the near-physical quark mass ensembles to the physical quark mass point.
We perform this analysis for all four gradient-flow scales with independent LECs for each scale as well as a similar parametrization of the $\e_{a,{\rm ch}}$ parameter.

\begin{figure}
\includegraphics[width=0.49\textwidth]{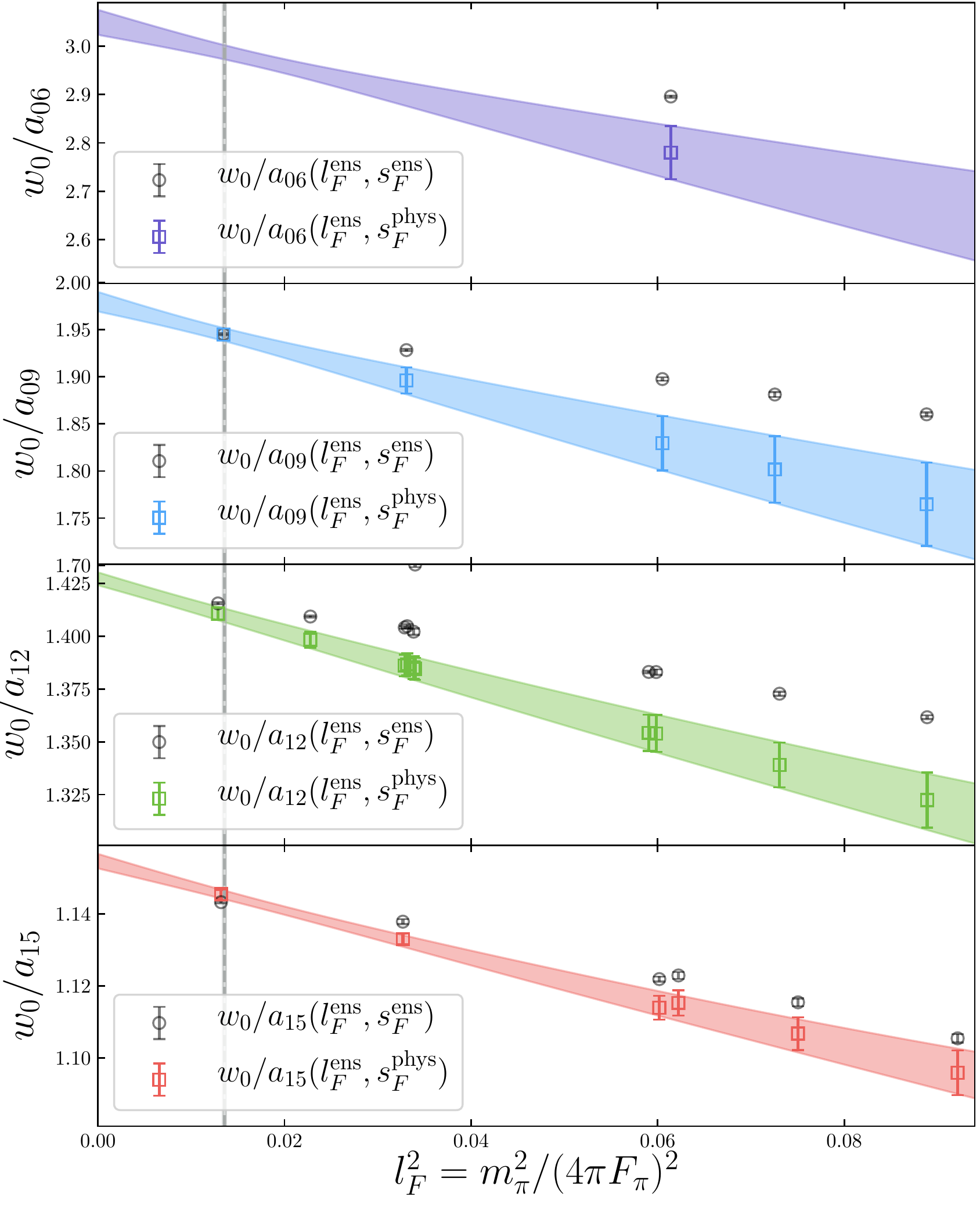}
\caption{\label{fig:gf_interpolation}
Interpolation of $w_0 / a$ values to the physical $l_F$ and $s_F$ point.
The open circles are the raw data from \tabref{tab:lattice_fits} and the open squares have been shifted to the infinite volume and $s_F^{\rm phys}$ point.
The vertical gray line is at $l_F^{2,\rm phys}$.
}
\end{figure}

When we perform the interpolation for each lattice spacing separately, we utilize this same expression except that we set all parameters proportional to any power of $\e_{a,{\rm ch}}$ to zero.  When the individual interpolations are used, the resulting values of $a_{15}$, $a_{12}$ and $a_{09}$ are compatible with those from the global analysis well within 1 standard deviation.  In \figref{fig:gf_interpolation}, we show an example interpolation of $w_{0,\rm orig}/a$ using the global analysis of all ensembles to the $l_F^{\rm phys}$ and $s_F^{\rm phys}$ point for each lattice spacing.
The open circles show the raw values of $w_0/a$ while the open squares show the values shifted to $\frac{w_0}{a}(l_F^{\rm ens}, s_F^{\rm phys}, \e_a^{\rm ens})$ using the resultant parameters determined in the global analysis.
%-------------------------------------------------------------------------------

A quark-mass independent determination of the lattice spacing can be made by using the determination of $\sqrt{t_0}$, \eqnref{eq:t0_result} or $w_0$, \eqnref{eq:w0_result} at the physical point, combined with the physical-quark mass interpolated values of $t_0/a^2$ or $w_0/a$ from either the original or improved values of these gradient flow scales.
Each choice represents a different scheme for setting the lattice spacing.
The continuum extrapolated value of some observable quantity, using any of these schemes, should agree in the continuum limit, while at finite lattice spacing, the results can be substantially different, as is evident in \figref{fig:extrapolation}.

In \tabref{tab:a_fm}, we provide the determination of the lattice spacing for each bare coupling, expressed in terms of the approximate lattice spacing.
It is interesting to note that the determination of the lattice spacing with $\sqrt{t_{0,\rm orig}}$ is substantially different than with the other gradient-flow scales, while the scale determined with from the three remaining auxiliary scales are very similar.

%-------------------------------------------------------------------------------
\begin{table}
\caption{\label{tab:a_fm}
The value of the lattice spacing in fm using the interpolation of the four different determinations of the gradient-flow scales, combined with the determination of $\sqrt{t_0}$ and $w_0$ at the physical point, \eqnref{eq:t0_result} and  \eqnref{eq:w0_result} respectively.
Any one scheme for determining the lattice spacing can be picked as a quark-mass independent scale setting that can be used to convert results from dimensionless lattice units to MeV.
The different schemes will result in different approaches to the continuum; however all choices should agree in the continuum limit.
}
\begin{ruledtabular}
\begin{tabular}{ccccc}
Scheme              & $a_{15}$/fm& $a_{12}$/fm& $a_{09}$/fm& $a_{06}$/fm\\
\hline
$t_{0,\rm orig}/a^2$& 0.1284(10)& 0.10788(83)& 0.08196(64)& 0.05564(44)\\
$t_{0,\rm imp}/a^2$ & 0.1428(10)& 0.11735(87)& 0.08632(65)& 0.05693(44)\\
$w_{0,\rm orig}/a$  & 0.1492(10)& 0.12126(87)& 0.08789(71)& 0.05717(51)\\
$w_{0,\rm imp}/a$   & 0.1505(10)& 0.12066(88)& 0.08730(70)& 0.05691(51)
\end{tabular}
\end{ruledtabular}
\end{table}

%----------------------------------------------------------
%    Summary
\section{Summary and Discussion \label{sec:summary}}
We have performed a precise scale setting with our MDWF on gradient-flowed HISQ action~\cite{Berkowitz:2017opd} achieving a total uncertainty of $\sim0.6\%$--$0.8\%$ for each lattice spacing, \tabref{tab:a_fm}.
The scale setting was performed by extrapolating the quantities $\sqrt{t_0}m_\O(l_F, s_F, \e_a, m_\pi L)$ and $w_0 m_\O(l_F, s_F, \e_a, m_\pi L)$ to the continuum ($\e_a\rightarrow0$), infinite volume ($m_\pi L\rightarrow\infty$) and physical quark mass limits ($l_F\rightarrow l_F^{\rm phys}$ and $s_F\rightarrow s_F^{\rm phys}$), and using the experimental determination of $m_\O$ to determine the scales $\sqrt{t_0}$ and $w_0$ in fm.
The values of $\sqrt{t_{0,\rm orig}}/a$, $\sqrt{t_{0,\rm imp}}/a$, $w_{0,\rm orig}/a$ and $w_{0,\rm imp}/a$ were interpolated to the infinite volume and physical quark mass limits for each lattice spacing, allowing for the quark-mass independent determination of $a$ for each bare coupling $\beta$, expressed in terms of the approximate lattice spacing; see \tabref{tab:a_fm}.

Of note, the approach to the continuum limit of $\sqrt{t_{0,\rm orig}}m_\O$ and $\sqrt{t_{0,\rm imp}}m_\O$ are quite different, \figref{fig:extrapolation}, with the use of $\sqrt{t_{0,\rm imp}}$ leading to an almost flat continuum extrapolation.  The two different extrapolations agree quite nicely in the continuum limit, as they must if all systematic uncertainties are under control.
In contrast, the use of the original and improved values of $w_0$ leads to very similar continuum extrapolations of $w_{0,\rm orig}m_\O$ and $w_{0,\rm imp}m_\O$, which also agree very nicely in the continuum limit.

We also observe that the use of $l_\O$ and $s_\O$ as small parameters to control the quark-mass interpolation are relatively heavily penalized as compared to the use of $l_F$ and $s_F$; see \tabref{tab:model_statistics} for an example.  We observe the same qualitative weighting with all choices of the gradient-flow scale.  Perhaps this is an indication that this parametrization is suboptimal.

Our final uncertainty using $w_0$ is dominated by the stochastic uncertainty, \eqnref{eq:w0_result}, providing a clear path to reducing the uncertainty by almost a factor of 3 before an improved understanding of the various systematic uncertainties becomes relevant.
At such a level of precision, a systematic study of the effect of isospin breaking on the scale setting, as has been performed by BMWc~\cite{Borsanyi:2020mff}, is likely required to retain full control of the uncertainty.
For $\sqrt{t_0}$, we observe the model-selection uncertainty is comparable to the stochastic uncertainty, \eqnref{eq:t0_result}, which arises from the different ways to parameterize the continuum extrapolation; see \eqnref{eq:final_t0w0_avg_eps2a}.
Therefore, additional results at $a\lesssim0.06$~fm will be required to obtain a fer-per-mille precision with $\sqrt{t_0}$.

The pursuit of our physics program of determining the nucleon elastic structure functions and improving the precision of our $g_A$ result~\cite{Chang:2018uxx,Berkowitz:2018gqe} will naturally lead to an improved scale setting precision.  The current precision is already expected to be subdominant for most of the results we will obtain, but a further improved precision is welcome.

%----------------------------------------------------------
%    ACKNOWLEDGEMENTS
\begin{acknowledgments}
We thank M.~Goltermann and O.~B\"{a}r for useful discussion and correspondence about the chiral corrections to $w_0$.
We thank Andrea Shindler for helpful discussions regarding reweighting.
We thank Peter Lepage for helpful correspondence regarding \texttt{lsqfit} and the log Gaussian Bayes factor.
We thank K.~Orginos for use of \texttt{wm\_chroma} that was used to compute some of the correlation functions used in this work.
We thank the MILC Collaboration for providing some of the HISQ configurations used in this work, and A. Bazavov, C. Detar and D. Toussaint for guidance on using their code to generate the new HISQ ensembles also used in this work.
We thank R. Sommer for helpful correspondence and encouragement to determine $t_0$ as well as $w_0$.

Computing time for this work was provided through the Innovative and Novel Computational Impact on Theory and Experiment (INCITE) program and the LLNL Multiprogrammatic and Institutional Computing program for Grand Challenge allocations on the LLNL supercomputers.
This research utilized the  NVIDIA GPU-accelerated Titan and Summit supercomputers at Oak Ridge Leadership Computing Facility at the Oak Ridge National Laboratory, which is supported by the Office of Science of the U.S. Department of Energy under Contract No. DE-AC05-00OR22725 as well as the Surface, RZHasGPU, Pascal, Lassen, and Sierra supercomputers at Lawrence Livermore National Laboratory.

The computations were performed utilizing \texttt{LALIBE}~\cite{lalibe} which utilizes the \texttt{Chroma} software suite~\cite{Edwards:2004sx} with \texttt{QUDA} solvers~\cite{Clark:2009wm,Babich:2011np} and HDF5~\cite{hdf5} for I/O~\cite{Kurth:2015mqa}.  They were efficiently managed with \texttt{METAQ}~\cite{Berkowitz:2017vcp,Berkowitz:2017xna} and status of tasks logged with EspressoDB~\cite{Chang:2019khk}.
The HMC was performed with the MILC Code~\cite{milc:code}, and for the ensembles new in this work, running on GPUs using \texttt{QUDA}.  The final extrapolation analysis utilized \texttt{gvar} v11.2~\cite{gvar:11.2} and \texttt{lsqfit} v11.5.1~\cite{lsqfit:11.5.1}. The analysis and data for this work can be found at this git repo: \url{https://github.com/callat-qcd/project_scale_setting_mdwf_hisq}.

This work was supported by the NVIDIA Corporation (MAC), the Alexander von Humboldt Foundation through a Feodor Lynen Research Fellowship (CK), the DFG and the NSFC Sino-German CRC110 (EB), the RIKEN Special Postdoctoral Researcher Program (ER), the U.S. Department of Energy, Office of Science, Office of Nuclear Physics under Awards No.~DE-AC02-05CH11231 (CCC, CK, BH, AWL), No.~DE-AC52-07NA27344 (DAB, DH, ASG, PV), No.~DE-FG02-93ER-40762 (EB), No.~DE-AC05-06OR23177 (CM); the Nuclear Physics Double Beta Decay Topical Collaboration (DAB, HMC, AN, AWL); the U.K. Science and Technology Facilities Council Grants No.~ST/S005781/1 and No.~ST/T000945/1 (CB); and the DOE Early Career Award Program (CCC, AWL).

\end{acknowledgments}

%\onecolumngrid
\appendix

\section{CHARM QUARK MASS REWEIGHTING \label{app:reweighting}}

The use of reweighting~\cite{Ferrenberg:1988yz} to estimate a correlation function with a slightly different sea-quark mass than the one simulated with is very common in LQCD; see for example Refs.~\cite{Aoki:2009ix,Luscher:2012av,Blum:2014tka}.  A nice discussion of mass reweighting, including single flavor reweighting is found in Refs.~\cite{Finkenrath:2013soa,Leder:2014ota}.

In our case, we are interested in reweighting the computation from the hybrid Monte Carlo (HMC) simulated charm quark mass, $m_c^{\rm HMC}$, to the physical quark mass, $m_c^{\rm phys}$ which requires an estimate of the ratio of the fermion determinant with the physical mass to the determinant with the HMC mass.
If the mass shift is $a\; \d m_c = m_c^{\rm phys} - m_c^{\rm HMC}$ then up to $\mathcal{O}(\delta m^2)$ this ratio can be written (including the quarter-root arising from rooted-staggered fermions)
\begin{equation}
	\label{eq:reweighting factor}
	w^4_i = \det\left[ \one + \delta m\; (D[U_i]+ m_c^{\rm phys})^{-1} \right]
\end{equation}
for each configuration $U_i$ and observables may be computed using the weight $w$,
\begin{equation}
	\label{eq:reweighting procedure}
	\langle O \rangle_{\rm phys} = \frac{ \langle O w \rangle_{\rm HMC}}{\langle w \rangle_{\rm HMC}}.
\end{equation}

We can use two methods to stochastically estimate $w$ for each configuration.  First, by rewriting the determinant as the exponential of a trace-log, one finds
\begin{equation}
	\label{eq:trace reweighting}
	w_i^{\rm tr} = \exp \frac{\delta m}{4} \tr \left[(D[U_i]+m_c^{\rm phys})^{-1}\right] + \mathcal{O}(\delta m^2)
\end{equation}
and we can use vectors of complex Gaussian noise $\eta$ to estimate the trace,
\begin{align}
	\label{eq:noisy trace}
	\tr \left[(D[U_i]+m_c^{\rm phys})^{-1}\right] &\approx \frac{1}{N_\eta} \sum_i \eta_i^{\dagger} (D[U_i]+m_c^{\rm phys})^{-1} \eta_i\,,
	\\
	\label{eq:gaussian noise}
	\eta &\sim \frac{1}{\pi^V} \exp - \eta^\dagger \eta \, ,
\end{align}
where $V$ is the size of each $\eta$ vector.

Alternatively, we may estimate the determinant in the reweighting factor \eqref{eq:reweighting factor} using the identity,
\begin{equation}
	\det A = \pi^{-V} \int \mathcal{D}\eta\; \mathcal{D}\eta^\dagger\; e^{-\eta^\dagger A^{-1} \eta},
\end{equation}
which is often used to implement pseudofermions.  Up to $\mathcal{O}(\delta m^2)$, this becomes
\begin{equation}
	w^4_i = \pi^{-V} \int \mathcal{D}\eta\; \mathcal{D}\eta^\dagger\; e^{-\eta^\dagger\eta} e^{\delta m\; \eta^\dagger (D[U_i]+m_c^{\rm phys})^{-1} \eta} \, ,
\end{equation}
which tells us to draw $\eta$ according to the same Gaussian \eqref{eq:gaussian noise} and estimate
\begin{equation}
	\label{eq:pseudofermion reweighting}
	w^{\rm ps}_i = \left( \frac{1}{N_\eta} \sum_j e^{\delta m\; \eta_j^\dagger (D[U_i]+m_c^{\rm phys})^{-1} \eta_j}\right)^{1/4}.
\end{equation}

Both the trace method \eqref{eq:trace reweighting} and the pseudofermion method \eqref{eq:pseudofermion reweighting} are only valid to $\mathcal{O}(\d m^2)$; when they agree we assume those corrections are under control.
In order to stabilize the numerical estimate of the reweighting factors, it is also common to split the reweighting factor into a product of reweighting factors where each is computed with a fraction of the full mass shift~\cite{Hasenbusch:2001ne,Hasenfratz:2008fg,Liu:2012gm}.
For example, with a simulated mass of $m_1$ and target mass of $m_1+\D m$, one could use two steps of $\D m/2$ and estimate the reweighting factor with the trace method,
\begin{align}\label{eq:reweight_product}
	w_i^{\rm tr} &= w^{\rm tr}_{i,1} w^{\rm tr}_{i,2}\, ,
\\
	w^{\rm tr}_{i,1} &=
	\exp \frac{\Delta m / 2}{4} \frac{1}{N_{\eta}} \sum_j \eta^\dagger_j \left(D[U_i] + m_1 + \frac{\Delta m}{2}\right)^{-1} \eta_j\, ,
\nonumber\\
	w^{\rm tr}_{i,2} &=
	\exp \frac{\Delta m / 2}{4} \frac{1}{N_{\theta}} \sum_j \theta^\dagger_j \left(D[U_i] + m_1 + \Delta m\right)^{-1} \theta_j \, ,
\nonumber
\end{align}
using independently sampled complex Gaussian noise $\eta$ and $\theta$.
Of course, one may split the shift $\D m$ into finer steps if needed, for increased computational cost.

The reweighting factor accounts for a change in the action and is exponential in the spacetime volume.
This can lead to numerical under- or overflow.
As a cure, we factor out the average reweighting factor.
Recognizing the trace of the inverse Dirac operator on a configuration $U_i$ as the scalar quark density times the lattice volume,
\begin{equation}
	\left(V \bar{q} q\right)_i = \tr \left[(D[U_i]+m_q)^{-1}\right] \, ,
\end{equation}
we can rescale $w$, shifting by the ensemble average $V \langle\bar{c}c\rangle$ computed via \eqref{eq:noisy trace}.
For example, rescaling the trace method \eqref{eq:trace reweighting} gives
\begin{equation}
	w_i^{\rm tr} = \exp \frac{\d m}{4}\left(\tr\left[(D[U_i]+m_c^{\rm phys})^{-1}\right] - V\langle \bar{c}c\rangle\right);
\end{equation}
such a rescaling cancels exactly in the reweighting procedure \eqref{eq:reweighting procedure}.
A similar rescaling cures the pseudofermion method \eqref{eq:pseudofermion reweighting}.
If we split the mass shift as in \eqref{eq:reweight_product}, each factor of the weight may be independently so stabilized.

On the a06m310L ensemble the lattice volume and shift in the mass are $V = 72^3 \times 96$ and
\begin{align}
\d a_{06} m_c &= a_{06}m_c^{\rm phys} - a_{06}m_c^{\rm HMC}
\nonumber\\&
    = -0.0281(4)\, .
\end{align}
While the shift in the mass is only about 10\% of the physical charm quark mass, it is of the order of the physical strange quark mass.
In order to stabilize the numerical estimate of the reweighting factors, we split this mass shift into ten equal steps, and for each step, we used $N_\eta = 128$ independent Gaussian random noise vectors.
For each step in the reweighting, we used the same Naik value of $\e_N=-0.0533$ as was used in the original HMC.  This ensures that the Dirac operator only differs from one mass to the next by the quark mass itself.
As the Naik term is used to improve the approach to the continuum limit, this is a valid choice to make as it results in a slightly different approach to the continuum than if one simulated at the physical charm quark mass with the optimized Naik value for that mass.

\begin{figure}
\includegraphics[width=\columnwidth]{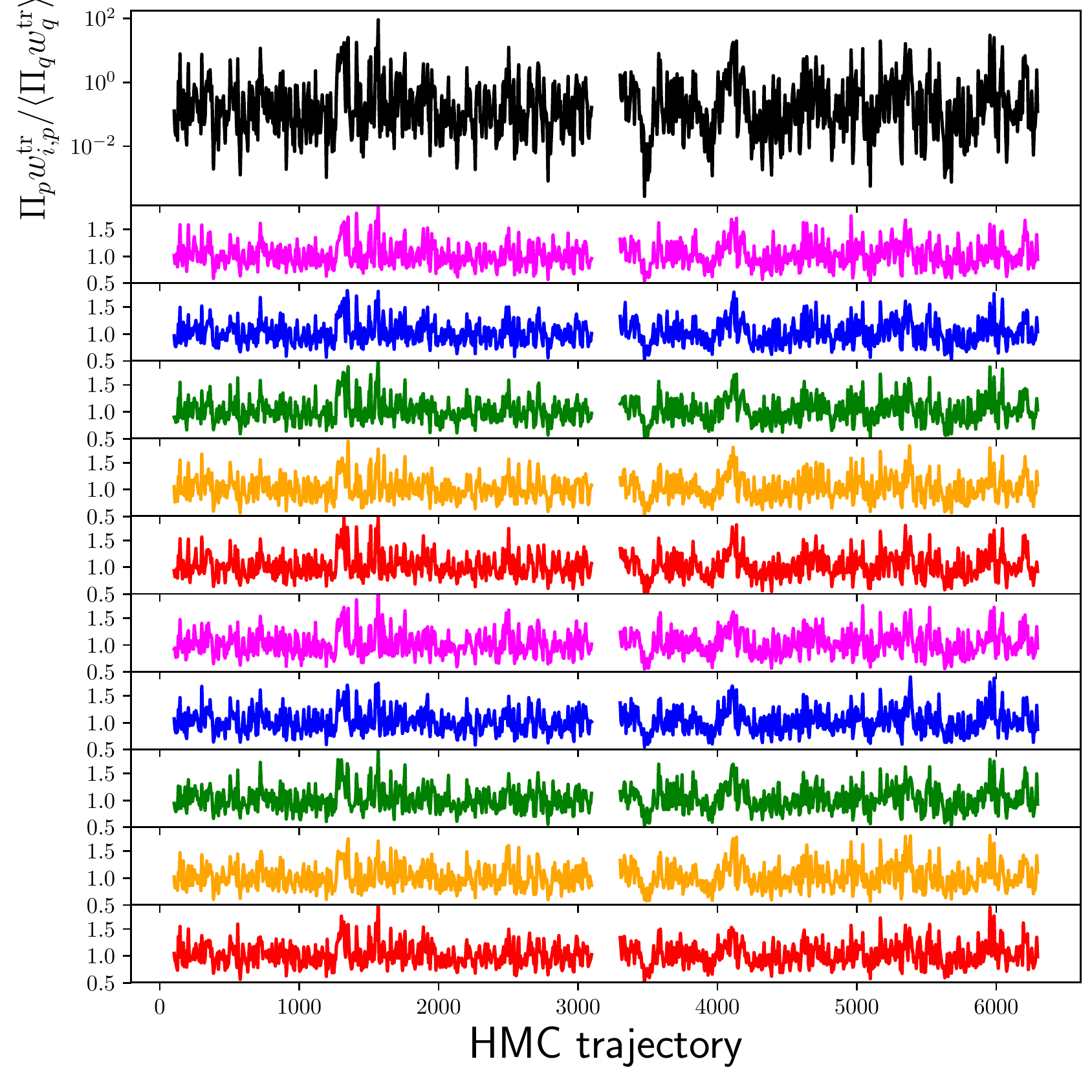}
\caption{\label{fig:rw_histo}
Distribution of reweighting factors, $r$, for the a06m310L ensemble.  The reweighting was performed with ten equal steps of the mass difference from $a_{06}m_c^{\rm HMC}=0.286$ to $a_{06}m_c^{\rm phys}=0.2579$.  The reweighting factors from each step are shown in the bottom ten subpanels with the product reweighting factor shown in the top panel.
The a06m310L ensemble was generated with two different streams of equal length.  For more information on the ensembles, see Ref.~\cite{Miller:2020xhy}.
}
\end{figure}

In \figref{fig:rw_histo}, we show the reweighting factors for each of the mass steps, with the bottom panel having a mass $a_{06}m_c = 0.28319$ closest to the HMC mass and the second from top panel having $a_{06}m_c^{\rm phys} = 0.2579$, scaled for numerical stability.  The top panel is the resulting product reweighting factor normalized by the average reweighting factor.  One observes that there are a few large reweighting factors of $\mathrm{O}(100)$.
We have verified that the trace estimation method \eqref{eq:trace reweighting} and the pseudofermion method \eqref{eq:pseudofermion reweighting} produce comparable normalized reweighting factors.
The large reweighting factors are likely due to the parent HMC distribution of configurations having a suboptimal overlap with the physical, target distribution.

%-------------------------------------------------------------------------------
% reweighting histogram
%-------------------------------------------------------------------------------

\subsection{Reweighted spectrum}
The next task is to understand how this reweighting impacts the extracted spectrum.
To aid in this discussion, we reiterate our strategy for fitting the correlation functions.
Because the noise of the omega baryon correlation function grows in Euclidean time, it is the most challenging to fit and so we focus our discussion on the omega.
Our strategy is to find a good quality fit to the correlation function for which the extracted ground state energy is stable against the number of states and the time-range used in the analysis.
For a given $t_{\rm min}$, we opt to chose the simplest model which satisfies this criteria, which amounts to picking the minimum number of excited states possible.

\begin{figure*}
\includegraphics[width=0.49\textwidth]{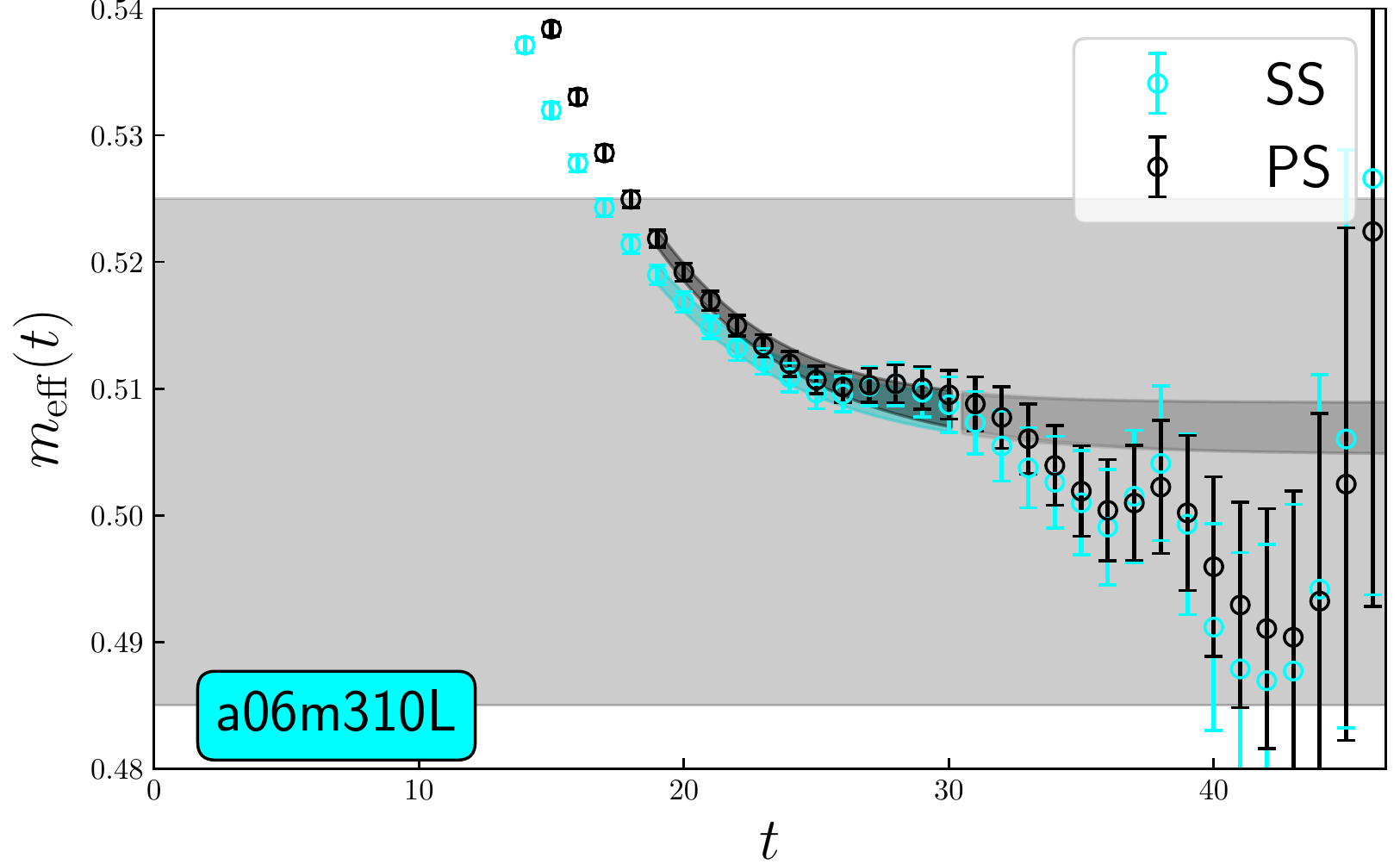}
\includegraphics[width=0.49\textwidth]{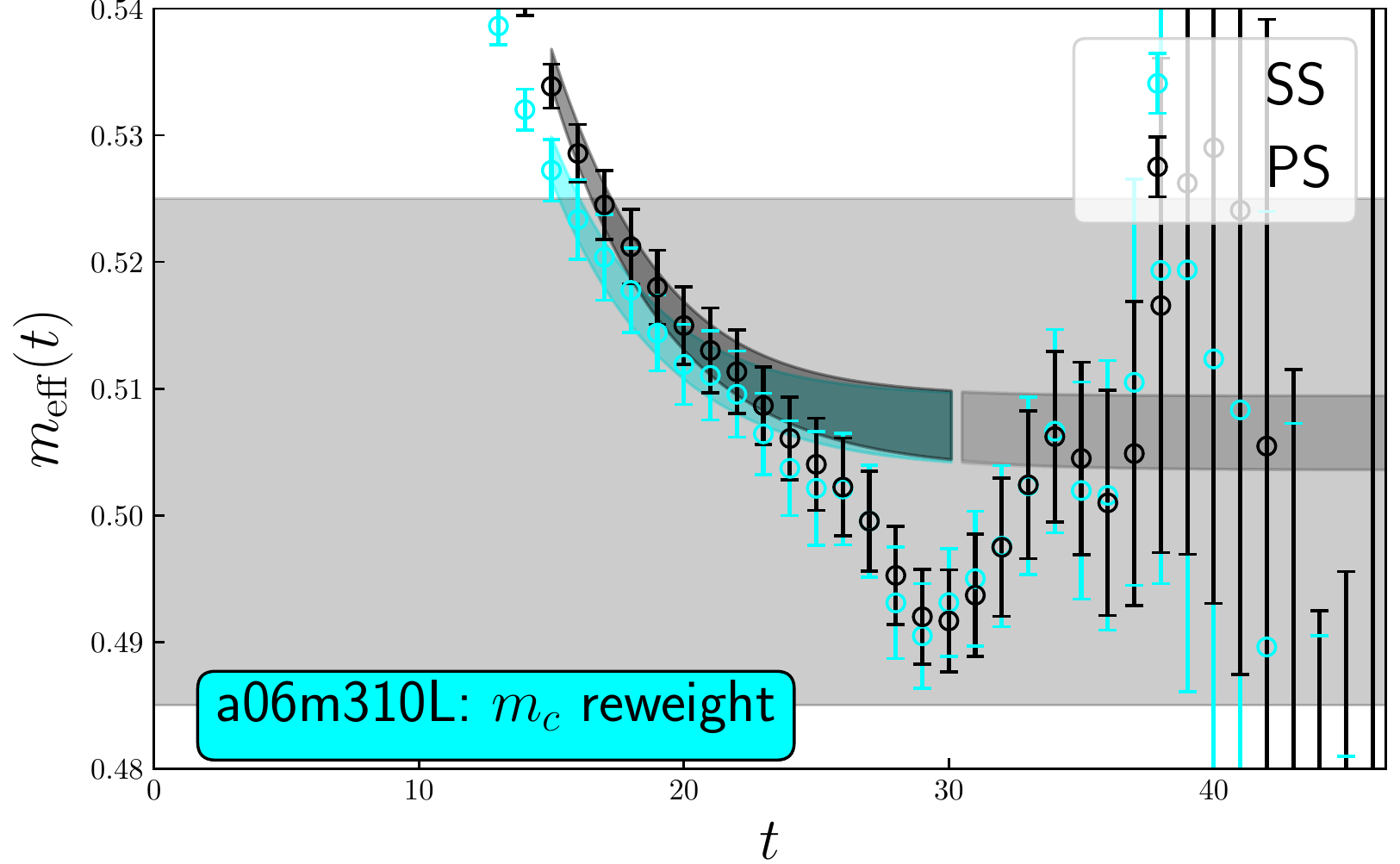}
\includegraphics[width=0.49\textwidth]{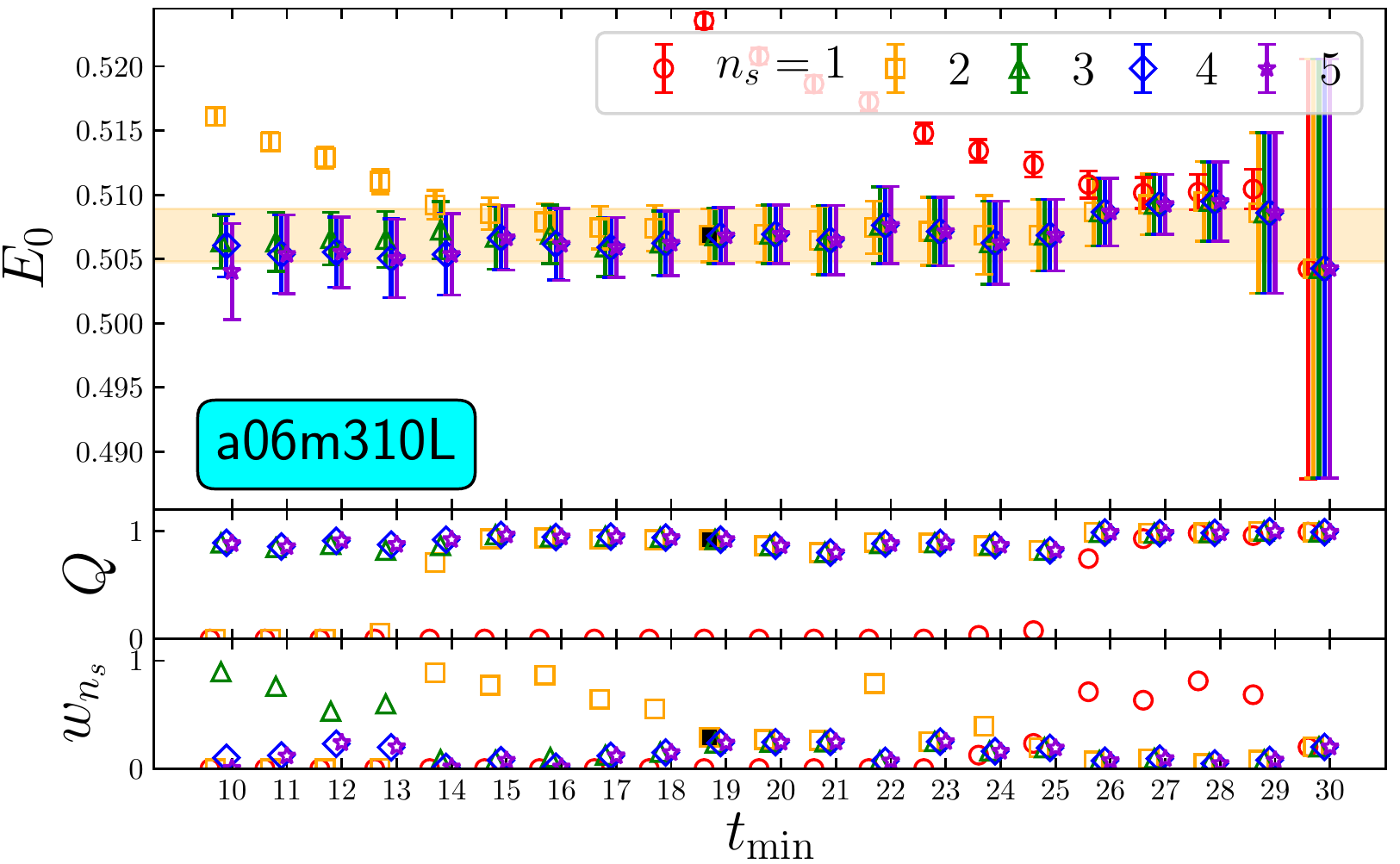}
\includegraphics[width=0.49\textwidth]{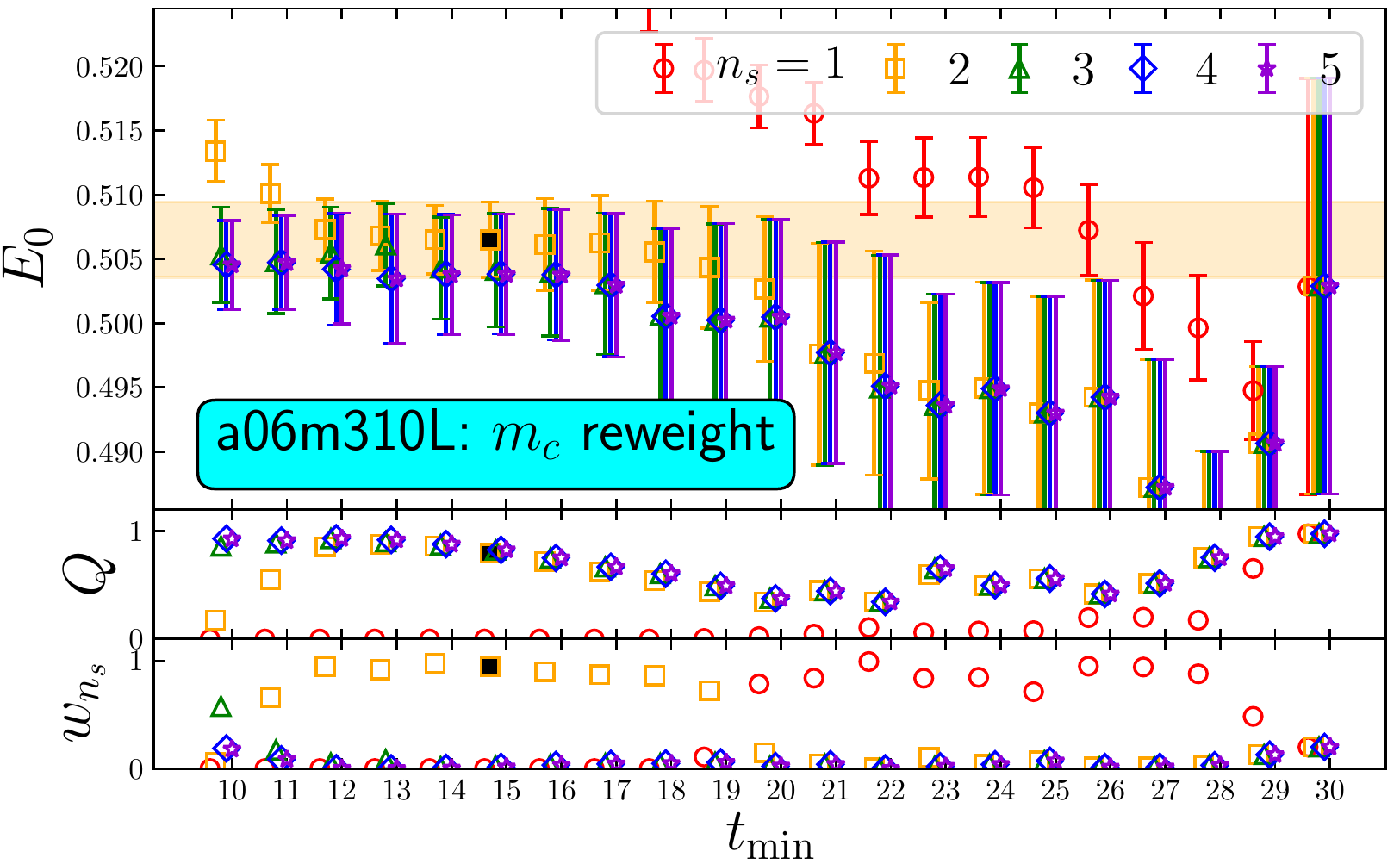}
\caption{\label{fig:stability_a06m310L}
Same as \figref{fig:stability_m135} for the a06m310L ensemble.}
\end{figure*}

The SS correlation functions are positive definite, therefore implying that the excited state contamination of the effective mass must come from above.  When examining the SS omega-baryon effective mass on the a06m310L ensemble, one observes that around $t=25$, the effective mass stops decreasing, and even increases a little.  Because this is not allowed for a positive definite correlation function, we can conclude this must be due to a correlated stochastic fluctuation; see \figref{fig:stability_a06m310L}.
In the rewighted effective mass, one observes more dramatic behavior of the effective mass beginning around the same time.
To be conservative, we set $t_{\rm max}=30$ in our analysis as this allows the analysis to be sensitive to these stochastic fluctuations, which fluctuate in the opposite direction between the unweighted and reweighted configurations.

\begin{figure*}
\includegraphics[width=\columnwidth]{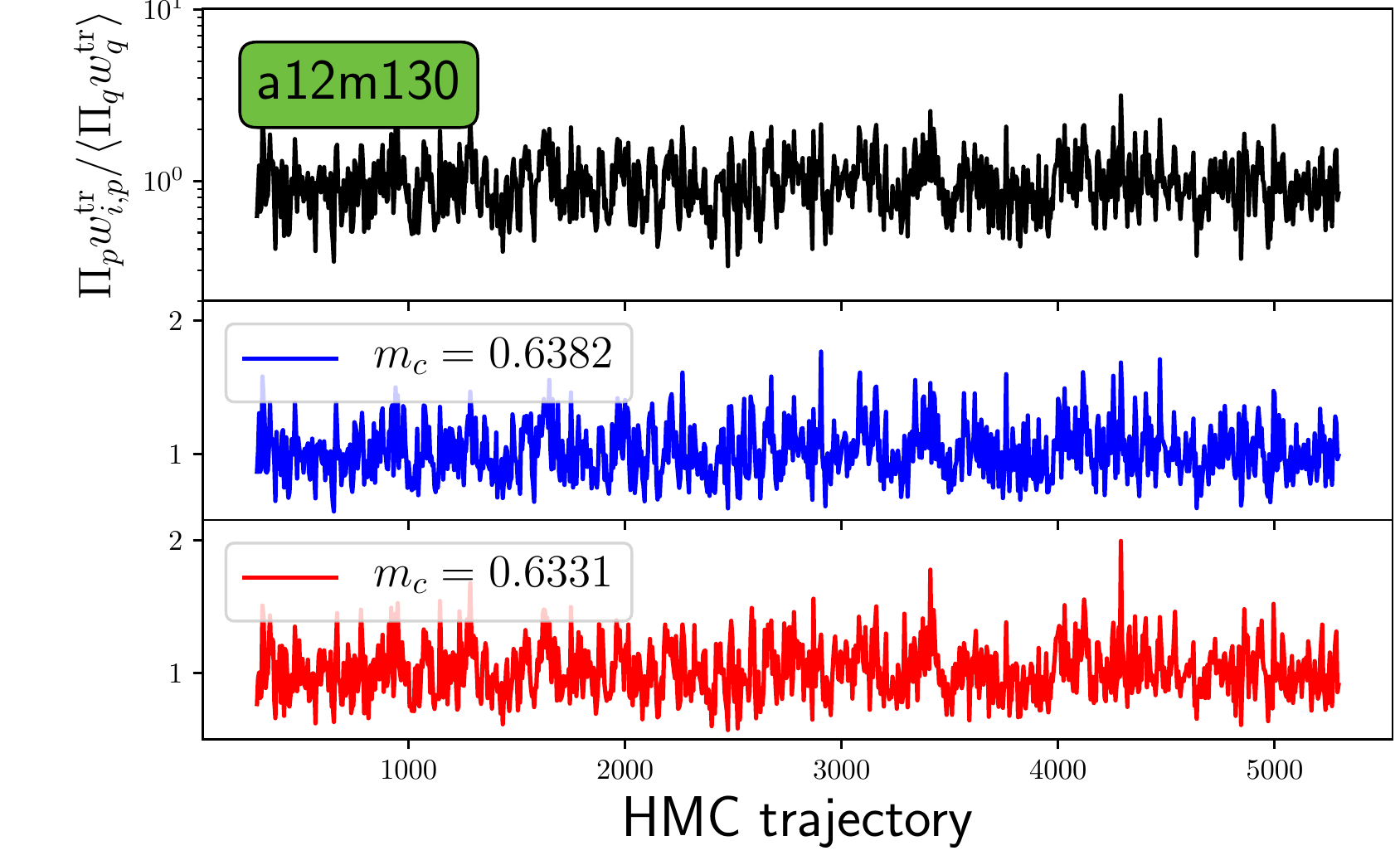}
\includegraphics[width=\columnwidth]{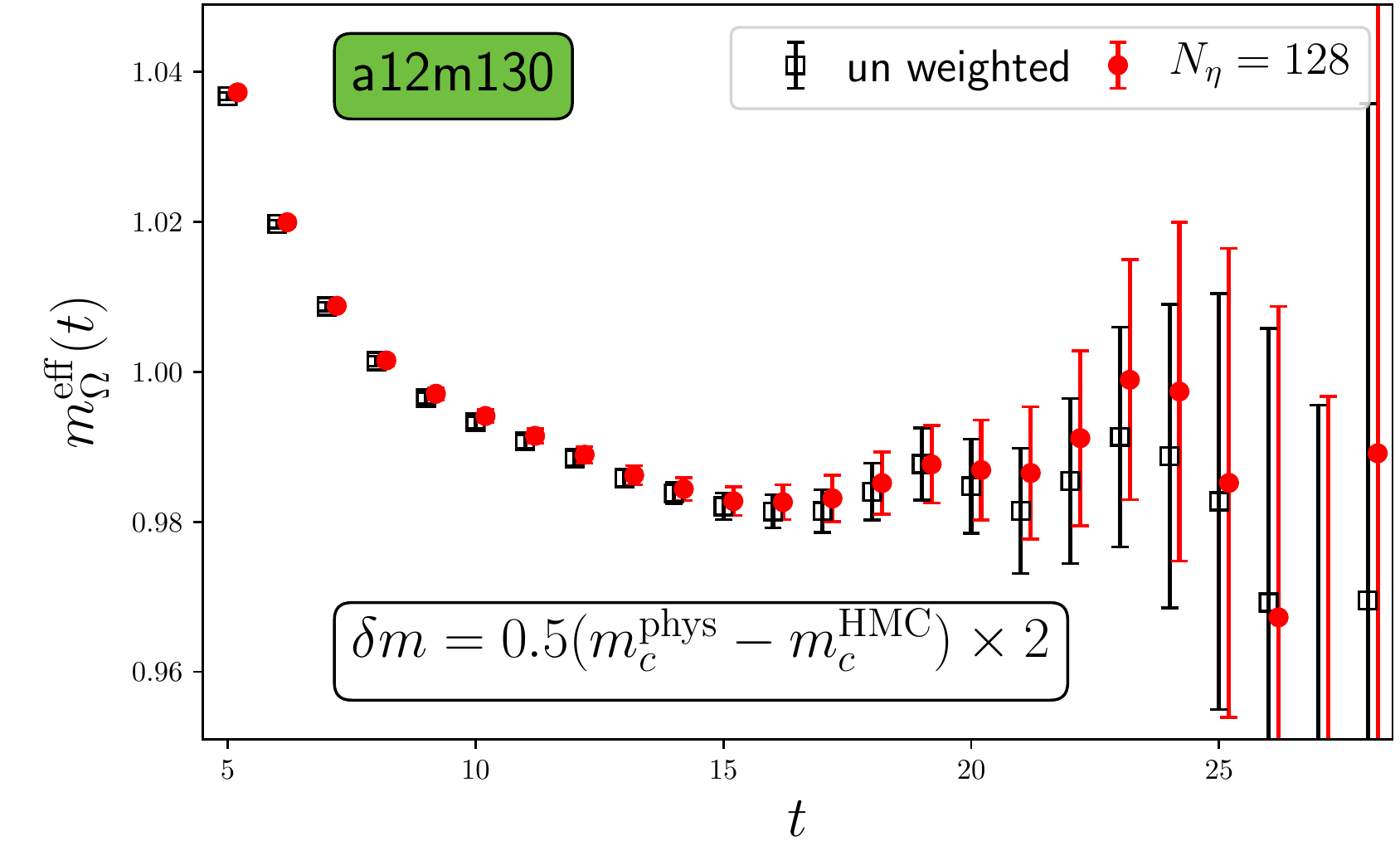}
\caption{\label{fig:a12m130_reweight}
The reweighting factors for the a12m130 ensemble (left) and the reweighted omega baryon effective mass (right).
}
\end{figure*}

As a comparison, we also show the reweighting factors and reweighted omega baryon effective mass on the a12m130 ensemble (\figref{fig:a12m130_reweight}) where the charm quark was 2\% different from its physical value.  In this case, the reweighting factors are much easier to estimate, and we do not observe any large values.

The lower panels in \figref{fig:stability_a06m310L} show the extracted ground state mass as a function of $t_{\rm min}$ and $n_{\rm state}$.  We observe that an $n_{\rm state}=2$ fit beginning at $t_{\rm min}=19$ and $15$ for the unweighted and reweighted correlation functions satisfies our optimization criteria.
These lead to the estimate value of $m_\O$ given in \eqnref{eq:mO_reweighting}.
In \tabref{tab:reweight}, we also show the reweighted value of $m_\pi$ and $m_K$ on this a06m310L ensemble.
While the pion and kaon have a statistically significant shift from the reweighting, when we use the reweighted values of $m_\pi$, $m_K$ and $m_\O$ from this ensemble, the final extrapolated value of $w_0 m_\O$ is within 1 standard deviation of the completely unweighted analysis.  As the a06m310L ensemble has the largest potential change from reweighting, we conclude that at the level of precision we currently have, our results are not sensitive to the slight mistuning of the charm quark mass from its physical value on each of the configurations.

%-------------------------------------------------------------------------------
% reweighting histogram
%-------------------------------------------------------------------------------

%-------------------------------------------------------------------------------
% reweighting histogram
%-------------------------------------------------------------------------------

%-------------------------------------------------------------------------------
% reweighting fit impacts table
\begin{table}
\caption{\label{tab:reweight}
The pion, kaon and omega masses on the a06m310L ensemble with and without reweighting as well as the correlated difference.
}
\begin{ruledtabular}
\begin{tabular}{c c c c}
State & Unweighted& Reweighted& Difference\\
\hline\\ [-0.3cm]
$am_\pi$    & $0.09456(06)$& $0.09419(14)$& $-0.00037(14)$\\
$am_K$      & $0.16204(07)$& $0.16173(16)$& $-0.00032(15)$\\
$am_\Omega$ & $0.5069(21)$& $0.5065(29)$& $-0.0004(34)$
\end{tabular}
\end{ruledtabular}
\end{table}

%-------------------------------------------------------------------------------

\section{MODELS INCLUDED IN AVERAGE}

We use $w_{0,\rm orig} m_\O$ as an example to demonstrate the model averaging and naming conventions for the various models.  The other gradient-flow scale studies have identical in form, extrapolation functions.  \tabref{tab:model_statistics} provides the various models and their relative weights for $w_{0,\rm orig} m_\O$.  Of note, in all four gradient-flow scale extrapolations, the extrapolations which make use of $l_\O$ and $s_\O$ versus $l_F$ and $s_F$, all have relatively smaller weights in the averaging, such that in all cases, these extrapolations could be dropped from the model averaging without effecting the final result.
A reprodution of this model averaging as well as for the other three gradient-flow scales can be obtained by downloading and running the associated analysis software at \url{https://github.com/callat-qcd/project_scale_setting_mdwf_hisq}.

\begin{table*}
\begin{ruledtabular} \centering
\caption{\label{tab:model_statistics} Models included in final model average of $w_{0,\rm orig}m_\O$.}
\begin{tabular}{ r c c c c l l}
Moddel & chi2/dof &   $Q$ &  logGBF& weight& $w_0 m_\O$& $w_0$[fm]\\
\hline
xpt\_n3lo\_alphaS\_F &  1.065   &  0.378&  64.919&  0.077&   1.448(10)  & 0.1709(12)\\
        xpt\_n3lo\_F &  1.070   &  0.372&  64.916&  0.077&  1.4467(91)  & 0.1707(11)\\
taylor\_n3lo\_alphaS\_F &  1.089   &  0.350&  64.863&  0.073&   1.449(10)  & 0.1709(12)\\
xpt\_n3lo\_FV\_alphaS\_F &  1.068   &  0.374&  64.861&  0.073&   1.448(10)  & 0.1709(12)\\
    xpt\_n3lo\_FV\_F &  1.074   &  0.367&  64.858&  0.073&  1.4467(91)  & 0.1707(11)\\
taylor\_n3lo\_FV\_alphaS\_F &  1.088   &  0.350&  64.857&  0.073&   1.449(10)  & 0.1709(12)\\
     taylor\_n3lo\_F &  1.094   &  0.343&  64.857&  0.073&  1.4470(90)  & 0.1707(11)\\
 taylor\_n3lo\_FV\_F &  1.094   &  0.343&  64.851&  0.072&  1.4470(90)  & 0.1707(11)\\
xpt\_n2lo\_alphaS\_F &  1.124   &  0.310&  64.610&  0.057&   1.449(10)  & 0.1710(12)\\
        xpt\_n2lo\_F &  1.130   &  0.304&  64.600&  0.056&  1.4475(91)  & 0.1708(11)\\
xpt\_n2lo\_FV\_alphaS\_F &  1.128   &  0.306&  64.551&  0.054&   1.449(10)  & 0.1710(12)\\
    xpt\_n2lo\_FV\_F &  1.134   &  0.299&  64.540&  0.053&  1.4475(91)  & 0.1708(11)\\
taylor\_n2lo\_alphaS\_F &  1.161   &  0.272&  64.410&  0.047&   1.450(10)  & 0.1711(12)\\
taylor\_n2lo\_FV\_alphaS\_F &  1.161   &  0.272&  64.403&  0.046&   1.450(10)  & 0.1711(12)\\
     taylor\_n2lo\_F &  1.168   &  0.265&  64.392&  0.046&  1.4480(90)  & 0.1708(11)\\
 taylor\_n2lo\_FV\_F &  1.167   &  0.266&  64.386&  0.045&  1.4480(90)  & 0.1708(11)\\
xpt\_n2lo\_FV\_alphaS\_O &  0.971   &  0.498&  59.864&  0.000&   1.441(14)  & 0.1700(16)\\
xpt\_n3lo\_FV\_alphaS\_O &  0.937   &  0.545&  59.813&  0.000&   1.440(13)  & 0.1699(16)\\
taylor\_n3lo\_alphaS\_O &  1.089   &  0.349&  59.792&  0.000&   1.444(13)  & 0.1704(15)\\
xpt\_n2lo\_alphaS\_O &  0.985   &  0.480&  59.789&  0.000&   1.441(13)  & 0.1701(16)\\
    xpt\_n3lo\_FV\_O &  0.937   &  0.545&  59.515&  0.000&   1.440(12)  & 0.1699(15)\\
taylor\_n3lo\_FV\_alphaS\_O &  1.089   &  0.349&  59.394&  0.000&   1.444(13)  & 0.1704(15)\\
xpt\_n3lo\_alphaS\_O &  0.949   &  0.529&  59.380&  0.000&   1.441(13)  & 0.1700(16)\\
    xpt\_n2lo\_FV\_O &  0.971   &  0.498&  59.234&  0.000&   1.441(12)  & 0.1700(14)\\
        xpt\_n2lo\_O &  0.985   &  0.480&  59.172&  0.000&   1.441(13)  & 0.1700(15)\\
     taylor\_n3lo\_O &  1.090   &  0.348&  58.874&  0.000&   1.444(12)  & 0.1703(14)\\
 taylor\_n3lo\_FV\_O &  1.090   &  0.349&  58.798&  0.000&   1.444(12)  & 0.1703(14)\\
taylor\_n2lo\_alphaS\_O &  1.131   &  0.303&  58.742&  0.000&   1.445(13)  & 0.1705(15)\\
 taylor\_n2lo\_FV\_O &  1.131   &  0.302&  58.607&  0.000&   1.444(12)  & 0.1704(14)\\
        xpt\_n3lo\_O &  0.949   &  0.528&  58.329&  0.000&   1.440(12)  & 0.1699(14)\\
taylor\_n2lo\_FV\_alphaS\_O &  1.131   &  0.303&  58.292&  0.000&   1.445(13)  & 0.1705(15)\\
     taylor\_n2lo\_O &  1.132   &  0.302&  58.070&  0.000&   1.444(12)  & 0.1704(14)\\
\hline
    Model average& &&&& 1.4481(98)(11)& 0.1709(12)(01)
\end{tabular}
\end{ruledtabular}
\end{table*}

A few example extrapolation formula are given below to demonstrate the naming convention:

\begin{widetext}
\begin{subequations}
\begin{align}
% Fpi\_lo\_fv
w_0 m_\O &=
    \underbrace{c_0}_\text{LO}
    +\underbrace{\d^\text{NLO}(l_F, s_F)}_\text{\color{teal} chiral NLO}
    +\underbrace{\d^{\text{NLO}}_{a,F}}_\text{\color{magenta} disc NLO}
& \texttt{[xpt\_nlo\_fv\_F]}
\\
% Fpi\_nlo\_log\_fv
w_0 m_\O &=
    \underbrace{c_0}_\text{LO}
    +\underbrace{\d^\text{NLO}(l_F, s_F)}_\text{\color{teal} chiral NLO}
    +\underbrace{\d^\text{\nxlo{2}}(l_F, s_F)}_\text{\color{teal} chiral \nxlo{2}}
    +\underbrace{\d^\text{\nxlo{2}}_{\ln}}_\text{\color{teal} chiral log \nxlo{2}}
& \texttt{[xpt\_n2lo\_fv\_F]} \nonumber\\
    & \phantom{=}
    +\underbrace{\d^{\text{NLO}}_{a,F}}_\text{\color{magenta} disc NLO}
    +\underbrace{\d^{\text{\nxlo{2}}}_{a,F}}_\text{\color{magenta} disc \nxlo{2}}
    +\underbrace{\d^\text{\nxlo{2}}_{L, F}}_\text{\color{orange} FV \nxlo{2}}
\\
% Fpi\_n2lo\_log\_log2\_fv
w_0 m_\O &=
    \underbrace{c_0}_\text{LO}
    +\underbrace{\d^\text{NLO}(l_F, s_F)}_\text{\color{teal} chiral NLO}
    +\underbrace{\d^\text{\nxlo{2}}(l_F, s_F)}_\text{\color{teal} chiral \nxlo{2}}
    +\underbrace{\d^\text{\nxlo{3}}(l_F, s_F)}_\text{\color{teal} chiral \nxlo{3}}
    +\underbrace{\d^\text{\nxlo{2}}_{\ln}}_\text{\color{teal} chiral log \nxlo{2}}
    +\underbrace{\d^\text{\nxlo{3}}_{\ln}}_\text{\color{teal} chiral logs \nxlo{3}}
& \texttt{[xpt\_n3lo\_fv\_F]} \nonumber\\
& \phantom{=}
    +\underbrace{\d^{\text{NLO}}_{a,F}}_\text{\color{magenta} disc NLO}
    +\underbrace{\d^{\text{\nxlo{2}}}_{a,F}}_\text{\color{magenta} disc \nxlo{2}}
    +\underbrace{\d^{\text{\nxlo{3}}}_{a,F}}_\text{\color{magenta} disc \nxlo{3}}
    +\underbrace{\d^\text{\nxlo{2}}_{L, F}}_\text{\color{orange} FV \nxlo{2}}
    +\underbrace{\d^\text{\nxlo{3}}_{L, F}}_\text{\color{orange} FV \nxlo{3}}
\end{align}
\end{subequations}
where the {\color{teal} chiral}, {\color{magenta} discretization} and {\color{orange} finite volume} corrections are defined in Eqs.~\eqref{eq:w0mO_ls}, \eqref{eq:w0mO_a} and \eqref{eq:w0mO_fv} respectively.
\end{widetext}

\bibliography{c51_bib}

%\onecolumngrid
\section{STABILITY PLOTS OF THE OMEGA GROUND STATE MASS \label{app:stability}}
\begin{figure*}
\includegraphics[width=0.49\textwidth]{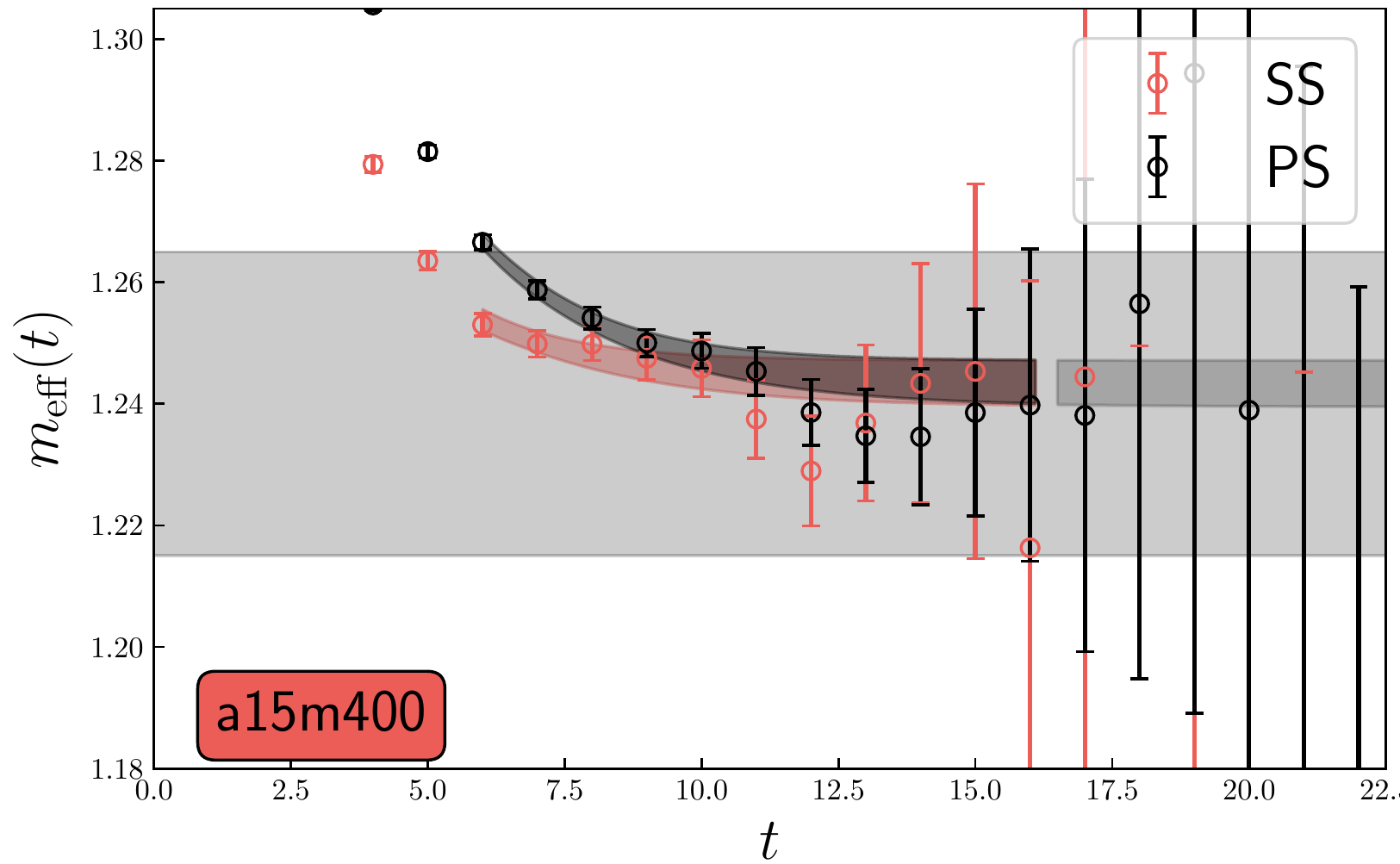}
\includegraphics[width=0.49\textwidth]{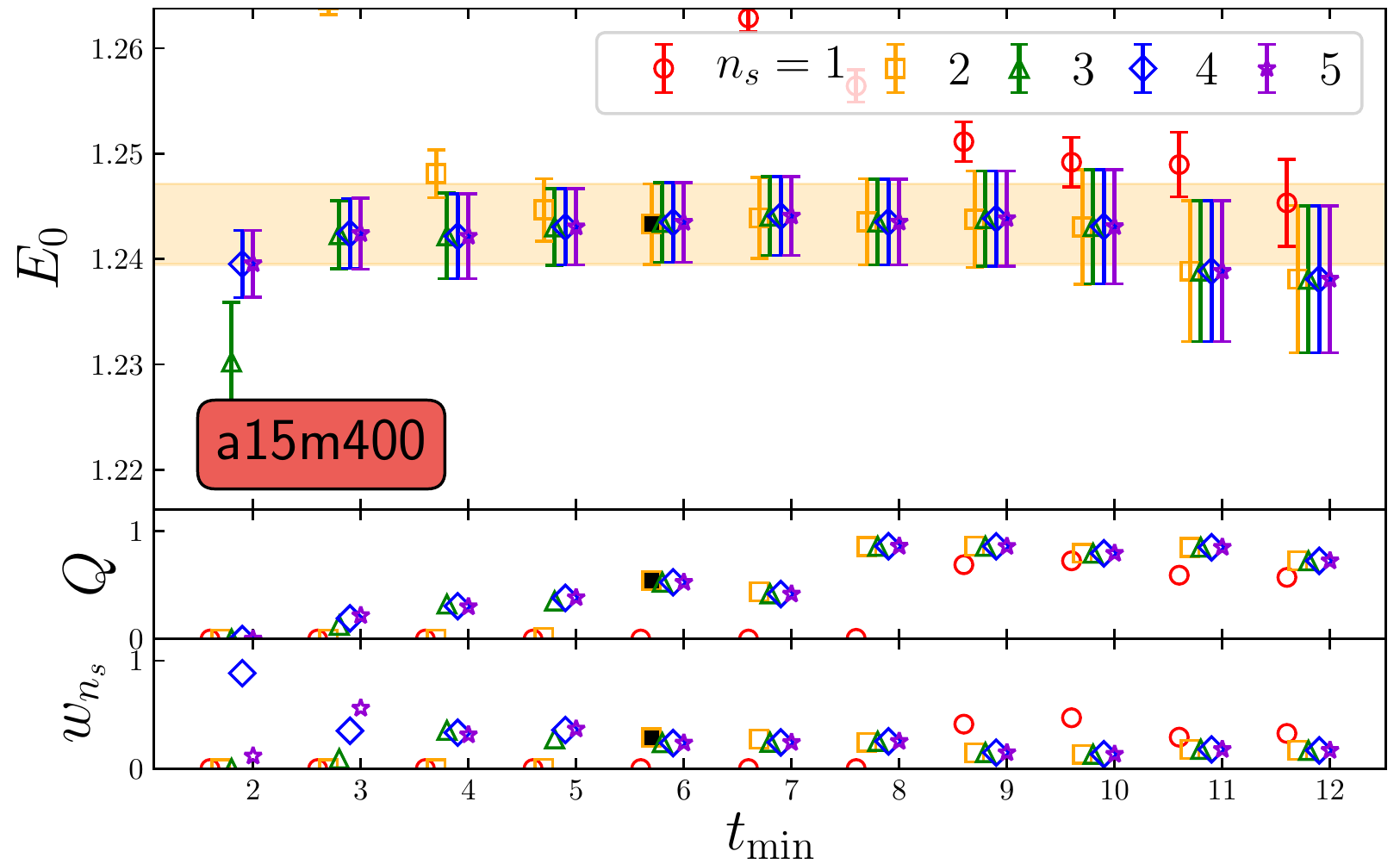}
\caption{\label{fig:stability_a15m400}
Same as \figref{fig:stability_m135} for the a15m400 ensemble.
}
\end{figure*}

\begin{figure*}
\includegraphics[width=0.49\textwidth]{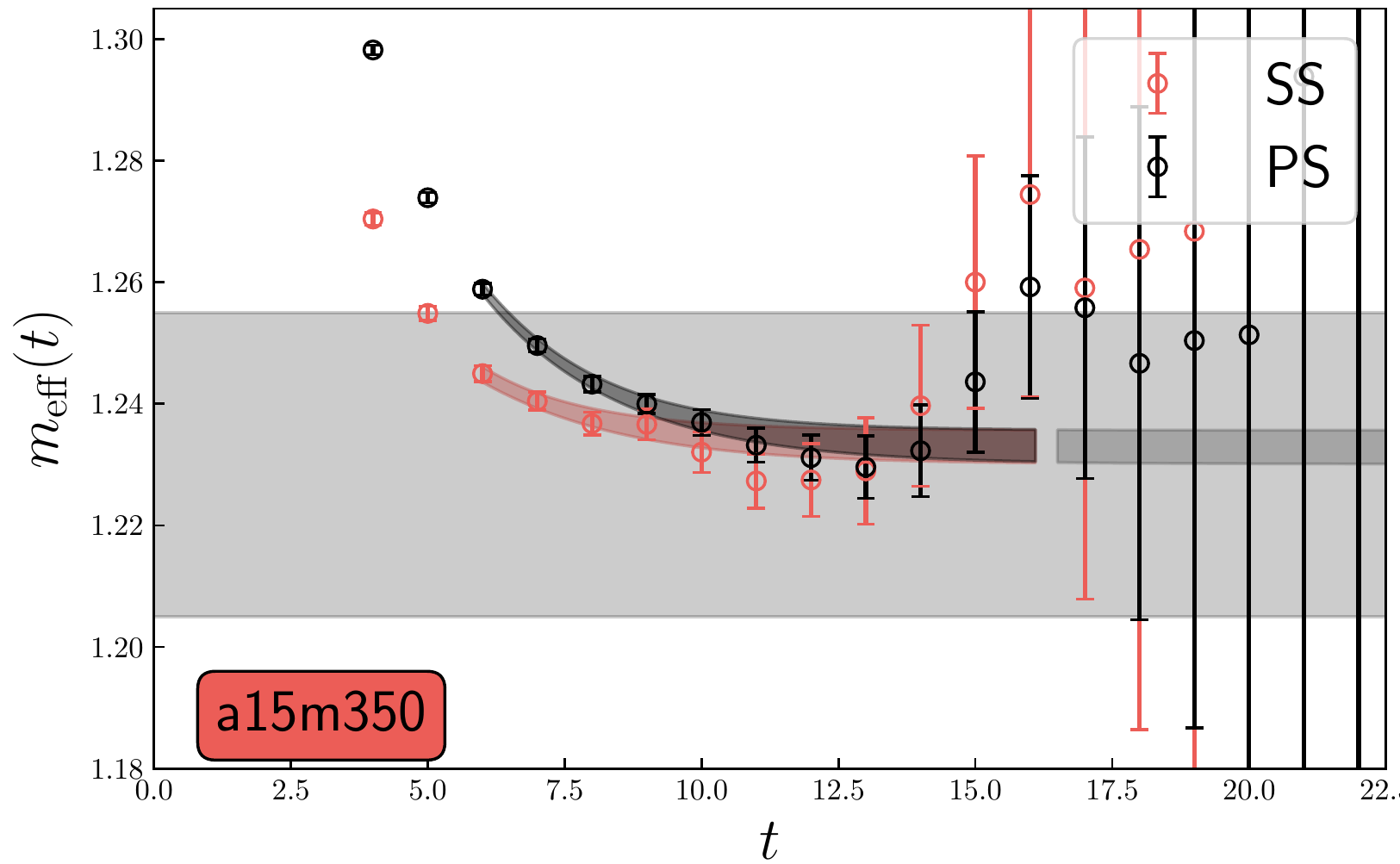}
\includegraphics[width=0.49\textwidth]{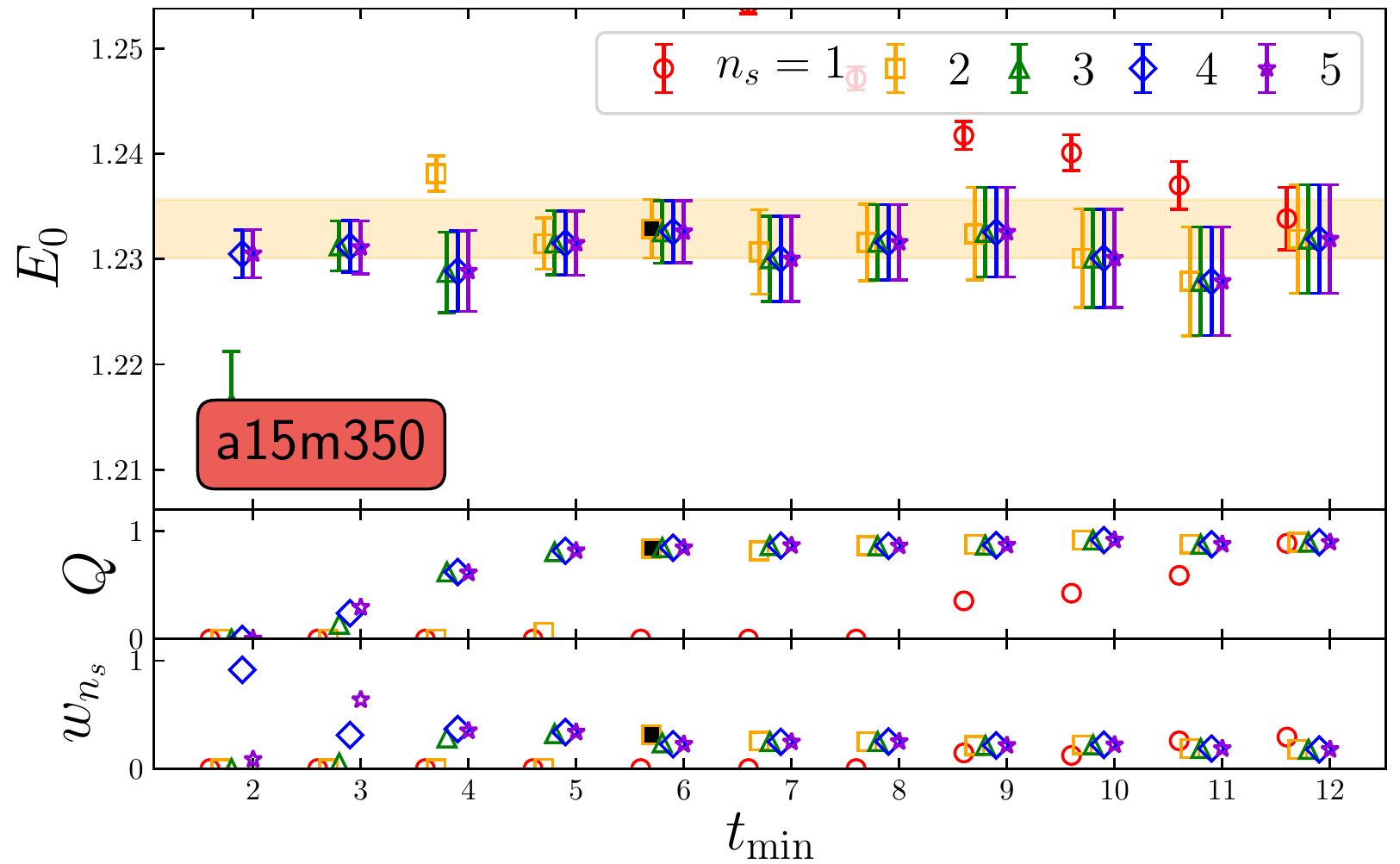}
\caption{\label{fig:stability_a15m350}
Same as \figref{fig:stability_m135} for the a15m350 ensemble.
}
\end{figure*}

\begin{figure*}
\includegraphics[width=0.49\textwidth]{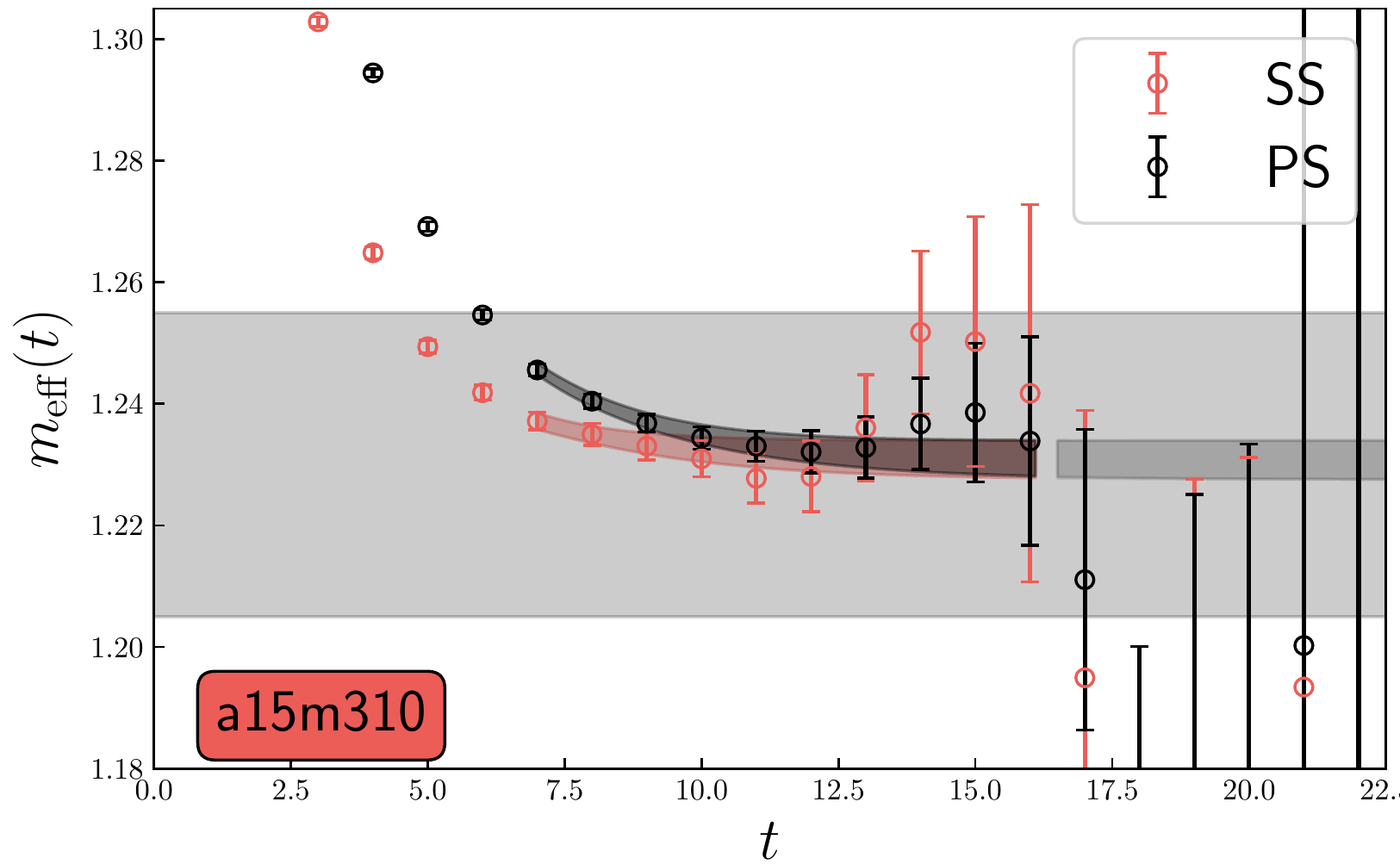}
\includegraphics[width=0.49\textwidth]{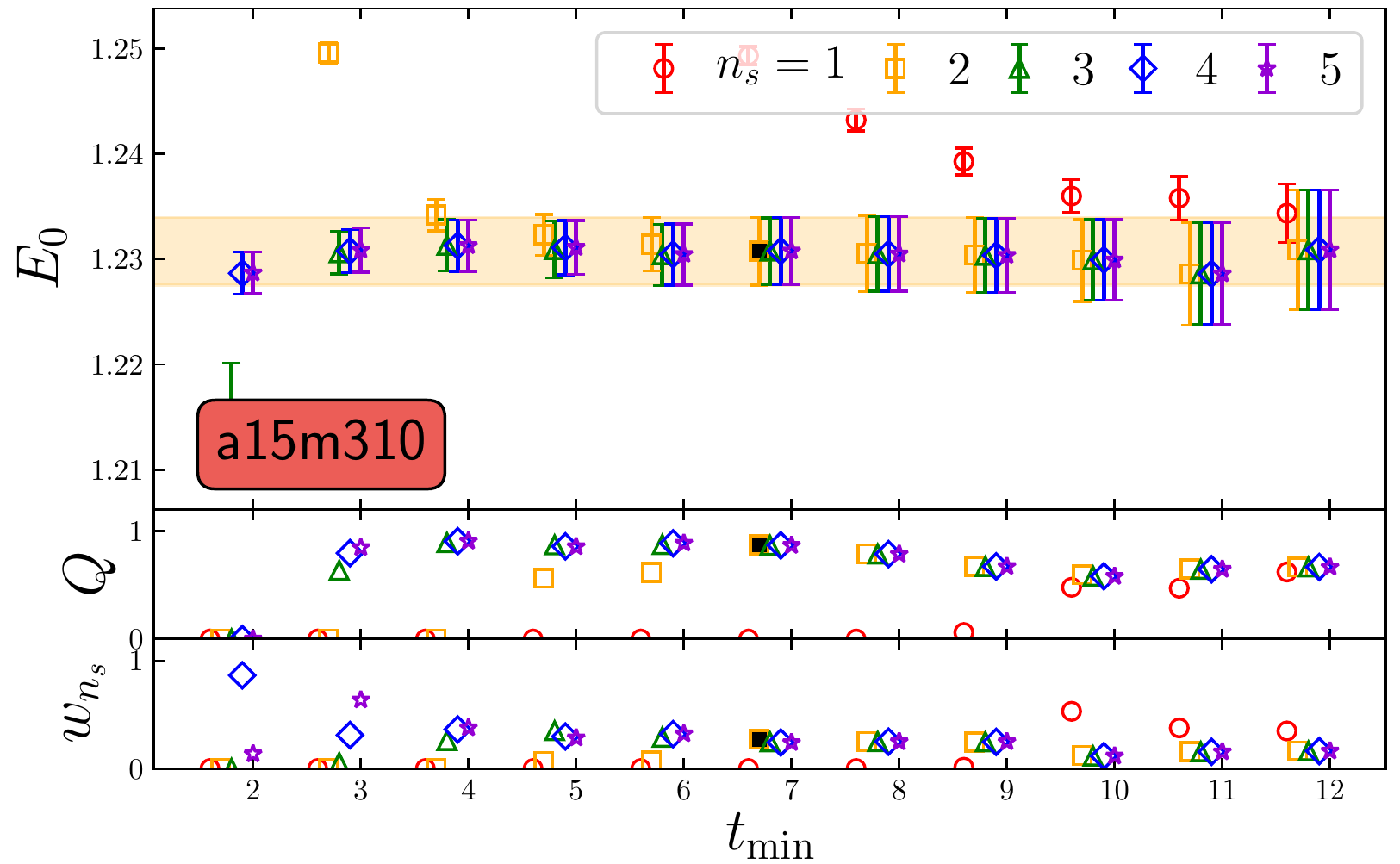}
\includegraphics[width=0.49\textwidth]{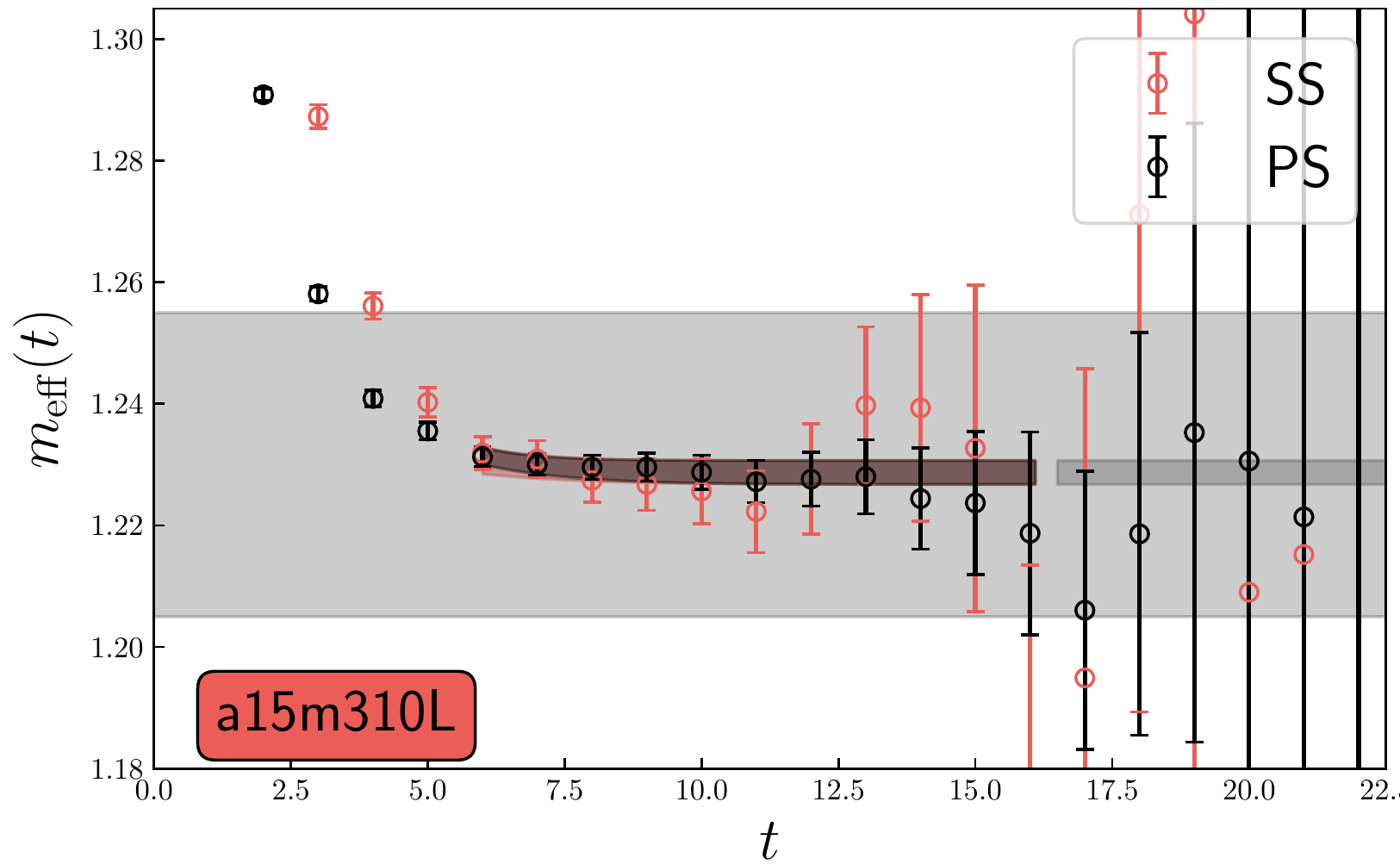}
\includegraphics[width=0.49\textwidth]{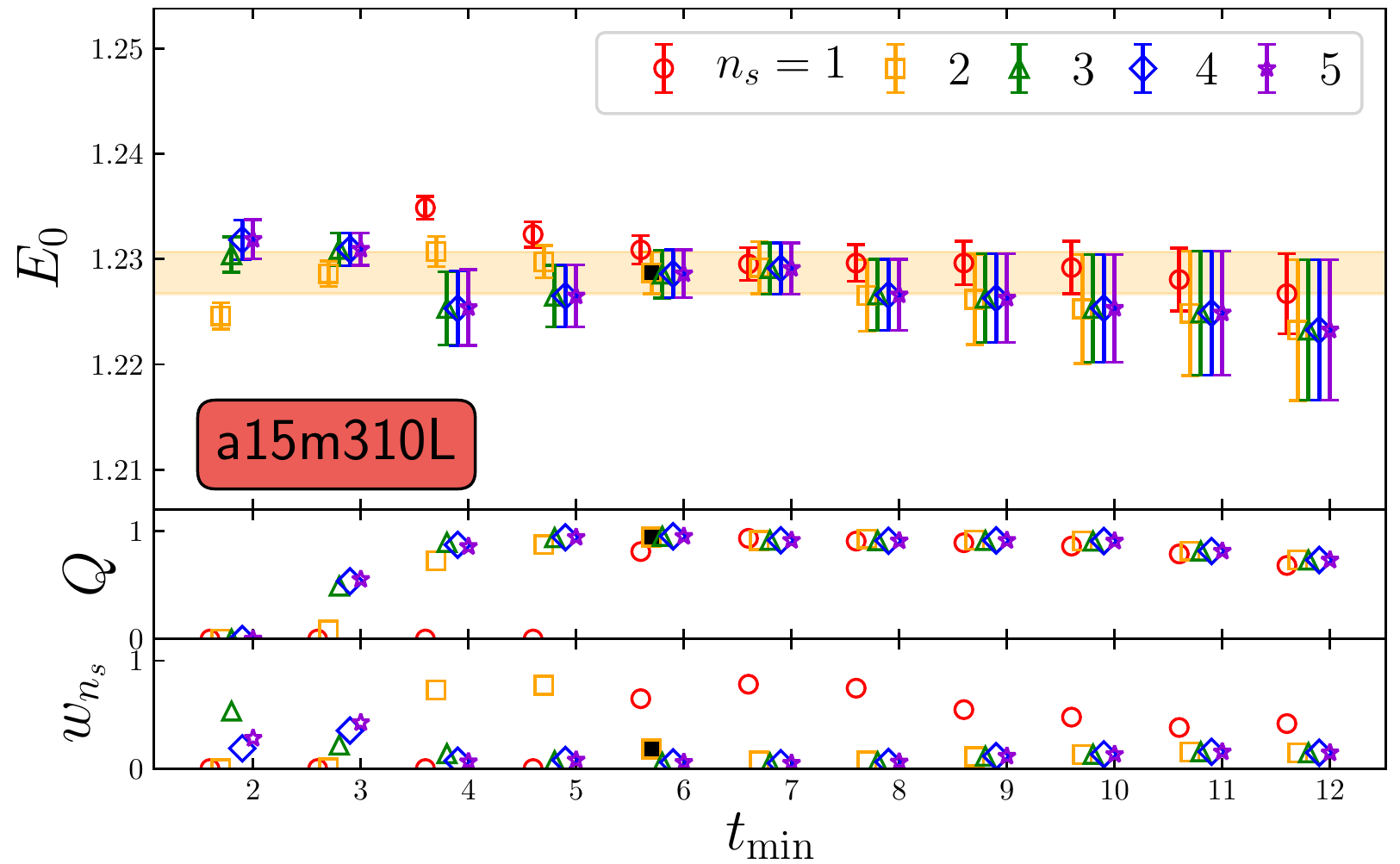}
\caption{\label{fig:stability_a15m310}
Same as \figref{fig:stability_m135} for the a15m310 and a15m310L ensembles.
}
\end{figure*}

\begin{figure*}
\includegraphics[width=0.49\textwidth]{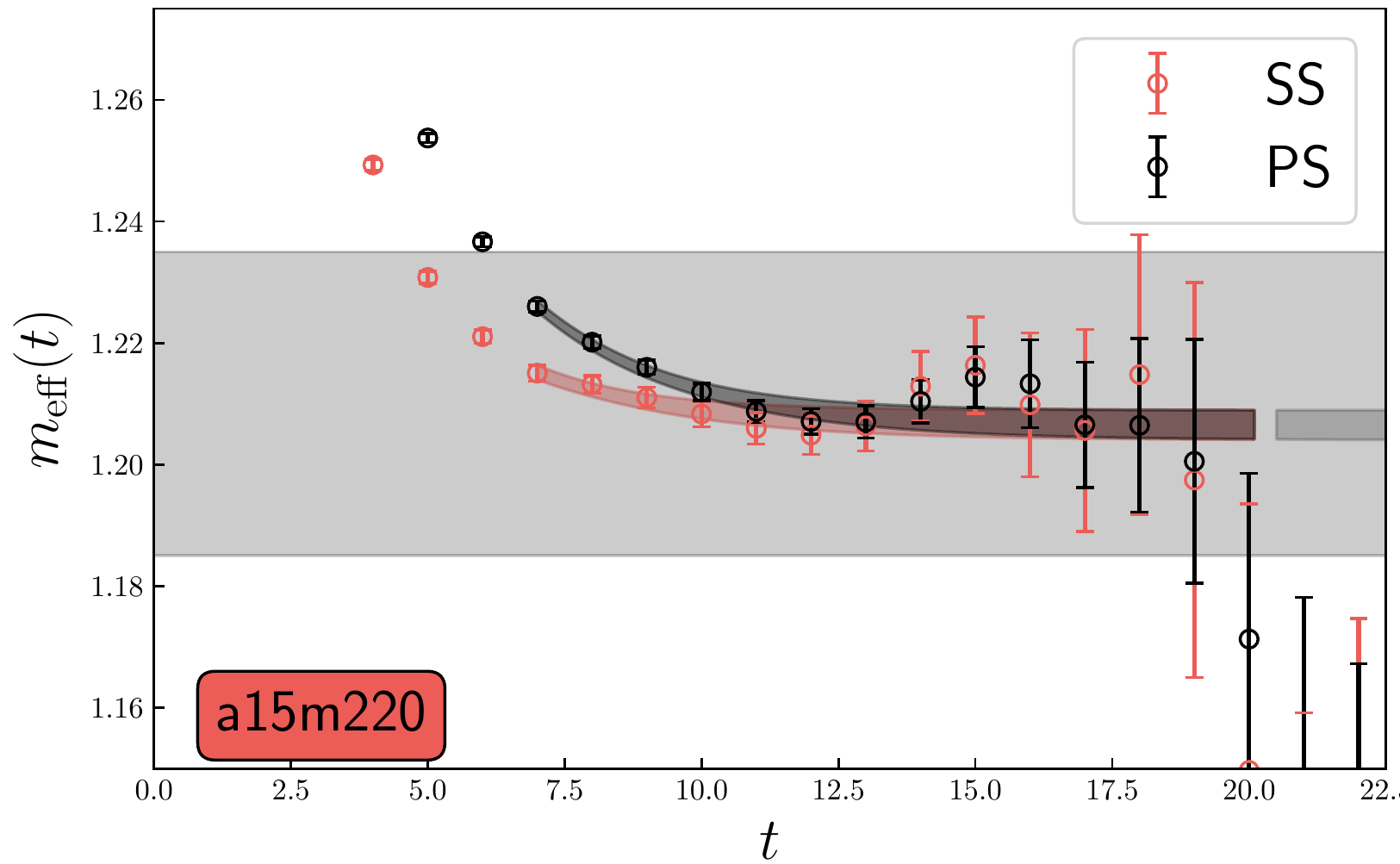}
\includegraphics[width=0.49\textwidth]{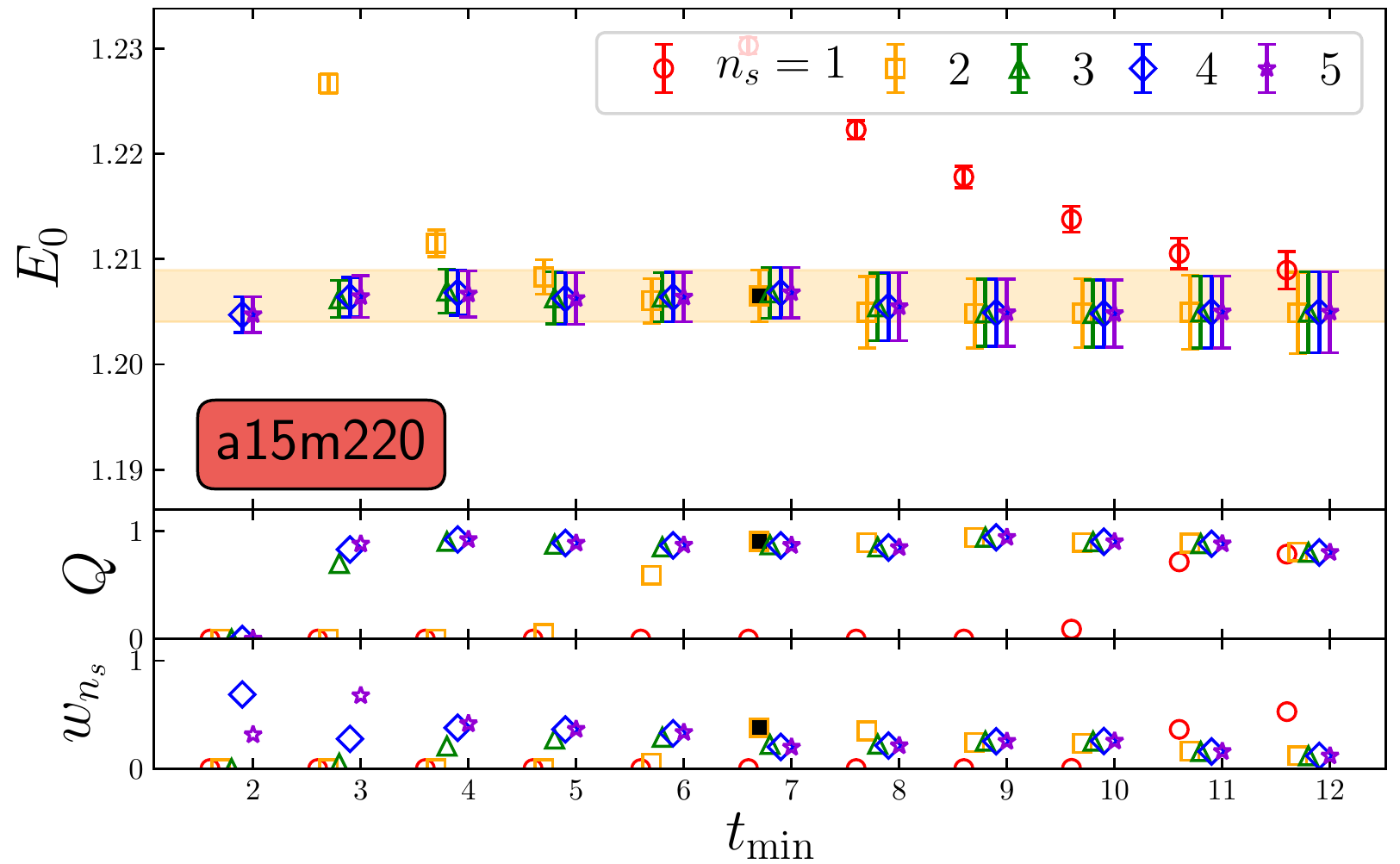}
\caption{\label{fig:stability_a15m220}
Same as \figref{fig:stability_m135} for the a15m220 ensemble.
}
\end{figure*}

\begin{figure*}
\includegraphics[width=0.49\textwidth]{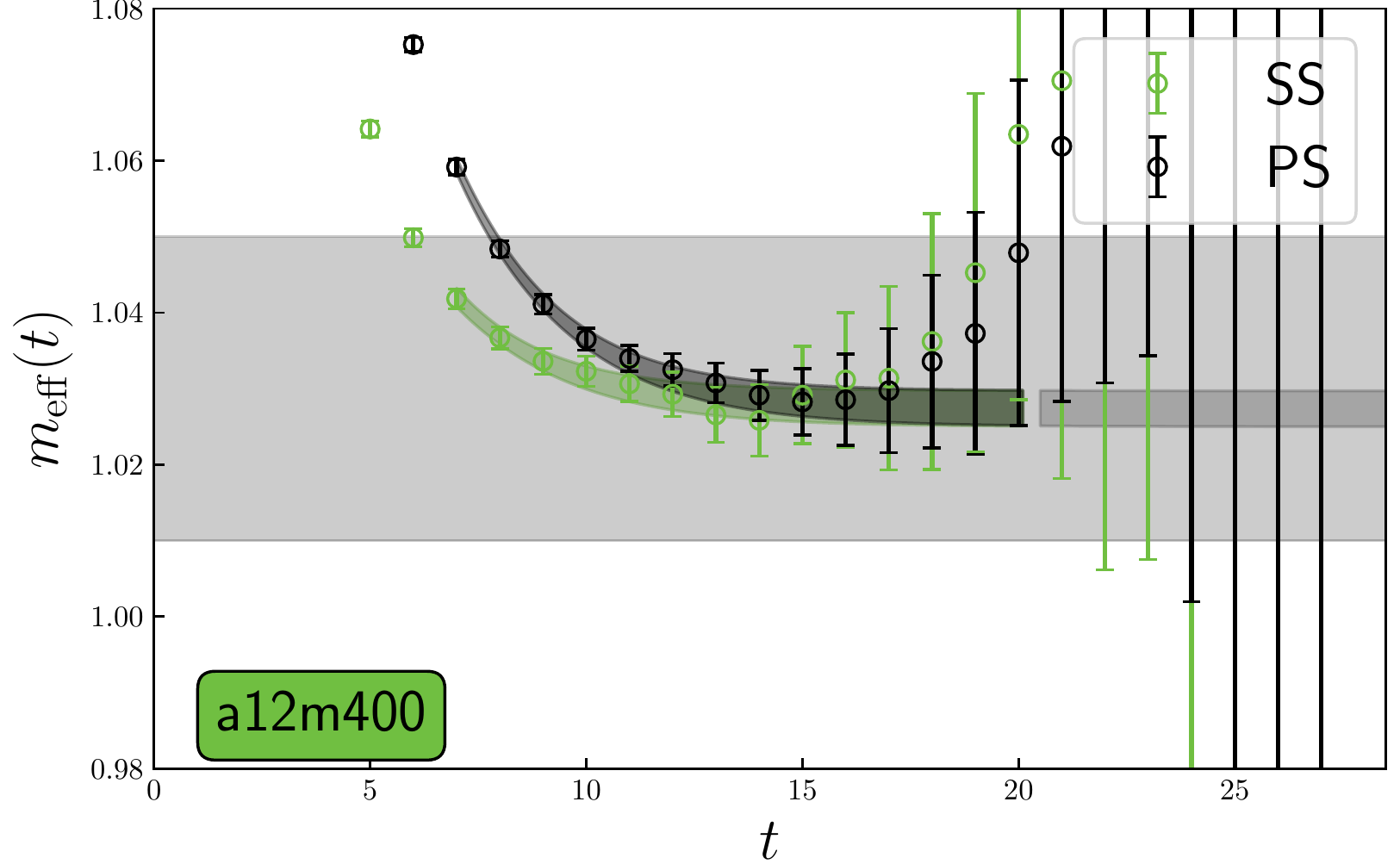}
\includegraphics[width=0.49\textwidth]{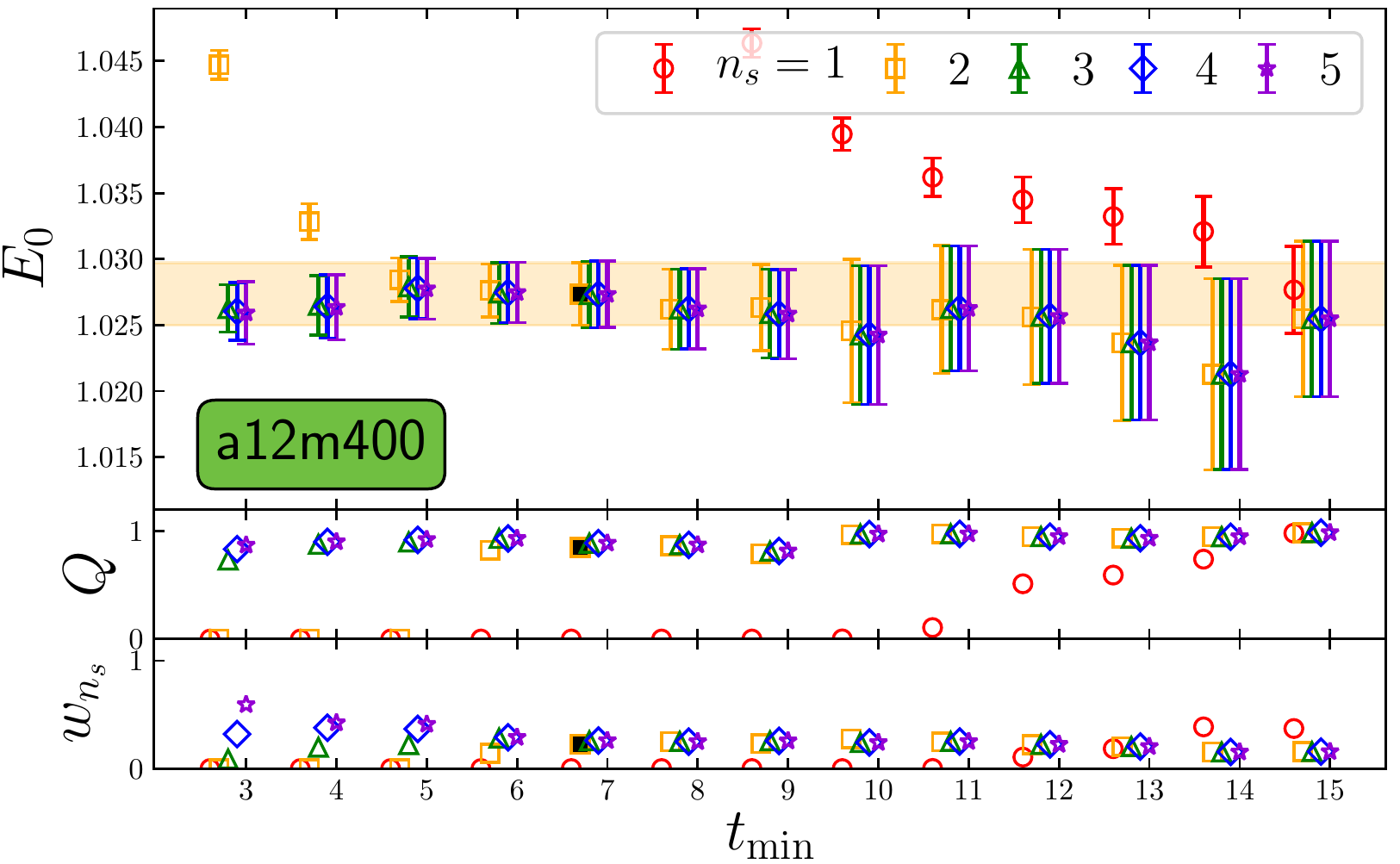}
\caption{\label{fig:stability_a12m400}
Same as \figref{fig:stability_m135} for the a12m400 ensemble.
}
\end{figure*}

\begin{figure*}
\includegraphics[width=0.49\textwidth]{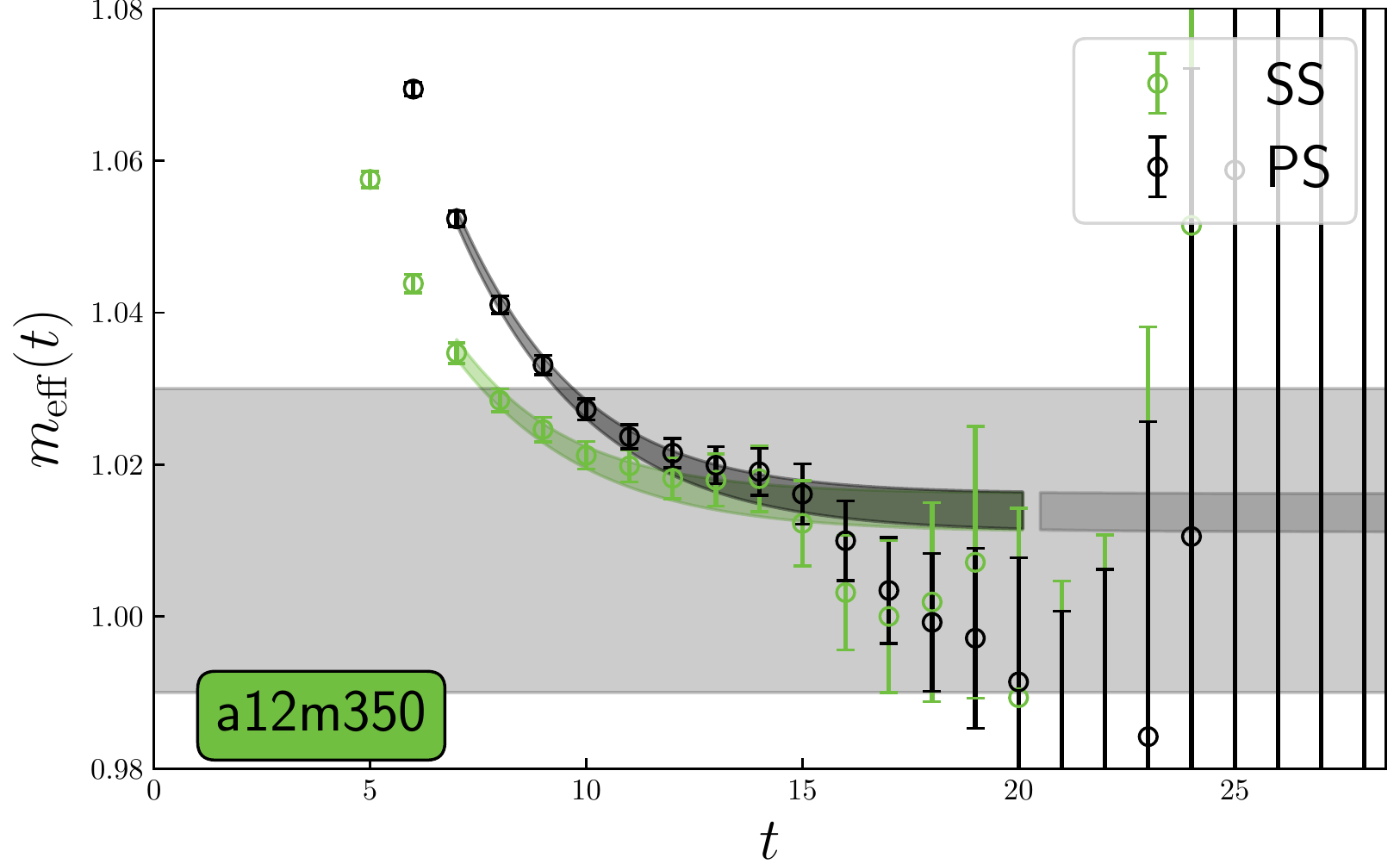}
\includegraphics[width=0.49\textwidth]{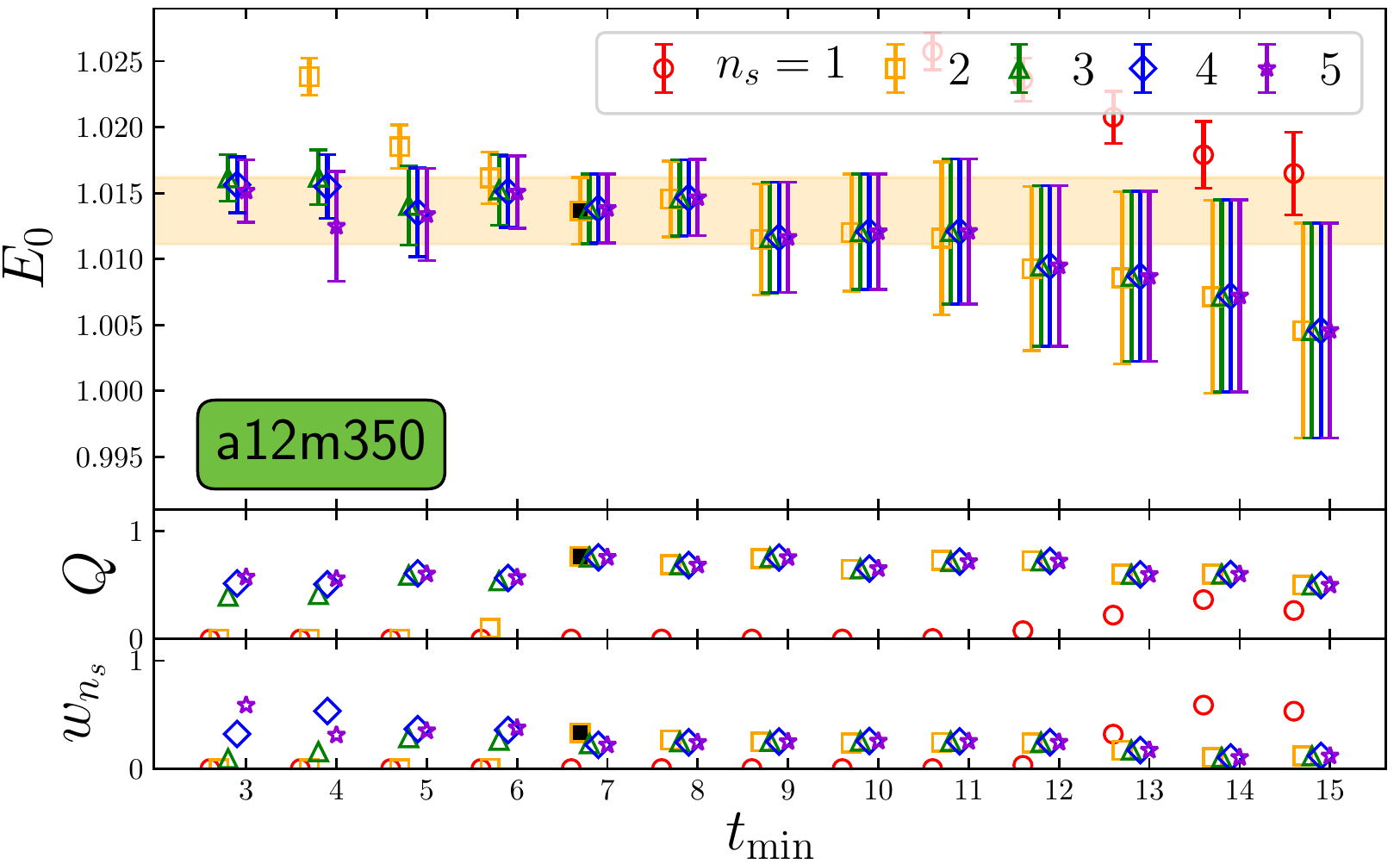}
\caption{\label{fig:stability_a12m350}
Same as \figref{fig:stability_m135} for the a12m350 ensemble.
}
\end{figure*}

\begin{figure*}
\includegraphics[width=0.49\textwidth]{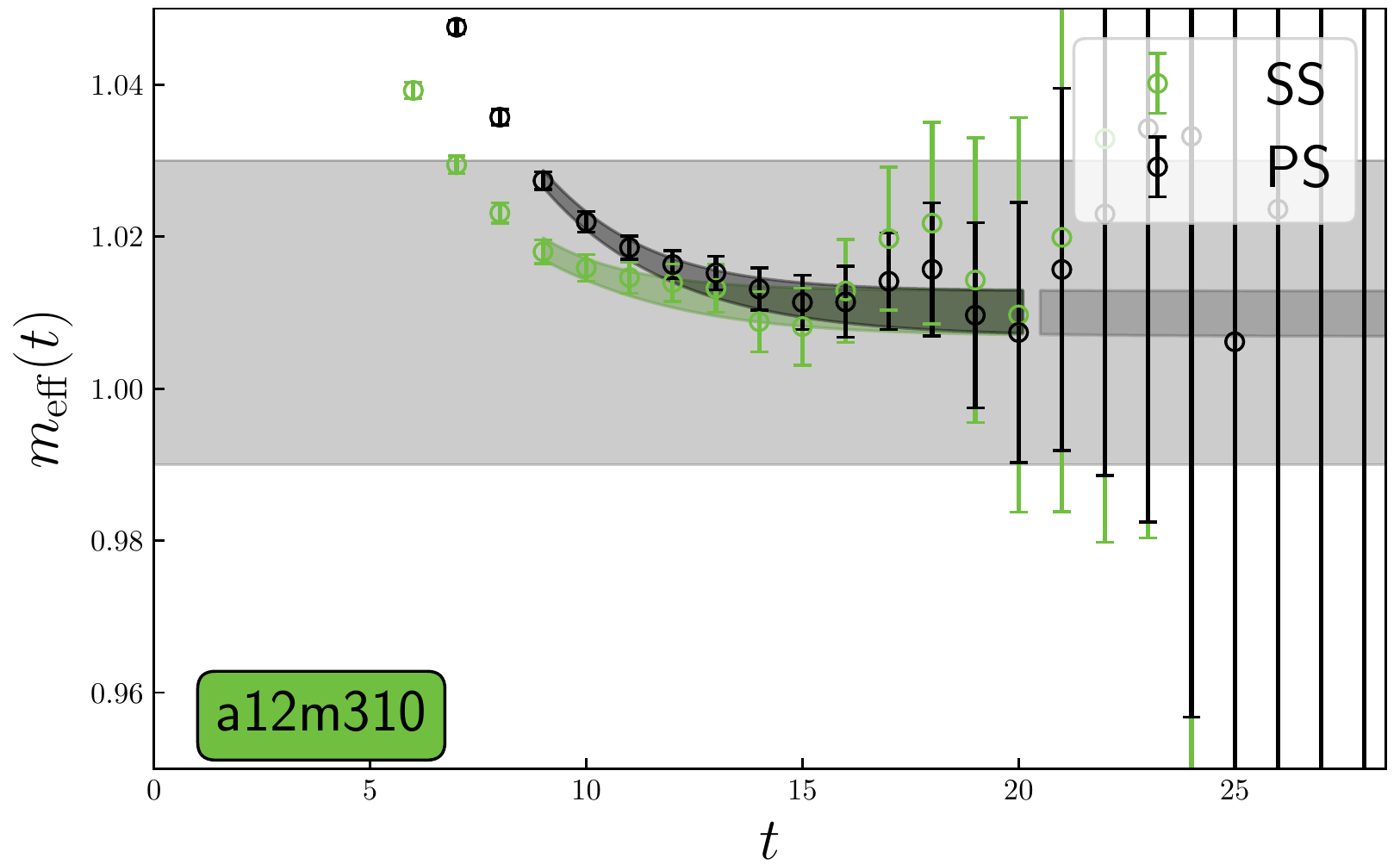}
\includegraphics[width=0.49\textwidth]{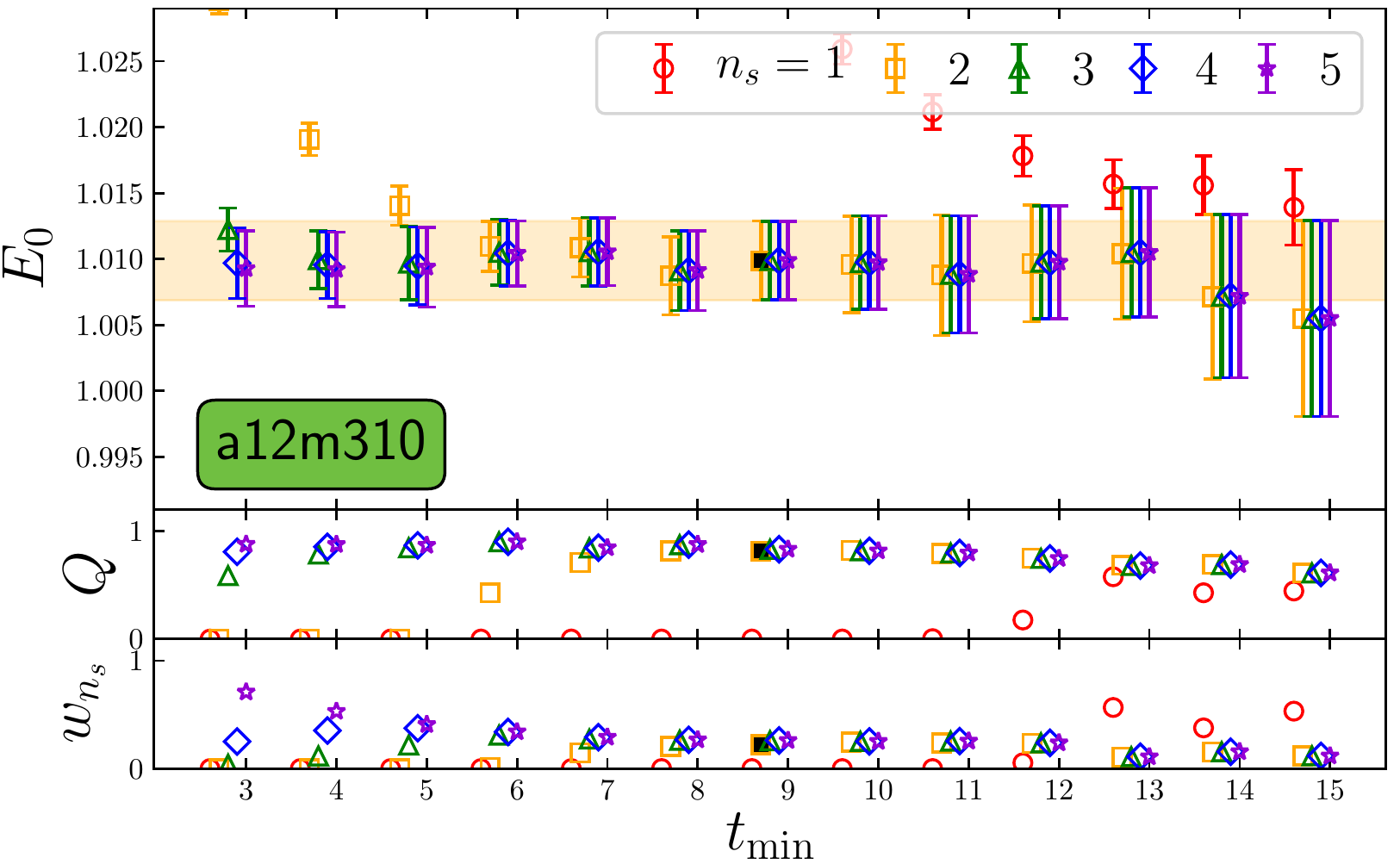}
\includegraphics[width=0.49\textwidth]{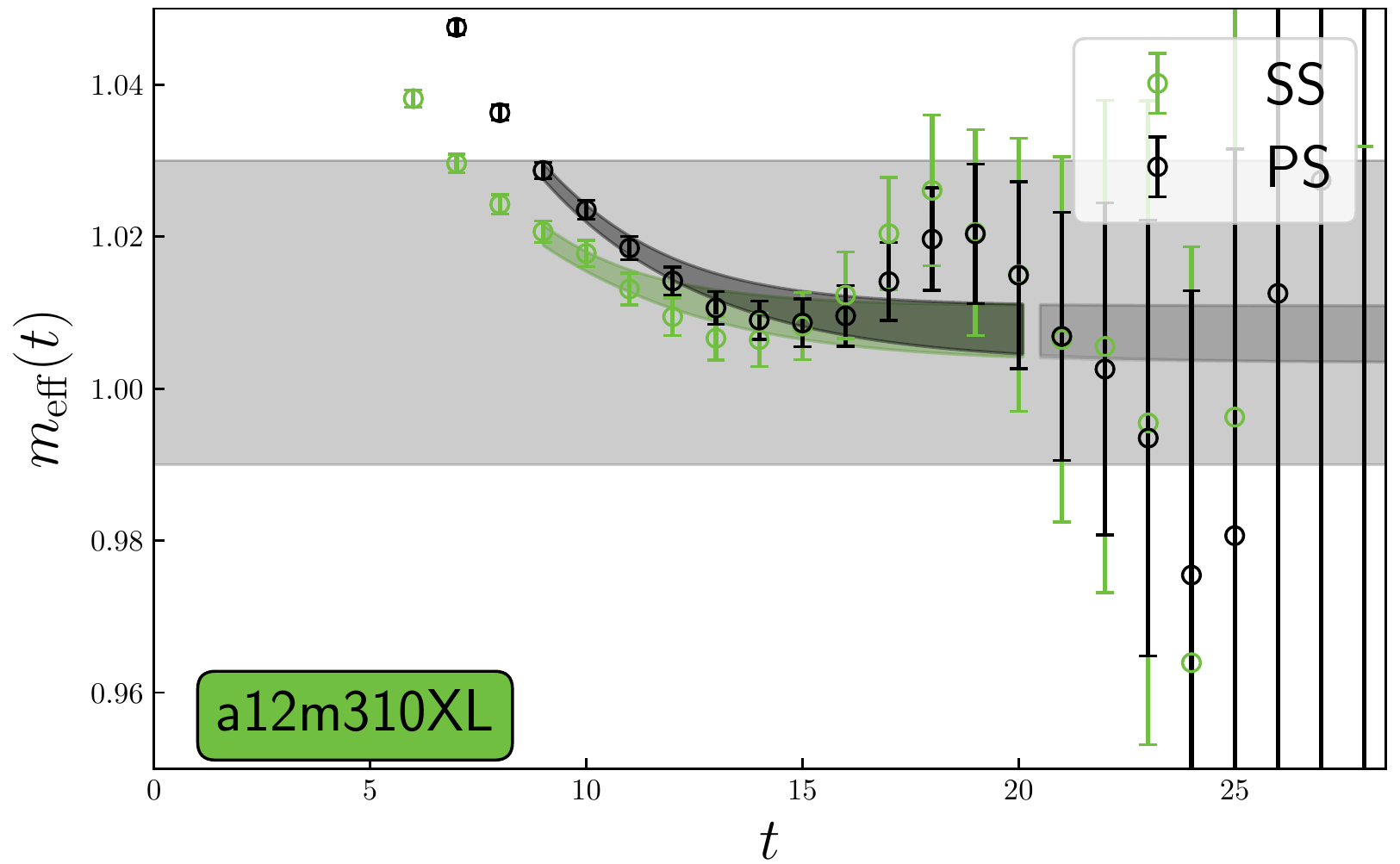}
\includegraphics[width=0.49\textwidth]{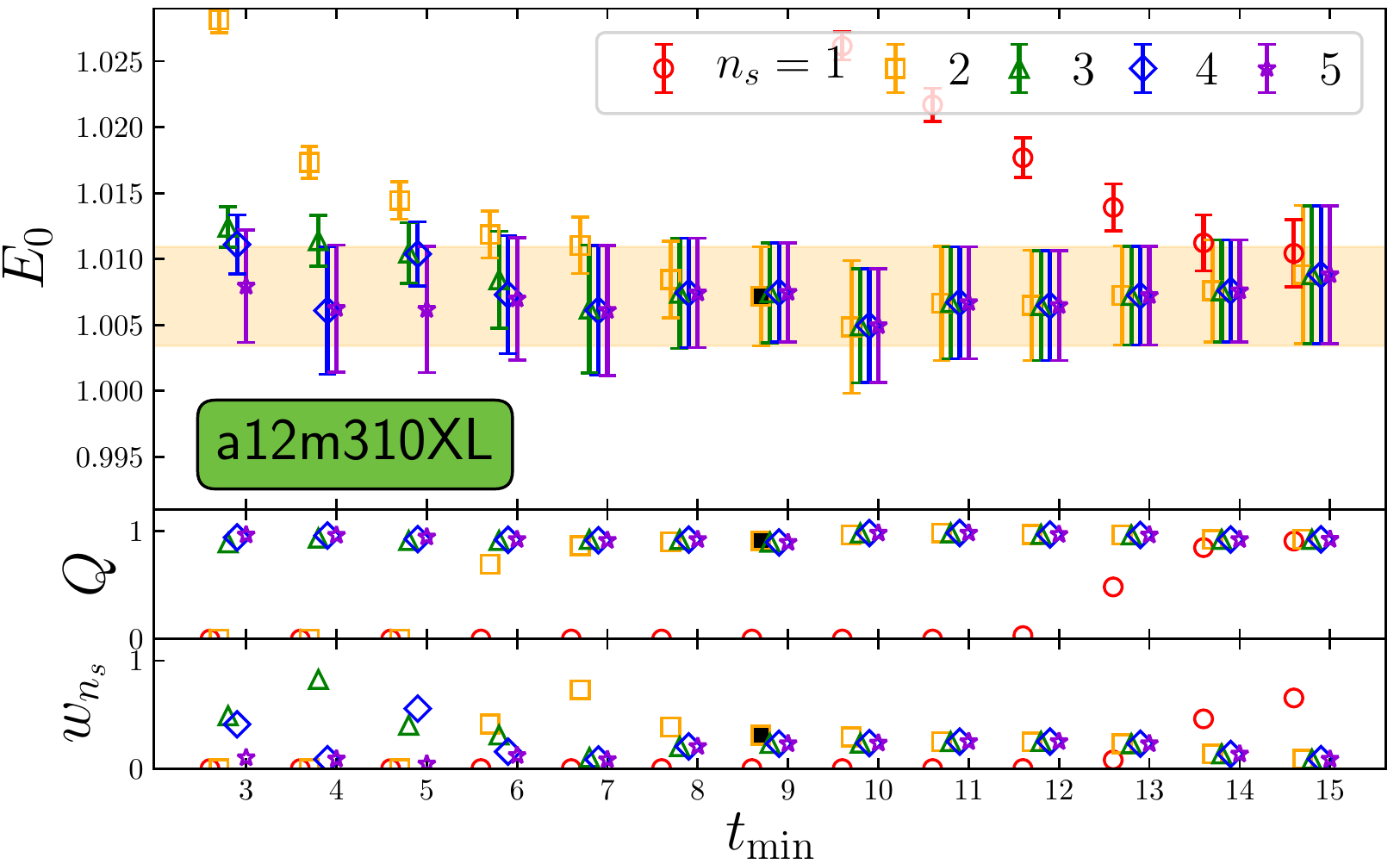}
\caption{\label{fig:stability_a12m310}
Same as \figref{fig:stability_m135} for the a12m310 and a12m310XL ensembles.
}
\end{figure*}

\begin{figure*}
\includegraphics[width=0.49\textwidth]{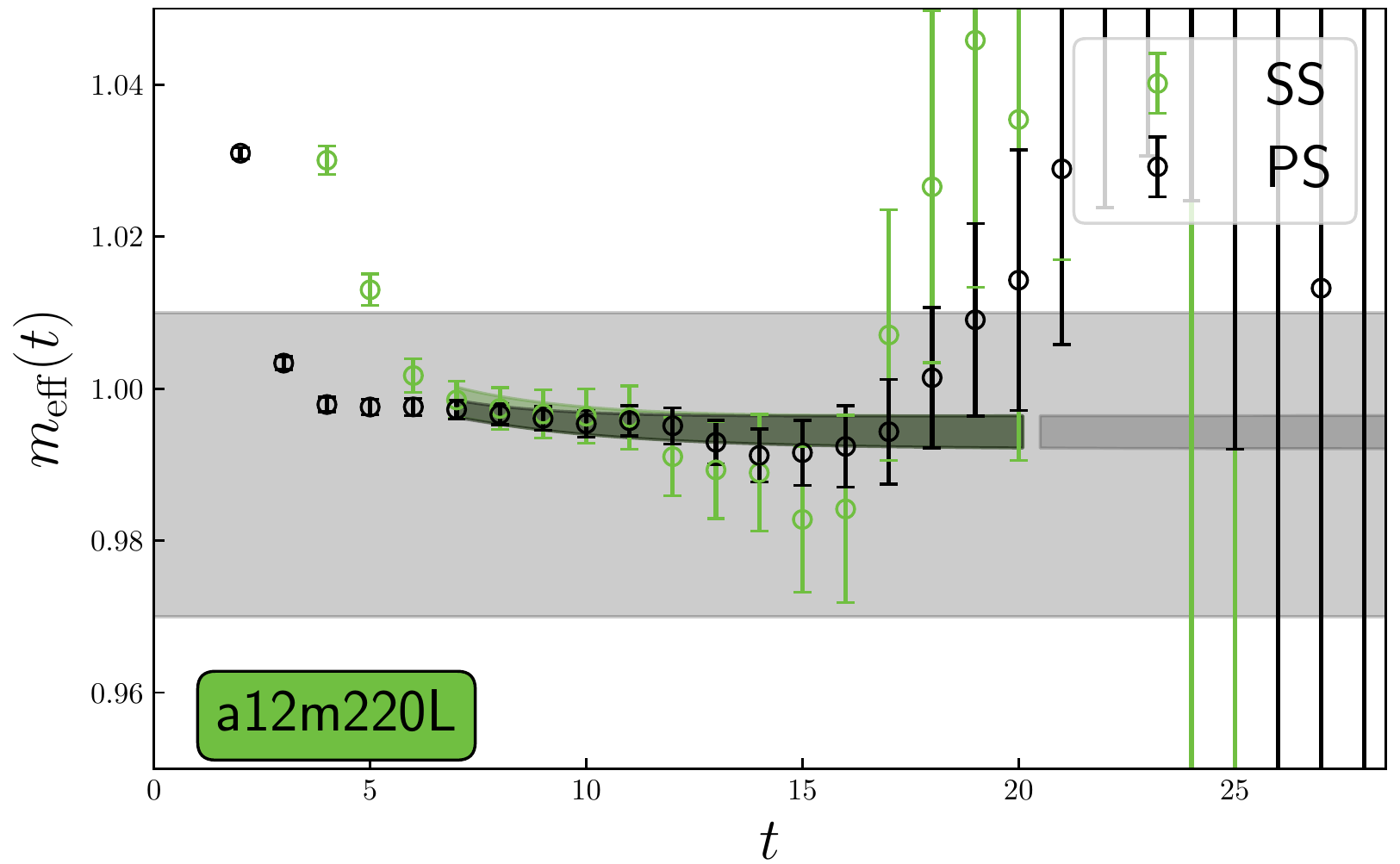}
\includegraphics[width=0.49\textwidth]{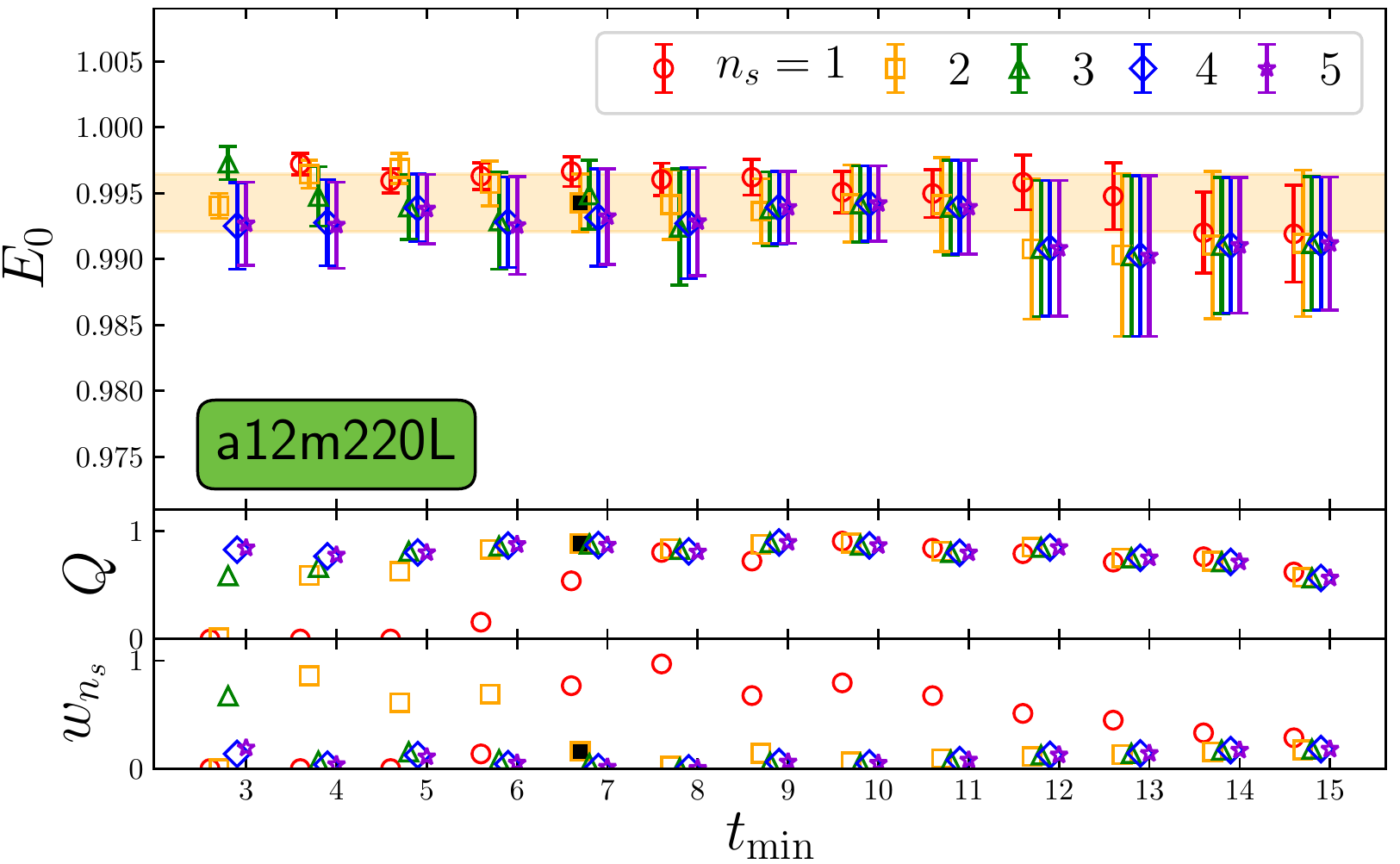}
\includegraphics[width=0.49\textwidth]{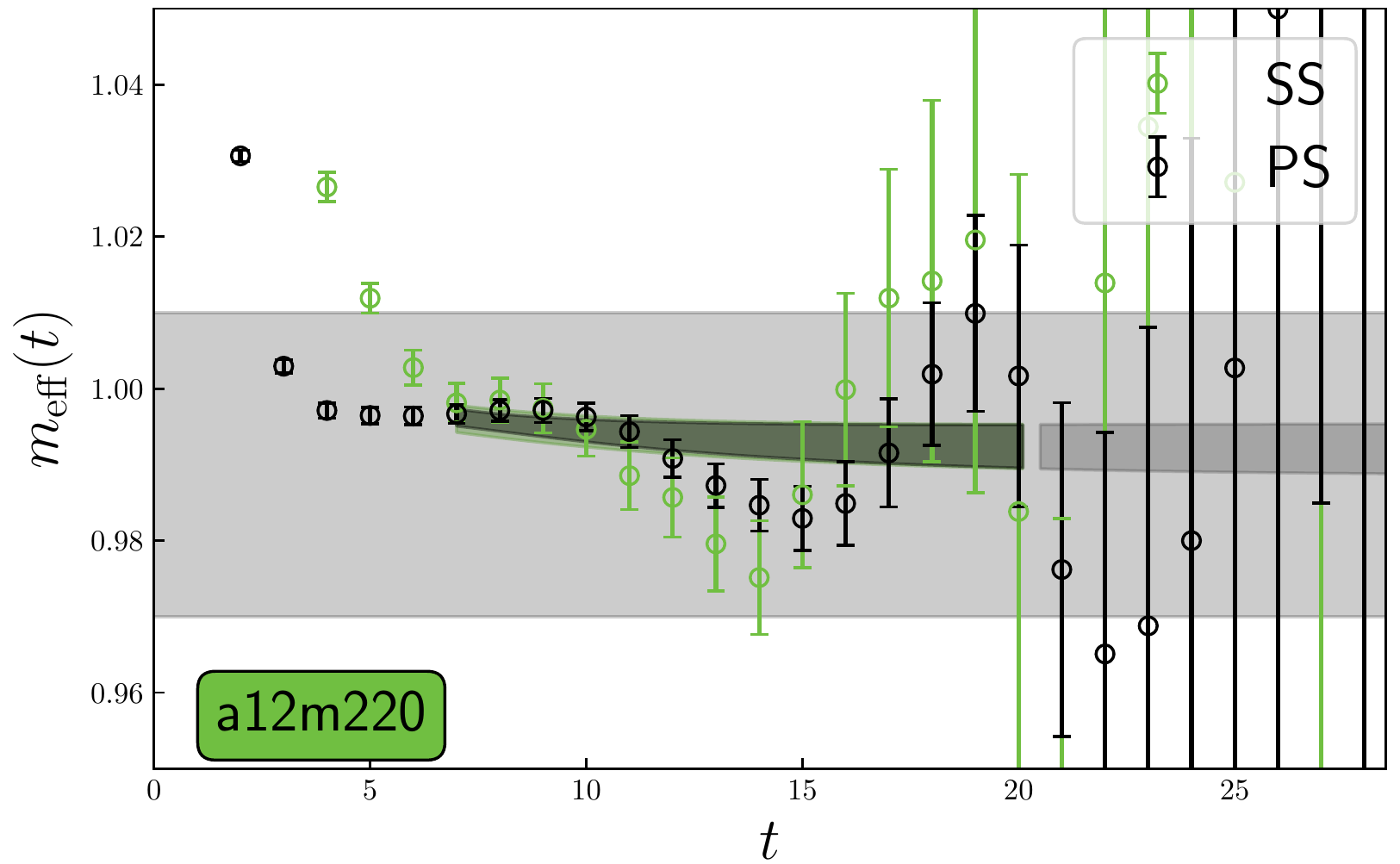}
\includegraphics[width=0.49\textwidth]{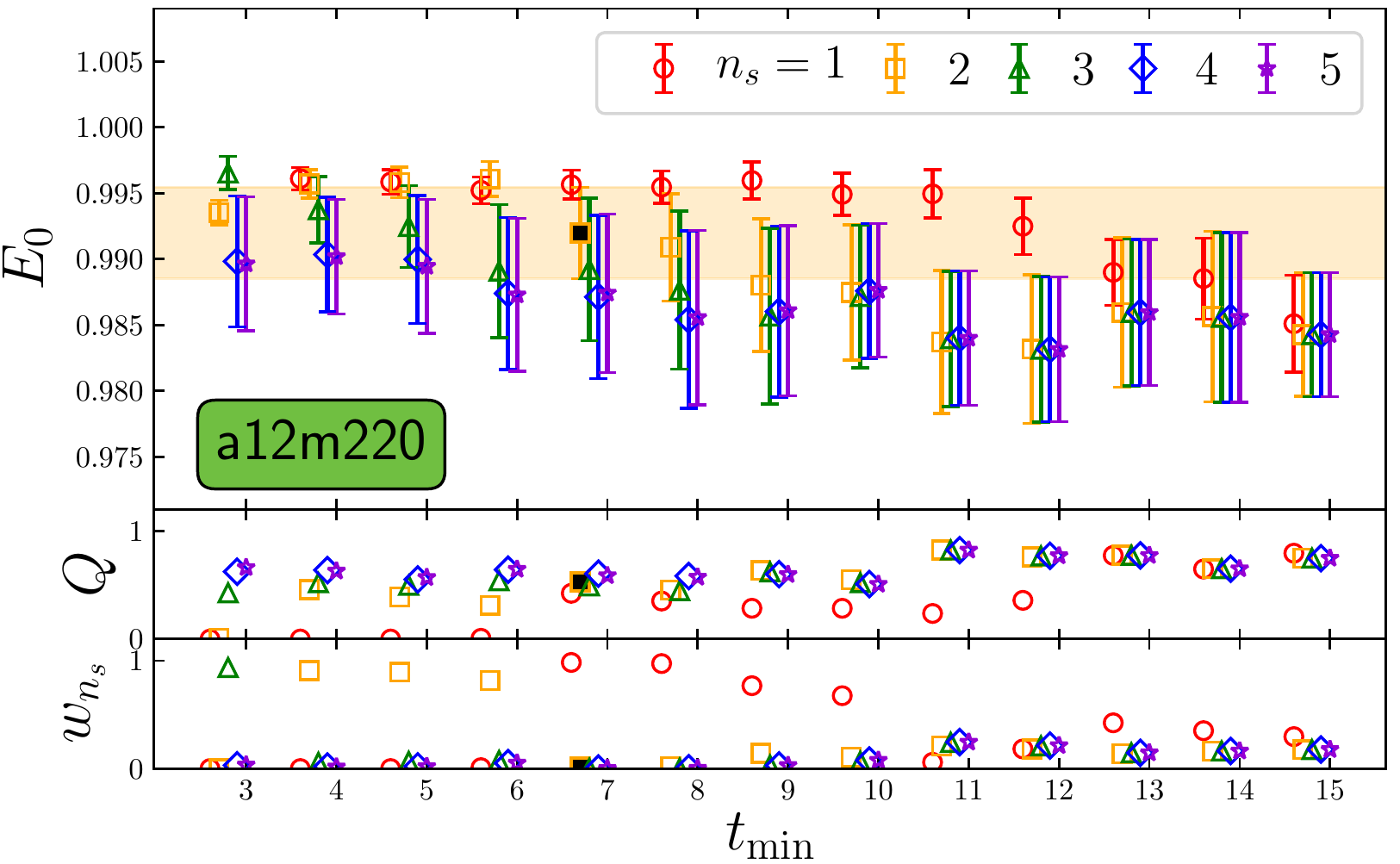}
\includegraphics[width=0.49\textwidth]{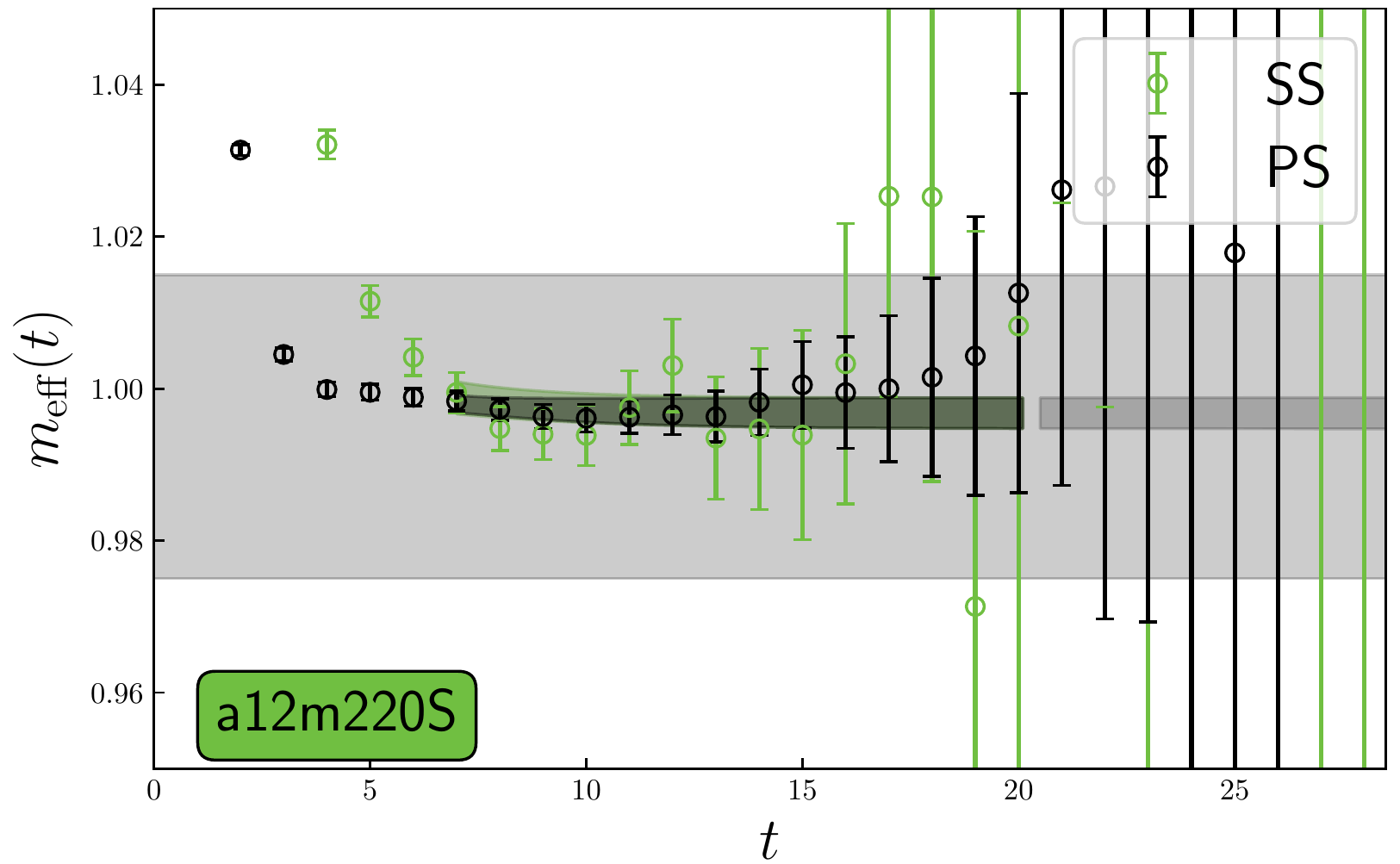}
\includegraphics[width=0.49\textwidth]{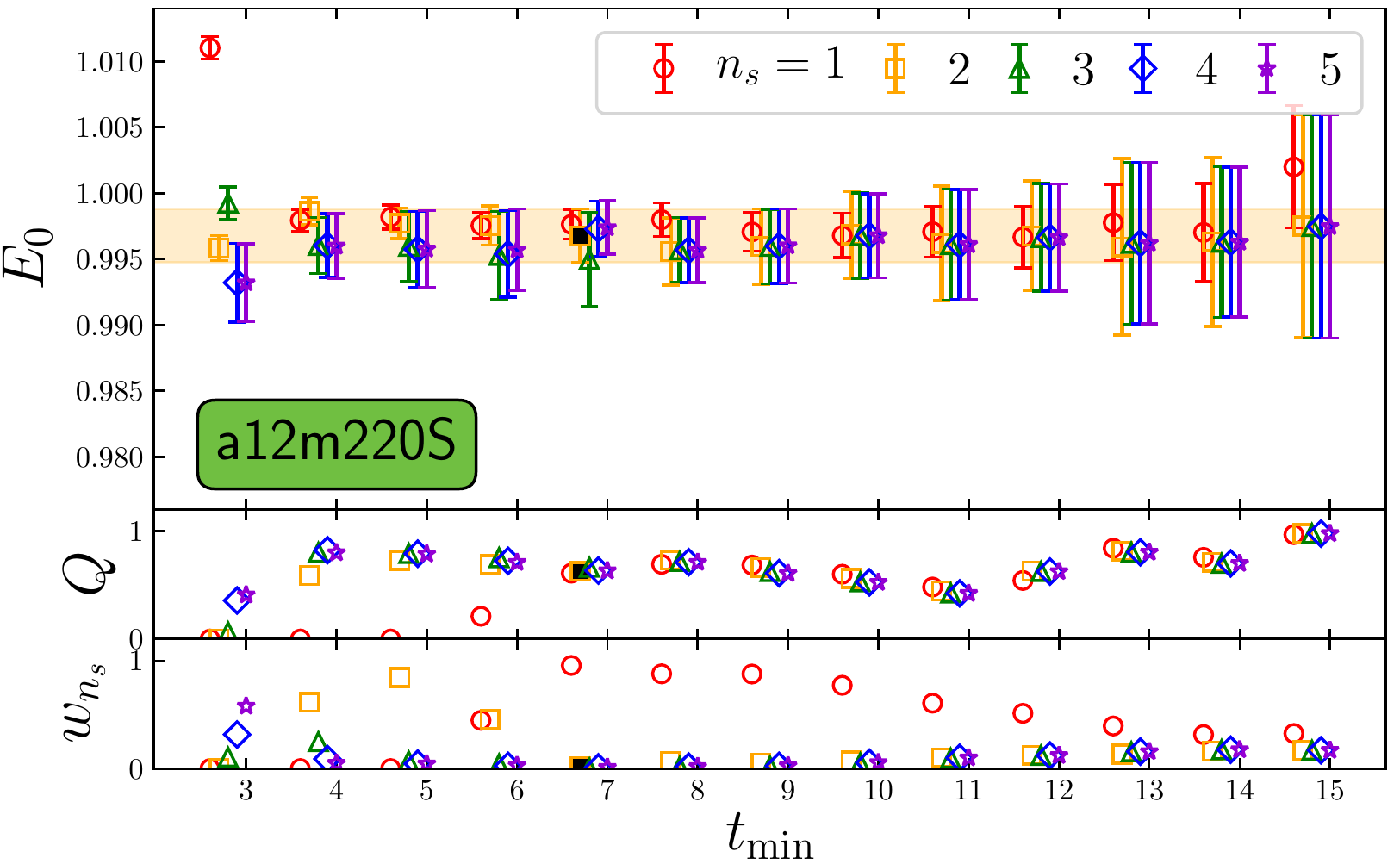}
\caption{\label{fig:stability_a12m220}
Same as \figref{fig:stability_m135} for the a12m220 ensembles.
}
\end{figure*}

\begin{figure*}
\includegraphics[width=0.49\textwidth]{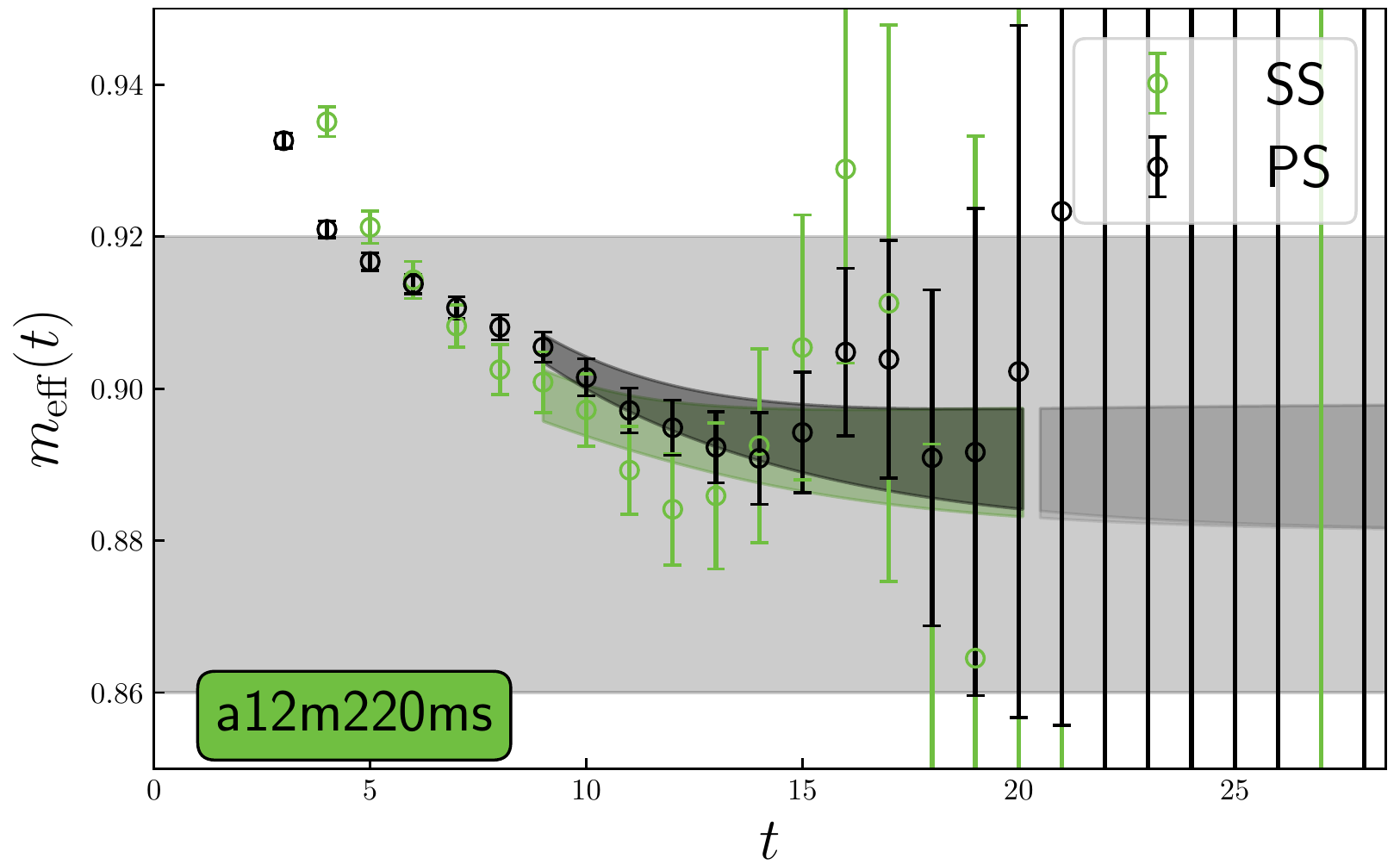}
\includegraphics[width=0.49\textwidth]{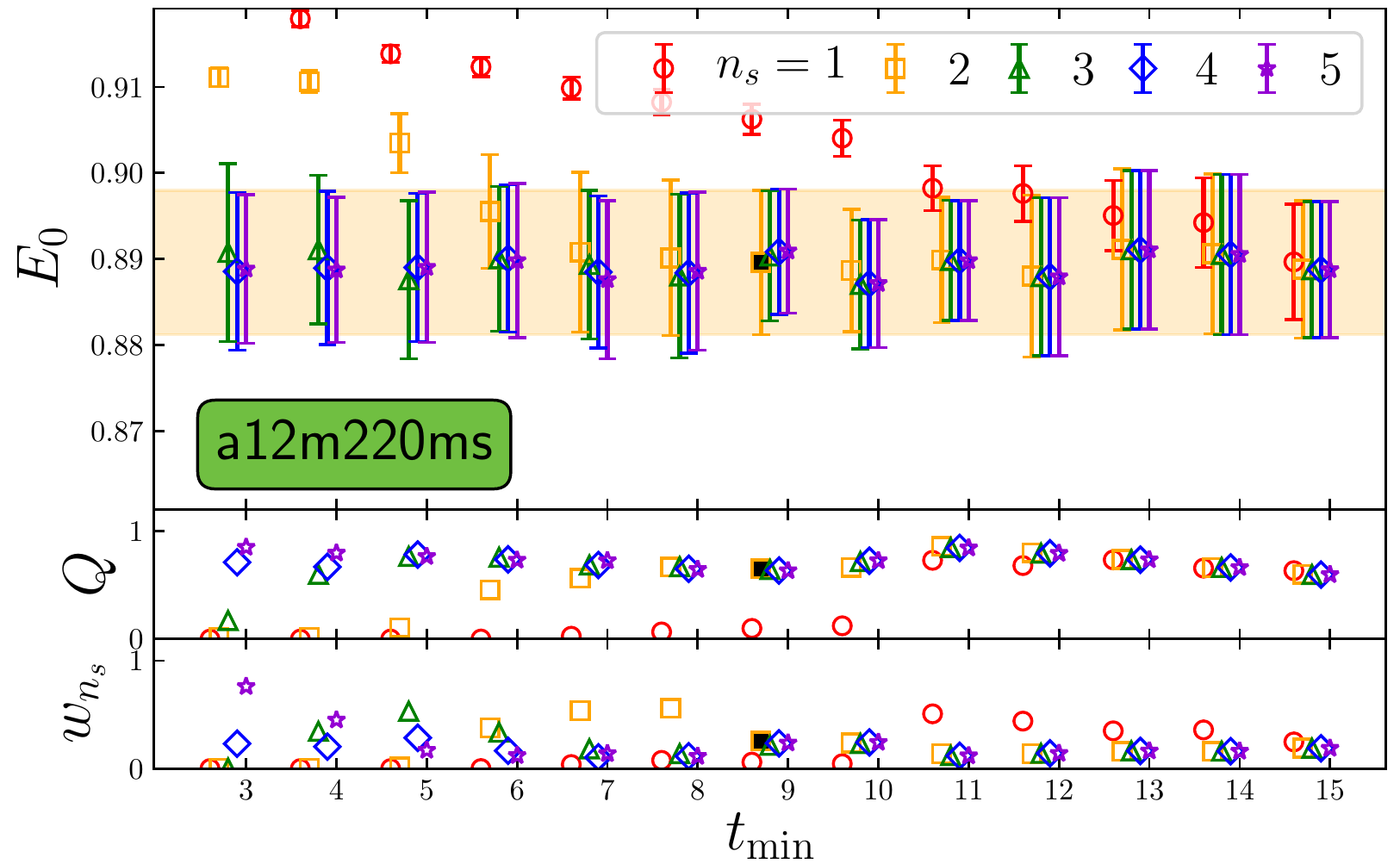}
\caption{\label{fig:stability_a12m220ms}
Same as \figref{fig:stability_m135} for the a12m220ms ensembles.
}
\end{figure*}

\begin{figure*}
\includegraphics[width=0.49\textwidth]{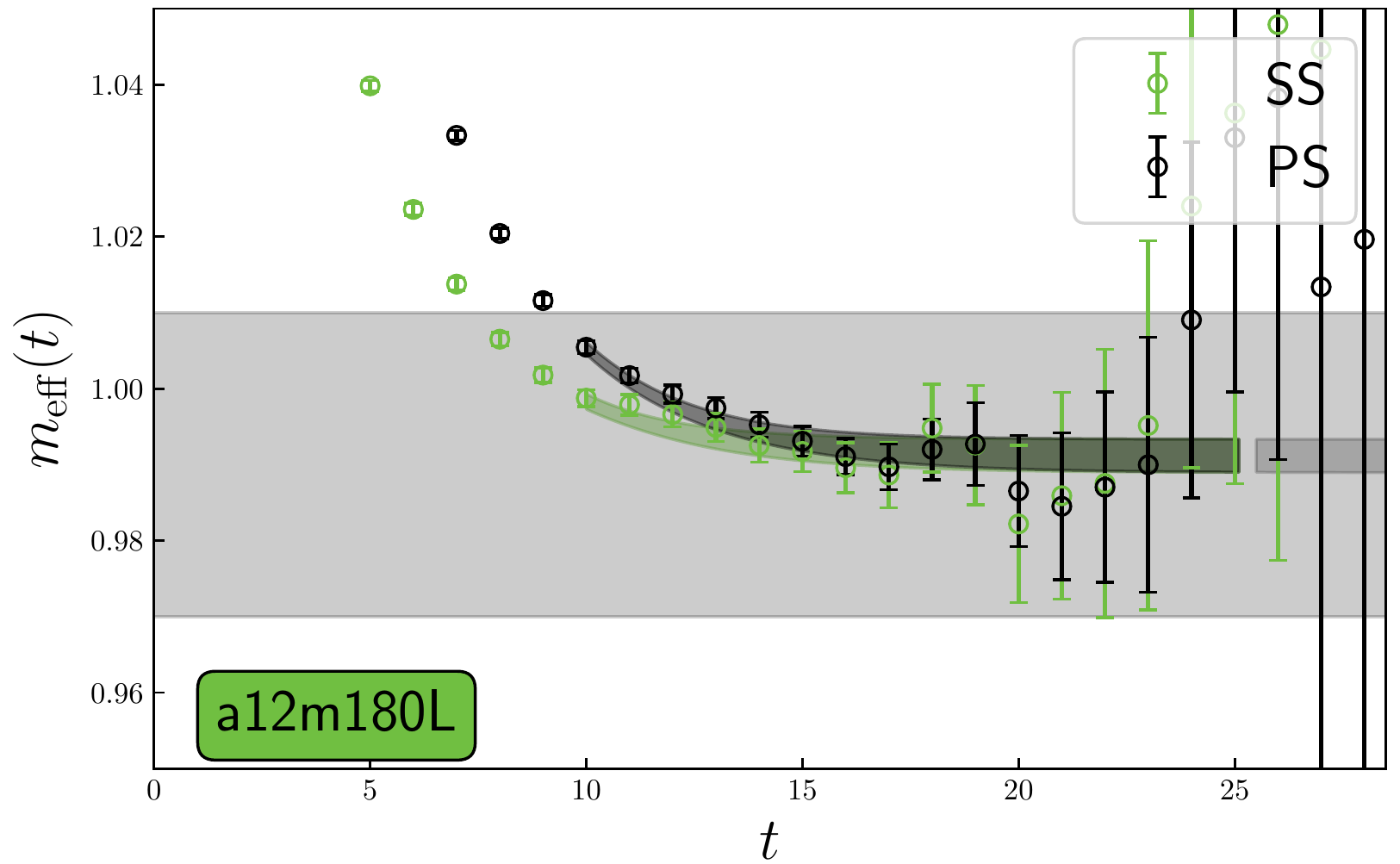}
\includegraphics[width=0.49\textwidth]{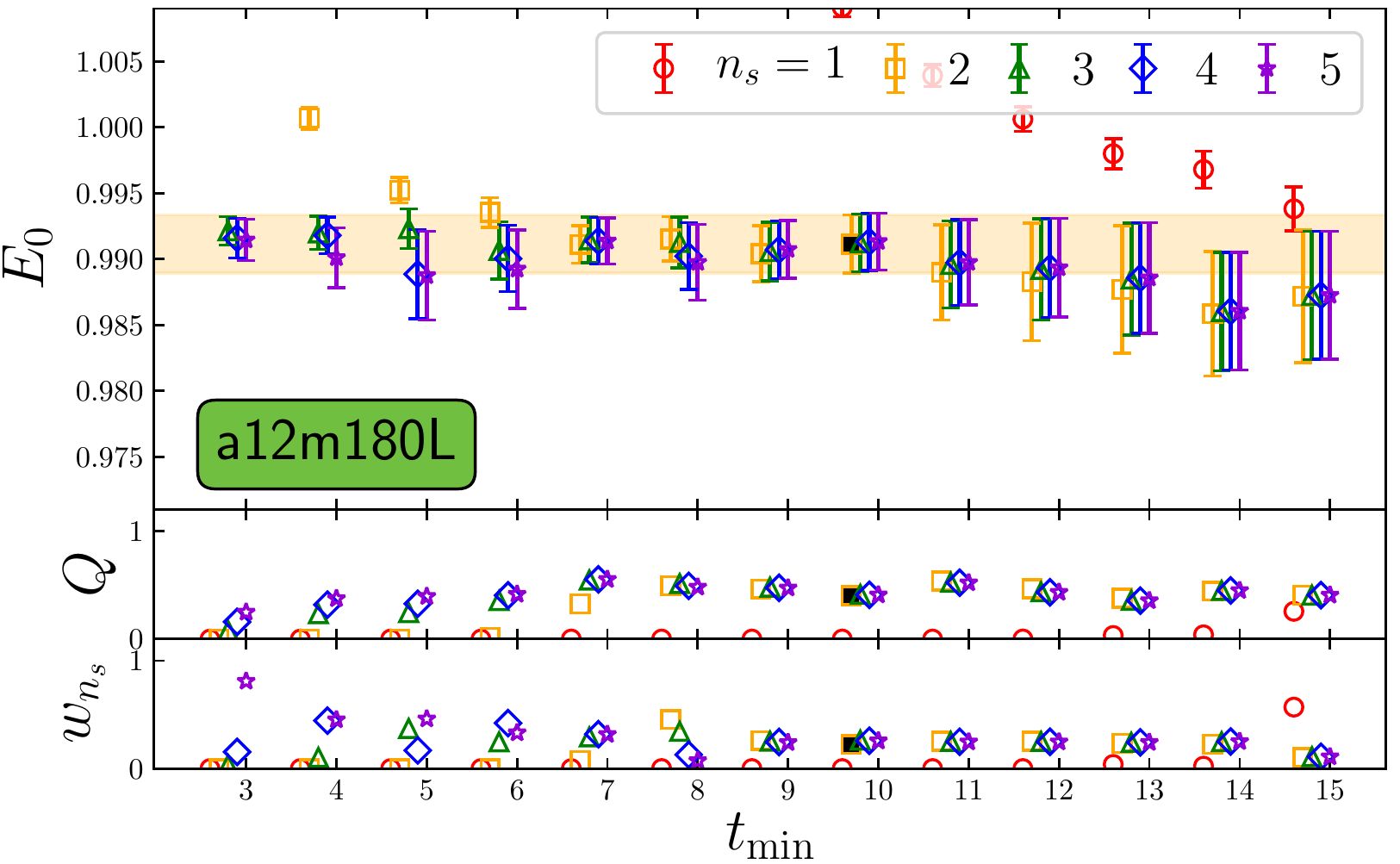}
\caption{\label{fig:stability_a12m180L}
Same as \figref{fig:stability_m135} for the a12m180L ensembles.
}
\end{figure*}

\begin{figure*}
\includegraphics[width=0.49\textwidth]{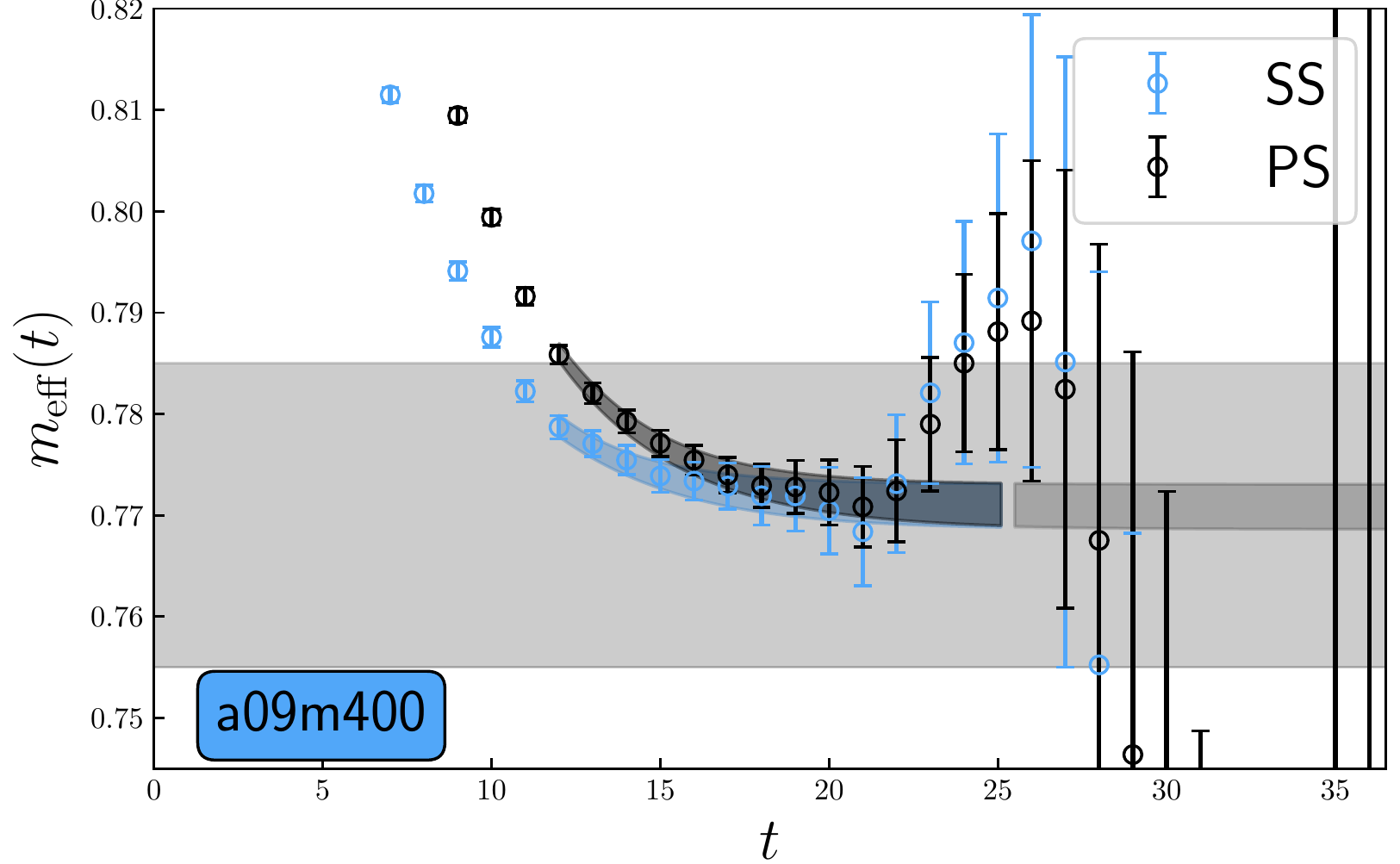}
\includegraphics[width=0.49\textwidth]{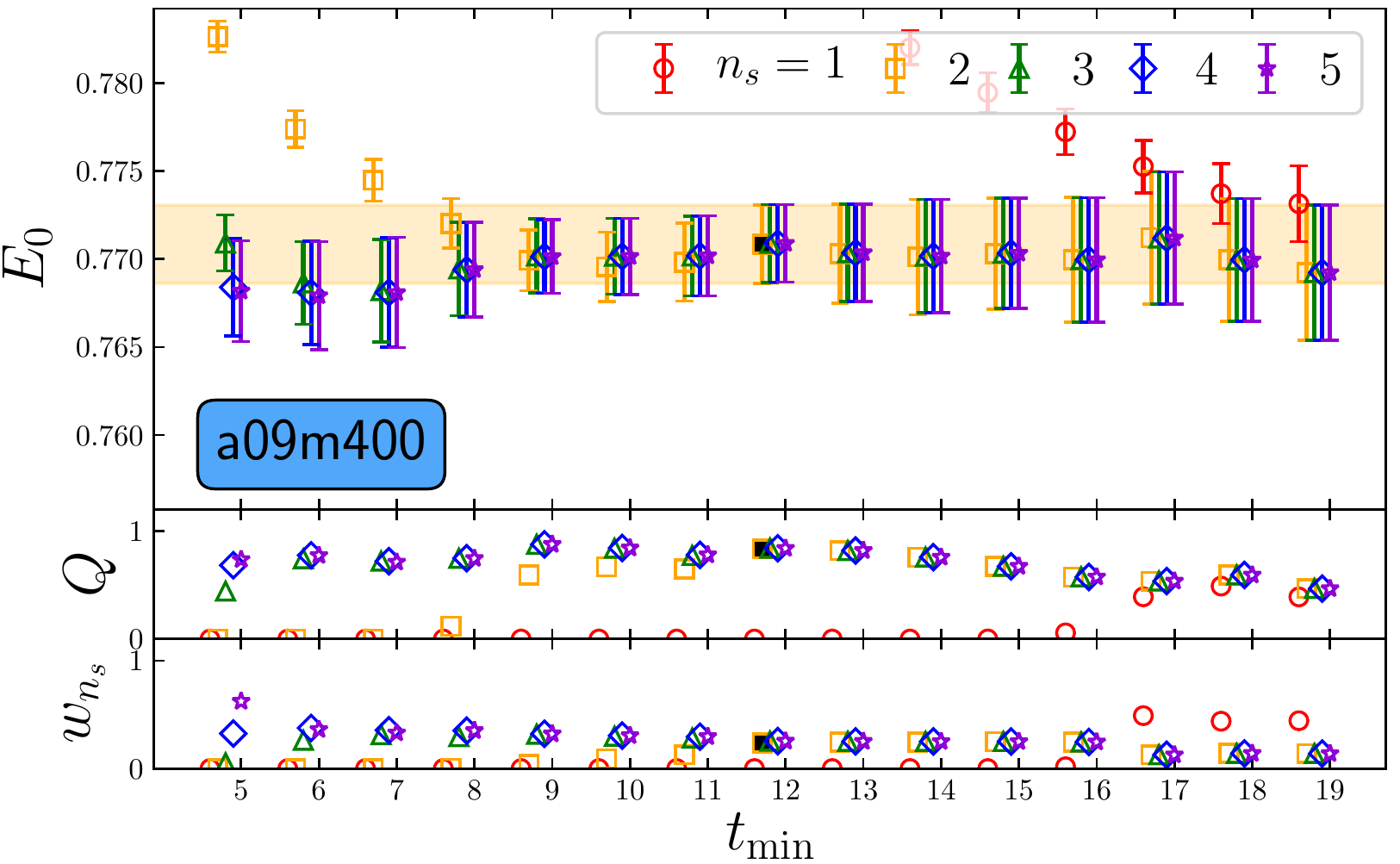}
\caption{\label{fig:stability_a09m400}
Same as \figref{fig:stability_m135} for the a09m400 ensemble.
}
\end{figure*}

\begin{figure*}
\includegraphics[width=0.49\textwidth]{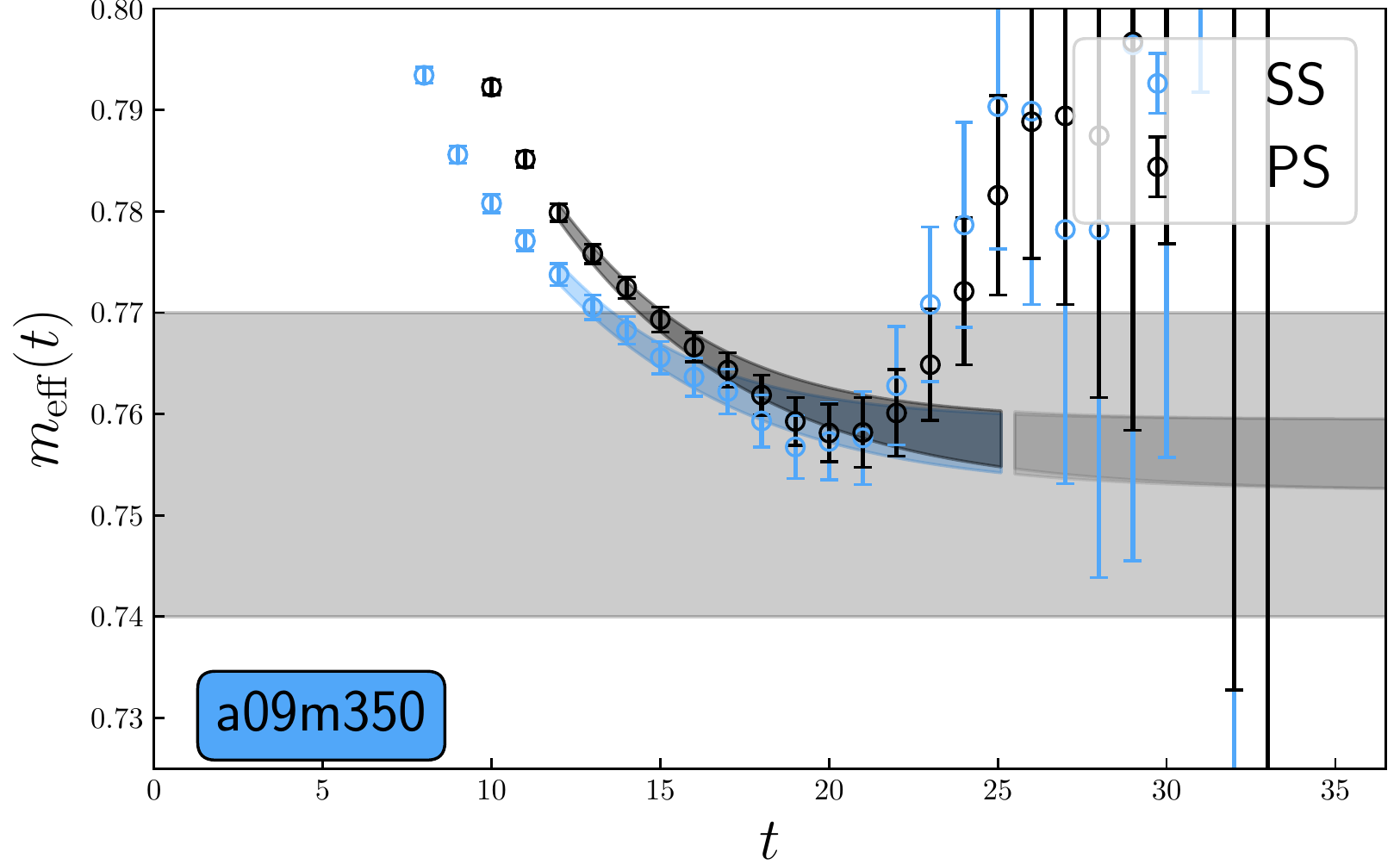}
\includegraphics[width=0.49\textwidth]{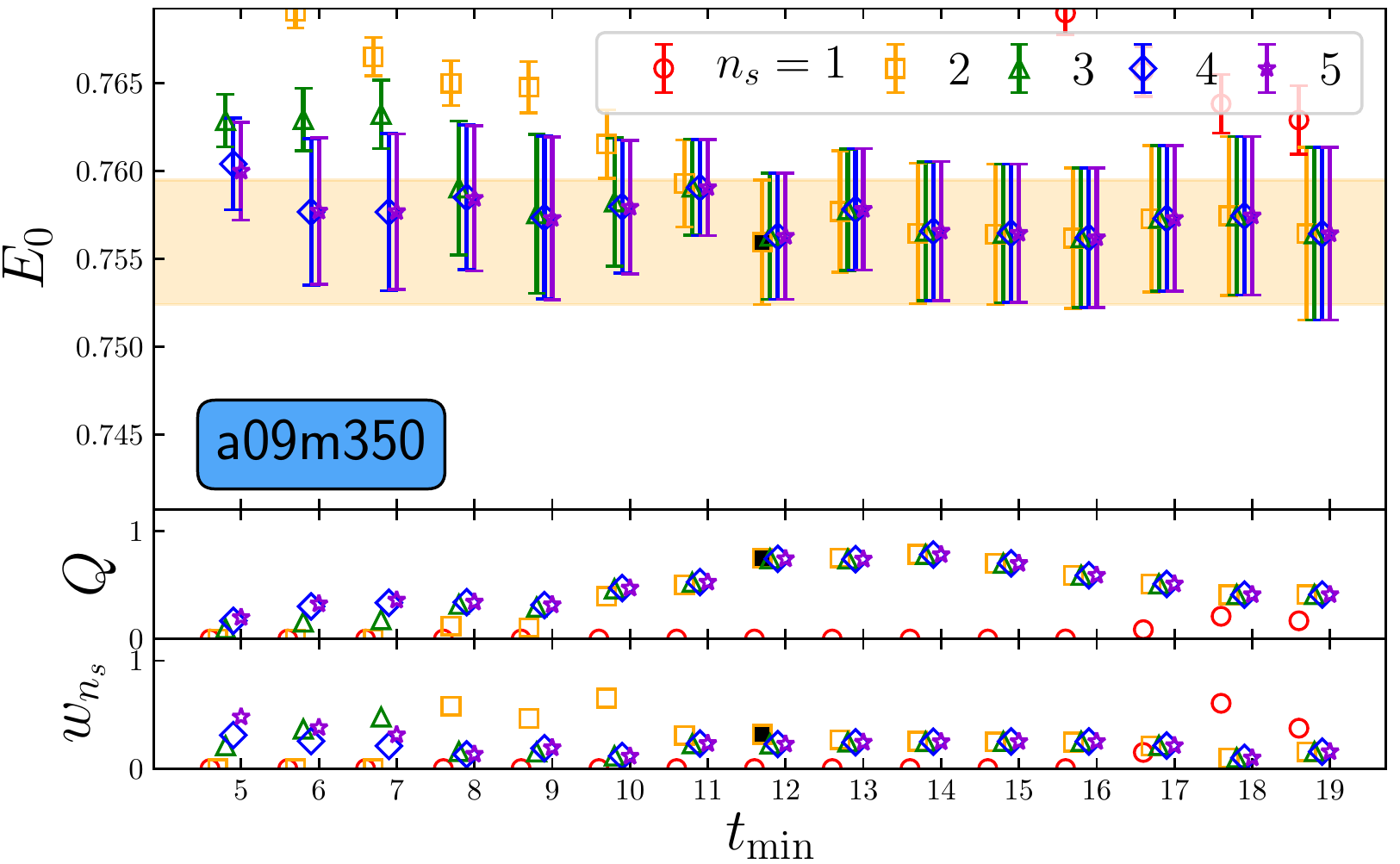}
\caption{\label{fig:stability_a09m350}
Same as \figref{fig:stability_m135} for the a09m350 ensemble.
}
\end{figure*}

\begin{figure*}
\includegraphics[width=0.49\textwidth]{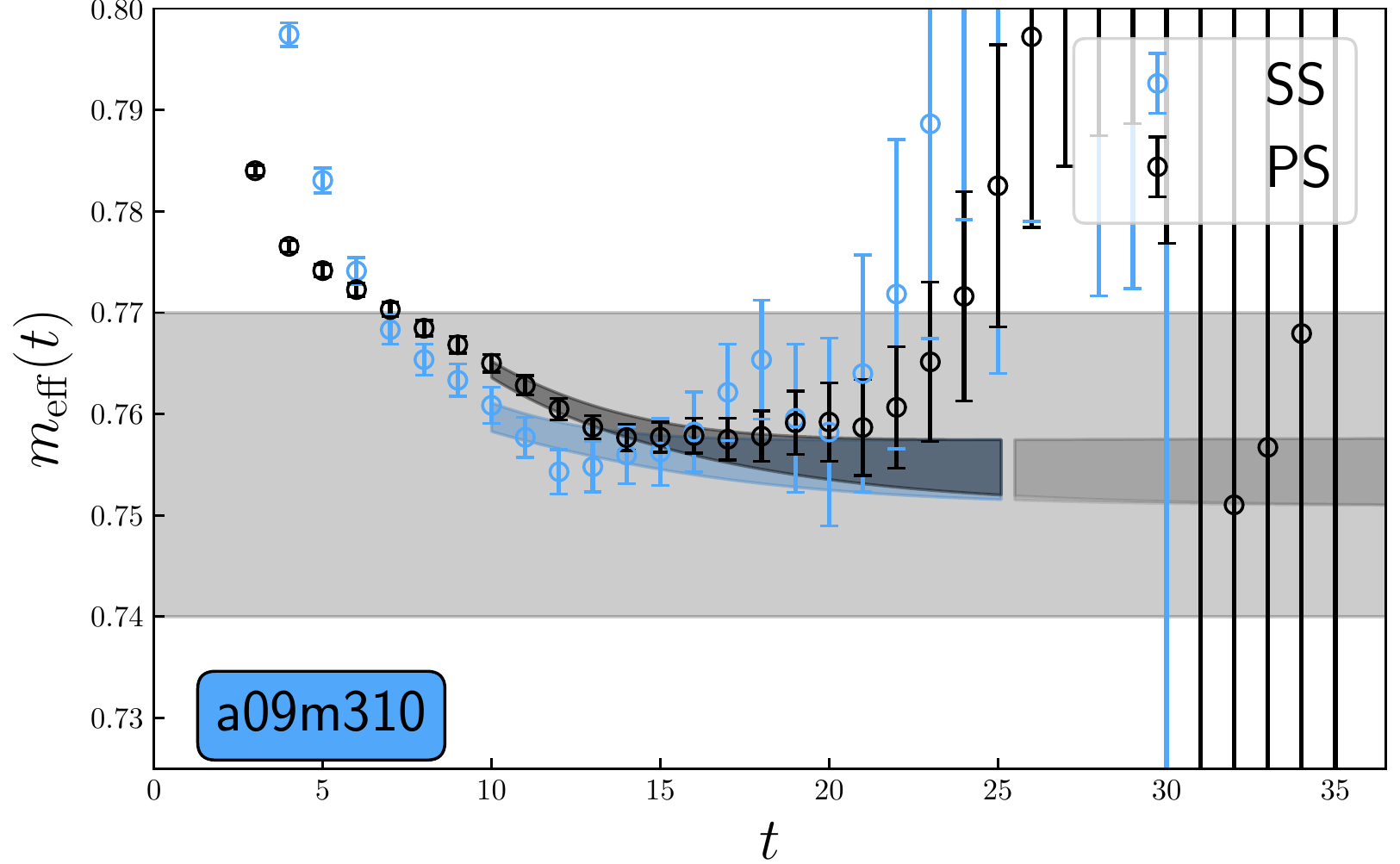}
\includegraphics[width=0.49\textwidth]{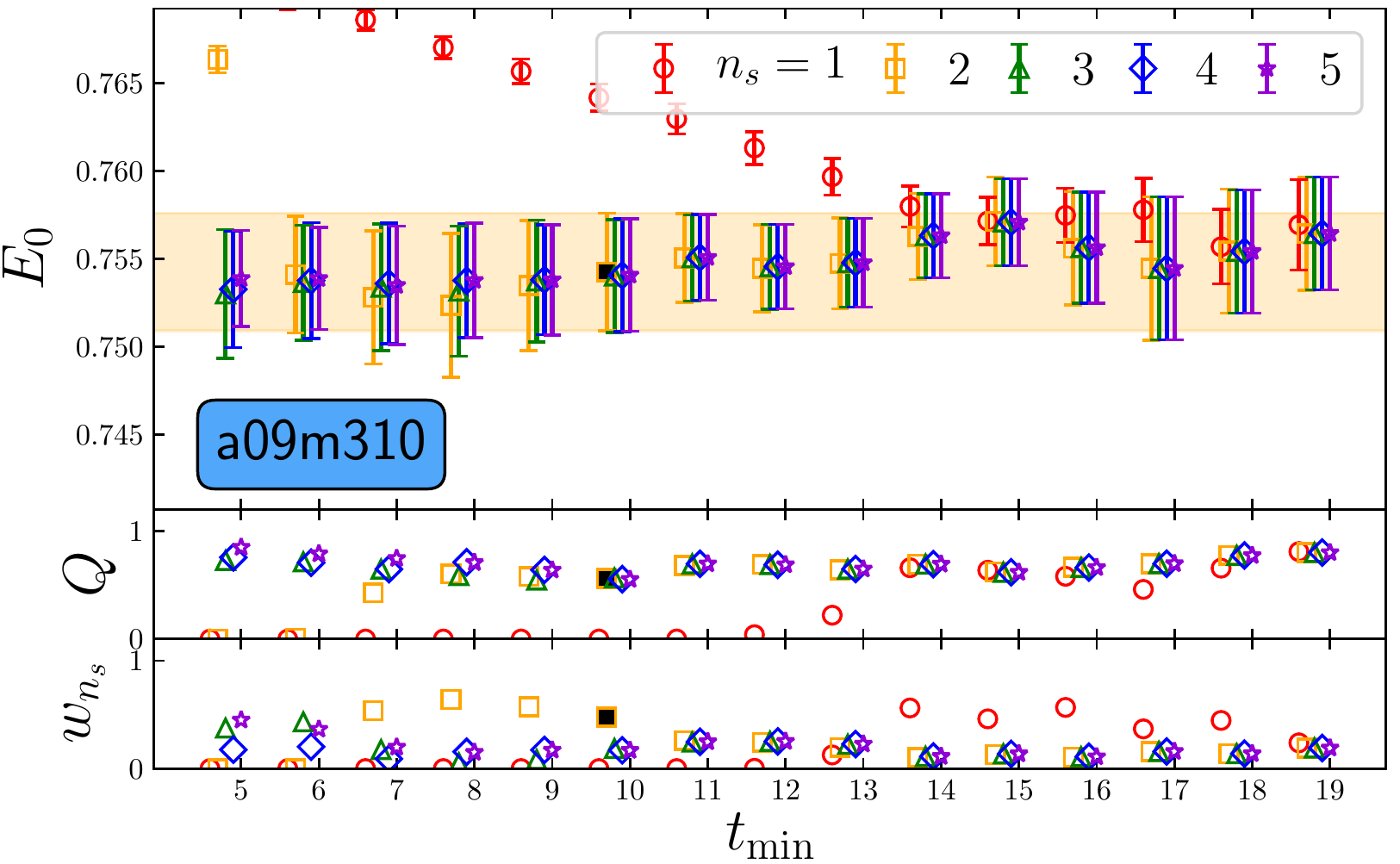}
\caption{\label{fig:stability_a09m310}
Same as \figref{fig:stability_m135} for the a09m310 ensemble.
}
\end{figure*}

\begin{figure*}
\includegraphics[width=0.49\textwidth]{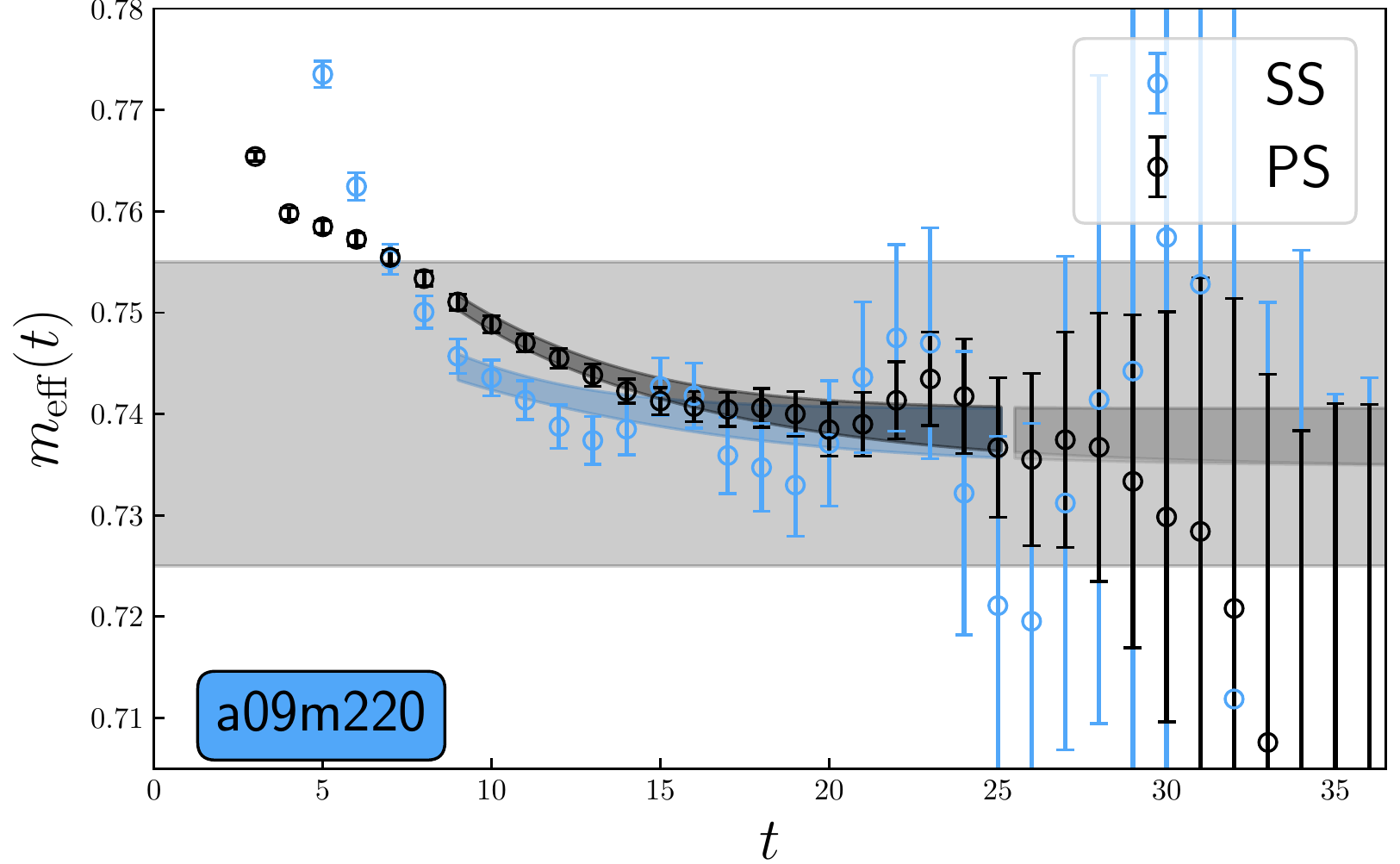}
\includegraphics[width=0.49\textwidth]{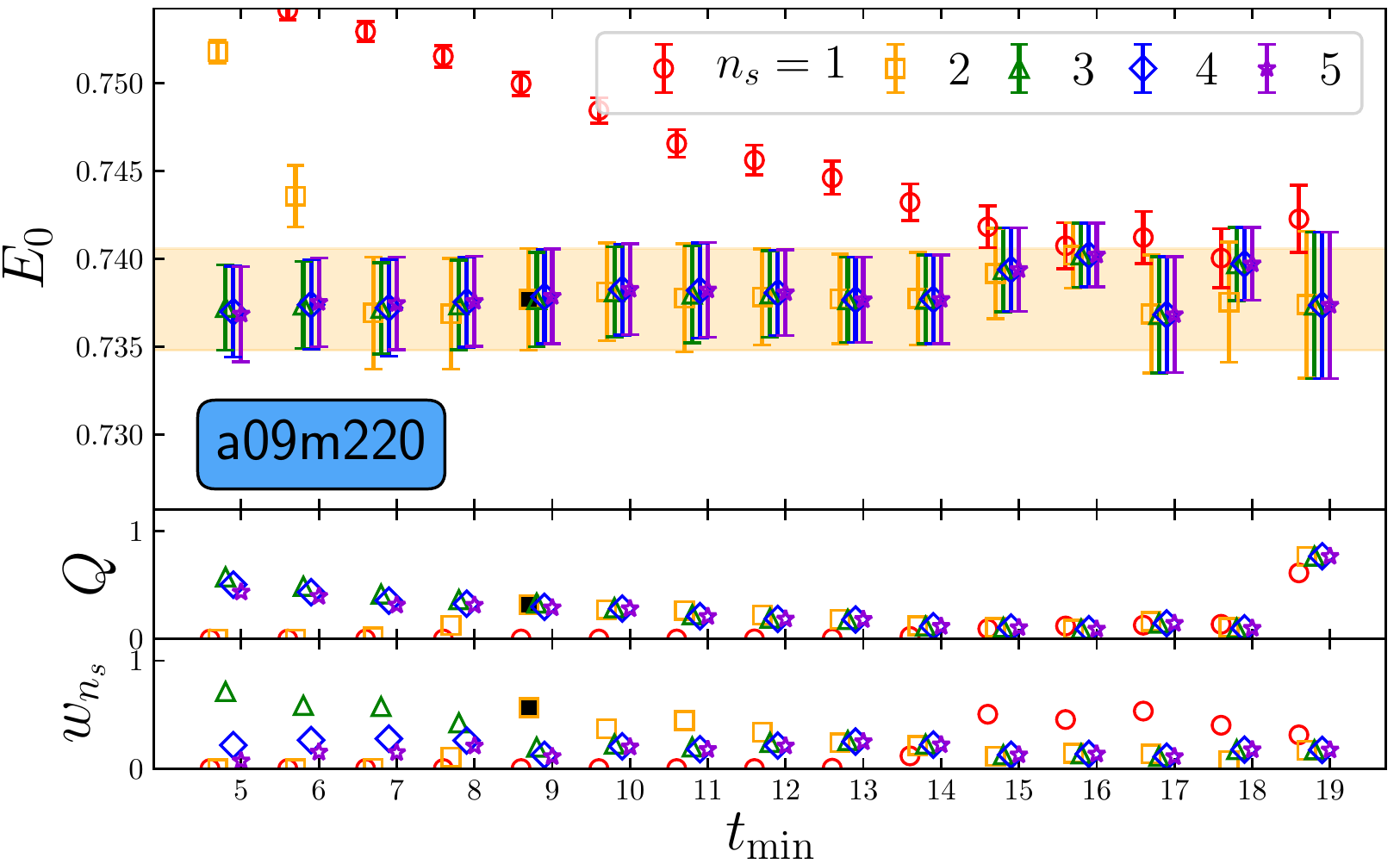}
\caption{\label{fig:stability_a09m220}
Same as \figref{fig:stability_m135} for the a09m220 ensemble.
}
\end{figure*}

Here we present the stability plots for the remaining Omega correlator fits used in our analysis, which are presented in \figref{fig:stability_a15m400}--\ref{fig:stability_a09m220}.

%-------------------------------------------------------------------------------
%\twocolumngrid

\end{document}